\documentclass[universe,article,accept,moreauthors]{Definitions/mdpi} 

\firstpage{1}
\pubvolume{12}
\issuenum{9}
\articlenumber{0}
\pubyear{2026}
\copyrightyear{2026}
\externaleditor{Tao Wu, Kuldeep Verma, Jia-Shu Niu}
\datereceived{30 May 2026} 
\daterevised{14 August 2026} % Comment out if no revised date
\dateaccepted{24 August 2026} 
\datepublished{ }

\DeclareRobustCommand{\ion}[2]{%
  \textup{#1\,\textsc{\lowercase{#2}}}%
}

\usepackage{caption}
\usepackage{xcolor}
\usepackage{soul}
\usepackage{cancel}

\newcommand{\rholeo}{\texorpdfstring{$\rho$}{rho}\,Leo}
\newcounter{letteredsub}[subsubsection]

\Title{A Decade of Radial-Velocity Monitoring of {\rholeo}: Moment~Analysis and Periodic Variability}
\Author{Vitalii Checha $^{1,}$*\orcidA{}, Anna Aret $^{1}$\orcidB{}, Indrek Kolka $^{1}$\orcidC{}, Tiina Liimets $^{1}$\orcidD{}, Veronika Mitrokhina $^{1}$\orcidE{}, Anni Kasikov $^{1}$\orcidF{}, Tõnis Eenmäe $^{1}$\orcidG{}, Sandipan P. D. Borthakur $^{1,2,3}$\orcidJ{} and Heleri Ramler $^{1}$\orcidH{}}

\AuthorNames{Vitalii Checha, Anna Aret, Indrek Kolka, Tiina Liimets, Veronika Mitrokhina, Anni Kasikov, Tõnis Eenmäe, Sandipan P. D. Borthakur and Heleri Ramler}
\address{%
$^{1}$ \quad Tartu Observatory, University of Tartu, Observatooriumi 1, 61602 T\~{o}ravere, Estonia; anna.aret@ut.ee (A.A.);  indrek.kolka@ut.ee (I.K.); tiina.liimets@ut.ee (T.L.); veronika.mitrokhina@ut.ee (V.K.); anni.kasikov@ut.ee~(A.K.); tonis.eenmae@ut.ee (T.E.); sandipan.borthakur@ut.ee (S.P.D.B.); heleri.ramler@ut.ee (H.R.)\\
$^2$ \quad Space Research Institute, Austrian Academy of Sciences, Schmiedlstrasse 6, 8042 Graz, Austria\\
$^3$ \quad Institute for Theoretical and Computation Physics, Graz University of Technology, Petersgasse 16, 8010~Graz,~Austria
}

\corres{Correspondence: vitalii.checha@ut.ee}

\abstract{We investigate the origin of long-term spectroscopic and photometric variability in the blue supergiant {\rholeo}, with particular emphasis on distinguishing between intrinsic pulsations and variability induced by a possible companion.
Our analysis is based on an 11.5-year spectroscopic time series obtained at Tartu Observatory, complemented by high-cadence, high-resolution spectroscopy from the Hertzsprung SONG telescope and space-based photometry from TESS.
We studied line-profile variability using normalised moments of the \ion{He}{I} $\lambda$6678, \ion{He}{I} $\lambda$5875, and \ion{Si}{III} $\lambda$4552 lines. Periodic signals were identified using the generalised Lomb--Scargle periodogram with iterative pre-whitening, and their temporal stability was examined with the weighted wavelet Z-transform.
We detect a persistent periodic signal at $P = 16.46$\,d in the first and third moments, present throughout the full observing interval, with a radial-velocity amplitude of 3.9\,km/s. This signal is also present in the SONG data and is visible in multiple spectral lines, indicating a global origin. Photometric observations reveal a dominant variability timescale near \hbox{$\approx$33\,d,} approximately twice the spectroscopic period.
The stable 16.46-day period, present throughout the entire observing interval, most likely results from non-radial pulsations of the supergiant. A binary origin of the signal is not excluded, but distinguishing between these scenarios is complicated by the supergiant’s complex variability pattern.}

\keyword{$\rho$ Leo; blue supergiants; stars: massive; stars: oscillations; spectroscopy; space-based photometry; time-series analysis}

\begin{document}

\section{Introduction}\label{Intro}
Asteroseismology provides a powerful framework for probing the internal structure of stars (see   \citep{2023Ap&SS.368..107B,2024ApJ...967L..39B,2024A&A...692R...1A,2025arXiv250908426S,2026arXiv260702775L,2025tasc.confE.121N} and references therein). Blue supergiants (BSGs) are of particular importance in this context, as their pulsational behaviour offers direct constraints on the internal physics and evolutionary pathways of massive stars. The relation between pulsation modes and the evolutionary stage of BSGs has been demonstrated by \citet{2013MNRAS.433.1246S}. The study of objects such as {\rholeo}, therefore, contributes to linking the observed surface variability with the underlying stellar structure and evolution.

{\rholeo} (HD~91316) is a slowly rotating B1\,Iab blue supergiant \citep{1997MNRAS.284..265H} with fundamental parameters of $M\approx 22\,M_\odot$ and $R\approx 32\,R_\odot$. In our previous work \citep{2026A&A...706A.200C}, we derived atmospheric parameters of $T_{\rm eff}= 23\,$kK, $\log g = 2.6$, and $v\sin$ i = 49\,km/s, which are consistent with earlier studies e.g., \citep{10.1111/j.1365-2966.2004.07799.x}. We also constrained the inclination angle of the star to be $21.7^\circ$, a parameter that had not been previously determined.

\textls[-15]{The stellar wind properties and mass-loss behaviour of {\rholeo} have been investigated in several studies e.g., \citep{2006A&A...446..279C,2020AstL...46..168K}, and evidence for non-radial pulsations has been reported~\citep{2016MNRAS.458.1604K,10.1093/mnras/sty308}. }
 
Line-profile variability in BSGs is generally multi-periodic, with variations on timescales from hours to weeks \citep{Cox1980,1989nos..book.....U}.  Such variations are typically associated with high-degree non-radial pulsations and with changing wind structures.  Detecting these modes reliably requires high signal-to-noise and dense time sampling, since many signals are quasi-periodic and can evolve or appear intermittently \citep{2007ARep...51..920K,2007AN....328.1170K}.

\textls[+15]{Massive stars have an exceptionally high incidence of binarity.  For example, \citet{2012Sci...337..444S} found that over 70\% of O-type stars will exchange mass with a companion during their evolution, and \citet{10.1111/j.1365-2966.2012.21317.x} reported that more than 82\% of stars above $\approx$16\,$M_\odot$ form close binary systems.  
Independent studies confirm similarly high multiplicity fractions, particularly for the most massive stars, where close binary systems dominate. The multiplicity fraction is observed to increase with stellar mass, making binarity a fundamental aspect of massive-star evolution \citep{2013ARA&A..51..269D,10.1111/j.1365-2966.2012.21317.x}. }

For evolved massive stars, such as blue and red supergiants, the observed binary fraction is typically lower, partly due to observational biases and the intrinsic difficulty of detecting companions in systems with strong stellar winds and large luminosity contrasts. Nevertheless, population studies suggest that a significant fraction of supergiants have experienced binary interaction or mergers during earlier evolutionary stages \citep{2017ApJS..230...15M}. These considerations motivate the investigation of possible binarity in objects such as {\rholeo}, where periodic variability may be influenced not only by pulsations but also by orbital motion or binary interaction.

The possibility that {\rholeo} is a binary has been discussed, for example, by \citet{2023A&A...677A.175W}. Lunar occultation in 1969 showed that star {\rholeo} is a binary \citep{1970A&A.....5..328D,1976A&A....48..245D}. Moreover, the Moon's altitude above the horizon at the moment of occultation was quite large, about 45 degrees, which supports the reliability of the result. However, \citet{2024AJ....168...28T} didn't confirm binarity based on speckle interferometry at the Southern Astrophysical Research (SOAR) Telescope.

The strength of our study lies in the systematic monitoring of {\rholeo} over an extended time span of 11.5 years, obtained with a single instrument and with good temporal sampling. This homogeneous dataset allows us to reliably investigate long-term variability patterns. We identify a persistent periodic signal with a period of 16.46 days and a radial-velocity amplitude of 3.9\,km/s. In addition, our analysis is complemented by high-resolution spectroscopy from the Stellar Observations Network Group (SONG) and space-based photometry from the Transiting Exoplanet Survey Satellite (TESS), providing an independent verification of the detected variability and a broader observational context.

\section{Observations and Data Processing}\label{S-obs}

This work is based on medium-resolution spectra obtained at Tartu Observatory (TO) over 11.5 years of monitoring, complemented by high-cadence, high-resolution spectroscopy from the SONG and space-based photometry from the TESS. The availability of contemporaneous TO and SONG observations, particularly during several overlapping nights, allows us to directly compare the results and assess the reliability and consistency of the derived variability patterns. Spectroscopic and photometric time series are presented in Appendix~\ref{timeseries}. A complete log of the TO observations is provided in Appendix~\ref{S-appendixobs}.

\subsection{TO Spectroscopy}\label{S-specobs}
We used the 1.5-m AZT-12 telescope at Tartu Observatory for long-term spectroscopic monitoring of {\rholeo}. We monitored the variability of the \ion{He}{I} 6678.151\,\AA\ line, a strong, unblended photospheric line in blue supergiants. The long-slit spectrograph ASP-32 mounted at the Cassegrain focus was employed with a diffraction grating of 1800\, lines\,mm$^{-1}$ \citep{2022A&A...658A.105F}. This setup provided spectra covering the wavelength range 6300--6730\,\AA{} with a spectral resolving power of R $\approx$ 10\,000. Typical exposure time about 300\,s, with values ranging from 120 to 800\,s, resulting in spectra with signal-to-noise ratio $S/N\approx300$--400.

Observations were carried out during 163 nights between 2014 and 2025, yielding a total of approximately 3500 spectra. Many of these nights were consecutive, with up to 80~spectra obtained per night, providing an excellent opportunity to investigate variability on a wide range of timescales. Individual observing sequences typically lasted between 2 and 4.5 h per night.

The data reduction was performed using the \textsc{IRAF v2.16}\endnote{IRAF was written at the National Optical Astronomy Observatory, which was operated by the Association of Universities for Research in Astronomy (AURA) under cooperative agreement with the National Science Foundation. IRAF Community Distribution is available at \url{https://iraf-community.github.io/} (accessed on 26.08.2026).} software package \cite{1986SPIE..627..733T,1993ASPC...52..173T} and standard procedures from the \textsc{noao}, \textsc{imred}, and \textsc{ccdred} packages. 
Wavelength calibration was carried out using ThAr lamp spectra obtained before and after the science exposures. The stability of the wavelength scale during each observing run was additionally verified using telluric absorption lines in the vicinity of 6570\,\AA. Heliocentric velocity corrections were applied to all spectra. The continuum normalisation was performed by fitting a linear function to the spectrum. 

\subsection{SONG Spectroscopic Observations}\label{SONG-spec}
We also used archival spectra obtained with the Hertzsprung SONG telescope and spectrograph at Observatorio del Teide located on Tenerife, Spain. We downloaded the SONG data, reduced with the standard pipeline, from the SODA archive\endnote{SODA archive \url{https://soda.phys.au.dk/} (accessed on 26.08.2026).}. The spectral regions of interest were extracted, corrected for heliocentric velocity, and normalised. The spectra were taken during the same period as our observations in the 2017 Season. A total of 488 spectra were collected between 5~January~2017 and 27~May~2017. The spectra cover the wavelength range 4400--6900\,\AA{}, with $S/N \approx 300$ and a spectral resolution of R~$\approx$~77\,000 and 90\,000. This allows us to investigate spectral variability using a larger set of spectral lines that form at different depths in the stellar atmosphere. However, \ion{He}{I} $\lambda$6678 line monitored at TO is not usable in the SONG spectra because it lies at the edge of a spectral order. Instead, we analysed \ion{He}{I} $\lambda$5876, which forms at a similar photospheric depth. Additionally, we examined the strong, clean \ion{Si}{III} $\lambda$4552 line, which probes different atmospheric layers and thus provides complementary information on stellar variability.

It is also important that the observations obtained at the SONG Observatory are contemporaneous with our TO data. Therefore, we also include them in our frequency analysis. The high spectral resolution of SONG enables detailed study of subtle line-profile variations, while the TO dataset ensures long-term temporal coverage.

\subsection{TESS Photometry}\label{tess-obs}
    We used space-based photometric observations of the {\rholeo} obtained by TESS in sectors 45 and 46 (from 6~November to 30~December~2021), and 72 (from 11~November to 7~December~2023) in the short-cadence mode (2 min exposures), for a total of about 52\,000 measurements covering 760 days. The corresponding light curves were extracted from the MAST\endnote{TESS data at MAST \url{https://archive.stsci.edu/missions-and-data/tess} (accessed on 26.08.2026) .} archive using standard pipeline ``SPOC'' \cite{2016SPIE.9913E..3EJ}, providing high-precision photometric time-series suitable for variability studies.

    It is important to note that these three sectors were observed with different TESS CCDs. As a consequence, the instrumental characteristics (i.e., sensitivity and response function) are not strictly identical between the sectors. This prevents a direct combination of the raw flux measurements for the purpose of analysing long-term variability across all sectors simultaneously, as potential offsets and systematic differences may introduce spurious~trends.

    To mitigate this effect, we normalised each sector independently by its mean flux level and subsequently combined the normalised light curves into a single dataset. This approach allows us to perform a consistent period analysis across all three sectors. However, such normalisation removes information about long-term variations and trends, making the dataset unsuitable for detecting long-period signals spanning multiple sectors. On the other hand, the analysis of short- and intermediate-period variability remains unaffected by this procedure. Therefore, the combined normalised light curve is well suited for identifying periodicities on timescales shorter than the individual sector lengths.

\section{Methods}\label{S-method}
\subsection{Characterisation of Spectral Lines}

To characterise the variability of the line profiles, we computed the first four line profile moments $M_0$--$M_3$ following the formalism introduced by \citep{1986MNRAS.219..111B,1986MNRAS.220..647B,1987MNRAS.224...41B} and further developed by \citet{2010aste.book.....A}:
\begin{align}
M_0 &= \sum_{i=1}^{N} (1 - F_i) \Delta v_i~, \label{eq:1} \\
M_1 &= \sum_{i=1}^{N} (1 - F_i)(v_i - v_0) \Delta v_i~, \label{eq:2} \\
M_2 &= \sum_{i=1}^{N} (1 - F_i)(v_i - v_0)^2 \Delta v_i~, \label{eq:3} \\
M_3 &= \sum_{i=1}^{N} (1 - F_i)(v_i - v_0)^3 \Delta v_i ~, \label{eq:4}
\end{align}
where $F_i$ is normalised flux value at wavelength $\lambda_i$, $v_i$ is the velocity corresponding to $\lambda_i$ with respect to the laboratory wavelength of the line and $\Delta v_i \equiv v_i - v_{i-1}$. The velocity zero point is taken to be $v_0 = 42$\,km/s, corresponding to the stellar systemic velocity determined by \citet{https://doi.org/10.1002/asna.200710776}. The summation interval was chosen as $\pm150$\,km/s relative to the centre of the line profile for each individual spectrum.
The zeroth moment $M_0$ corresponds to the equivalent width (EW), the first moment $M_1$ corresponds to the centroid of the line profile and traces radial-velocity variations, the second moment $M_2$ characterises the line width, and the third moment $M_3$ describes the line asymmetry (skewness). 

Although the choice of the velocity zero point does not affect the frequency analysis of the first and second moments, it is important for the third moment. If the third moment is calculated with respect to a fixed velocity zero point, a shift of the line profile produces a corresponding change in $M_3$, even when the shape of the profile remains unchanged. In other words, the third moment calculated about a fixed reference point is sensitive not only to changes in the line-profile asymmetry but also to radial-velocity variations traced by $M_1$. To isolate variations in the line-profile shape, we, therefore, calculate the central third moment, with the reference velocity set to the instantaneous centroid velocity $v_{\rm c}=M_1/M_0$ of the line profile for each individual observation:
\begin{align}
M_3 &= \sum_{i=1}^{N} (1 - F_i)(v_i - v_{\rm c})^3 \Delta v_i ~. \label{eq:4a}
\end{align}

We obtain the observed normalised moments $\langle v^j \rangle$ for j = 1, \ldots, 3 as $M_j /M_0$. These moments have velocity units (km/s)$^j$. 
A combined analysis of all moments enables the investigation of both persistent and transient components of variability within a unified~framework.

\subsection{Temporal Variability}

We analysed the temporal variability of the normalised spectroscopic moments and the TESS light curve using the generalised Lomb--Scargle (GLS) periodogram \citep{2009A&A...496..577Z}, Lomb--Scargle analysis with iterative pre-whitening (GLSp), implemented using the \textsc{LombScargle} class of the \textsc{astropy} Python package \citep{2022ApJ...935..167A}, and the weighted wavelet Z-transform (WWZ) \citep{1996AJ....112.1709F,2023Galax..11...69A,2024A&A...689A..35C,2025MNRAS.tmp.1744W}. The three methods provide complementary information on the observed variability. GLS is well suited to unevenly sampled data and provides a global characterisation of the dominant frequency content, but assumes that the detected signals remain coherent throughout the analysed time interval. GLSp extends this approach through iterative pre-whitening, allowing weaker frequency components to be recovered after removal of the dominant signals. However, the extracted frequencies may be affected by aliasing and by imperfect subtraction of closely spaced or non-stationary components. In contrast, WWZ provides a time--frequency representation that enables us to trace the persistence and relative strength of individual frequency components and to identify transient or quasi-periodic variability. Its interpretation may, however,  be complicated by sparse or highly non-uniform sampling and depends on the adopted balance between temporal and frequency resolution. Thus, GLS and GLSp are used to characterise the global frequency content and identify its individual components, whereas WWZ reveals their temporal~evolution.

\subsubsection{Generalised Lomb--Scargle}
The Lomb--Scargle method transforms an unevenly sampled time series into a power spectrum using the Lomb--Scargle periodogram. The method was introduced by Lomb in 1976 \citep{1976Ap&SS..39..447L} and further developed by Scargle in 1982 \citep{1982ApJ...263..835S}. Although the Lomb--Scargle periodogram can be viewed as a decomposition of the data into sinusoidal components, it is mathematically equivalent to a least-squares fit of sinusoidal models to the observations. The generalised Lomb--Scargle (GLS) periodogram was introduced by \citet{2009A&A...496..577Z}. Compared with the classical Lomb--Scargle formulation, GLS accounts for measurement uncertainties and includes a floating constant term in the sinusoidal fit, providing a more robust framework for the frequency analysis of unevenly sampled~observations.

For each dataset, the most significant frequencies were identified from the GLS power spectrum and visually inspected to verify that they were not artefacts of the time sampling. Their statistical significance was assessed using the Lomb--Scargle false-alarm probability (FAP), adopting a global FAP threshold of 0.1\%, estimated independently using both the analytic Baluev approximation and bootstrap resampling with 5000 iterations. Peaks exceeding the corresponding 0.1\% FAP significance level were considered statistically significant. The GLS periodograms were computed using the standard normalisation implemented in the \textsc{astropy.timeseries.LombScargle} routine; consequently, the periodogram power is expressed as a dimensionless normalised quantity.

\subsubsection{Lomb--Scargle with Iterative Prewhitening}

The generalised Lomb--Scargle periodogram with iterative pre-whitening (GLSp) is used to identify multiple periodic components in datasets containing more than one significant frequency. 
In such cases, the strongest component may dominate the periodogram, making weaker periodicities difficult to detect. Pre-whitening is an iterative procedure in which the dominant frequency is identified from the periodogram, modelled with a sinusoidal function, and subtracted from the data. The periodogram is then recomputed from the residuals to identify the strongest remaining peak. In our analysis, seven successive pre-whitening steps were applied to all~datasets.

The sinusoidal model after $N$ pre-whitening steps has the form:
\begin{align}
y_{\rm fit}(t)=\sum_{i=1}^{N} \label{eq:5}
A_i\sin\left(2\pi f_i t+\phi_i\right),
\end{align}
where $A_i$ is the amplitude, $f_i$ is the frequency, and $\phi_i$ is the phase of the $i$th component. The successive pre-whitening steps allow weaker periodic components that are initially masked by stronger variability to be revealed. The extracted GLSp components are not necessarily interpreted as individually statistically significant independent frequencies; rather, the procedure is used to characterise the complex frequency content of the observed variability and to identify recurrent periodicities for comparison between different diagnostics and~datasets.

\subsubsection{Weighted Wavelet Z-Transform}
The weighted wavelet Z-transform (WWZ) was introduced by \citet{1996AJ....112.1709F}. WWZ is a time-frequency analysis method that describes how the frequency content of a signal evolves with time. Unlike classical periodograms, which provide only the global frequency content of a dataset, WWZ produces a time-frequency map in which the wavelet power is evaluated as a function of both time and frequency. The method represents the signal locally by fitting sinusoidal wavelets weighted by a Gaussian window. As the window slides along the time axis, observations closest to its centre receive the largest weights, allowing us to track the evolution of the local spectral power with time. The balance between temporal and frequency resolution is controlled by the adopted decay constant and the time sampling. WWZ is, therefore, particularly well suited to identifying transient, evolving, or quasi-periodic signals and to tracking changes in their relative strength over time. In this work, WWZ complements the GLS and GLSp analyses by revealing the temporal evolution of the detected frequency components.

\section{Results}\label{S-res}
\subsection{Analysis of the Spectroscopic Data}\label{S-specana}

\subsubsection{Long-Term TO Data}\label{S-spec-All}
The long-term TO dataset comprises 3492 spectra collected during 163 nights between 21 January 2014 and 1 June 2025.
Such an extended time span with reasonably good coverage within each observing season allows us to probe a wide range of periods in our analysis. We are able to investigate variability in the star over timescales ranging from as short as 15 min to as long as 4150 days (the full duration of the TO spectroscopic observations). However, the stellar type must be taken into account--{\rholeo} is a quasi-periodic $\alpha$\,-Cygni-type variable. This implies that we do not expect stable periods persisting over such long timescales. Instead, we anticipate periods that appear and disappear, with their values evolving over time. Moreover, we show that different periods can mask one another, preventing individual periods from being clearly distinguished in phase diagrams.

\begin{figure}[H]
%    \isPreprints{\centering}{}
    \includegraphics[width=1\linewidth]{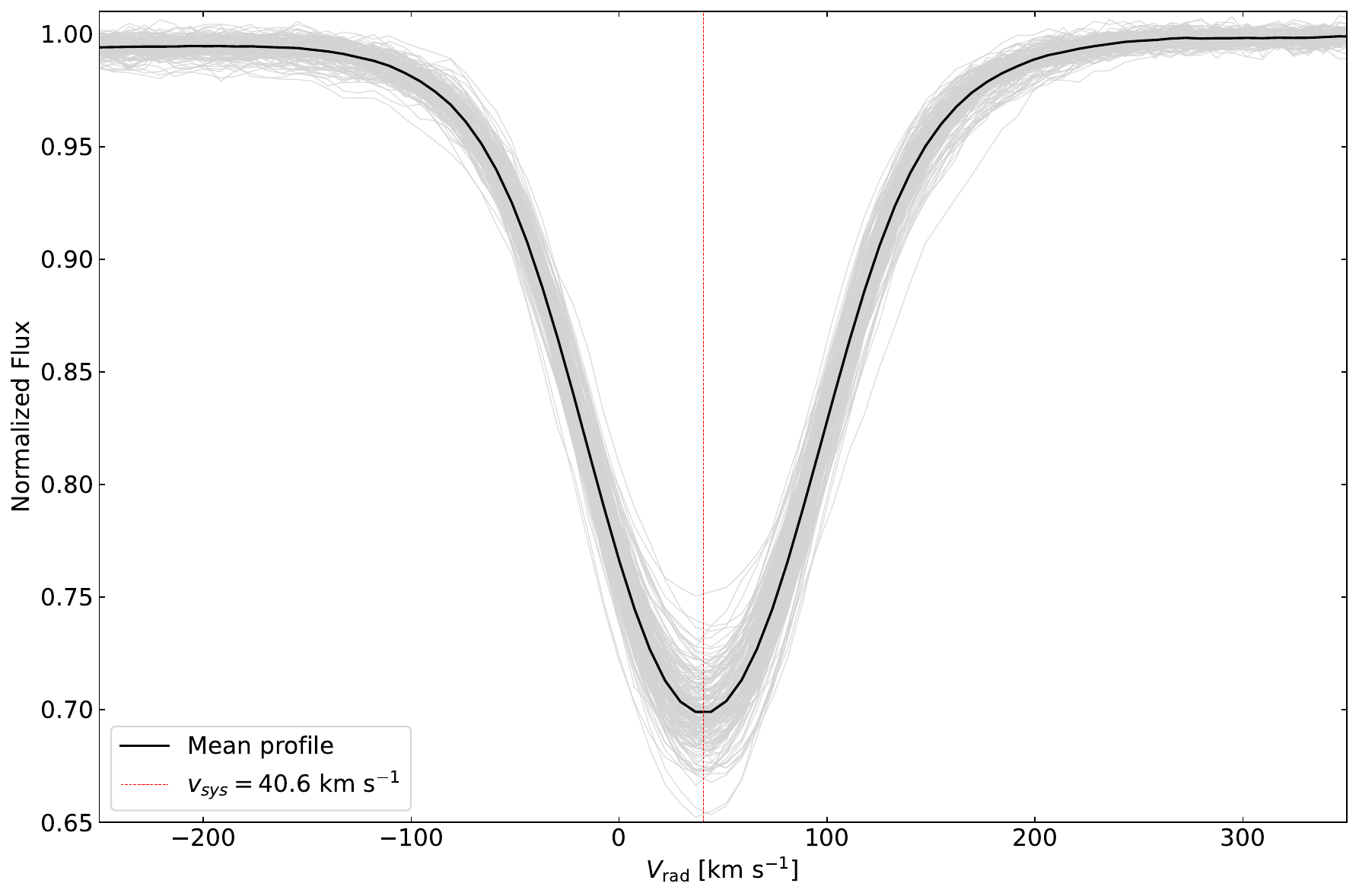}
    \caption{Variability of the \ion{He}{I} $\lambda$6678 line profile during the 11.5-year TO observing period. The black solid line shows the mean line profile. To illustrate the overall line-profile variability, the first observed spectrum from each of the 163 observing nights is overplotted in grey. The red dotted line marks the systemic radial velocity determined in this work from the mean value of the first moment.}
    \label{fig:l_p_12}
\end{figure}

\begin{figure}[H]
%    \isPreprints{\centering}{}
    \includegraphics[width=0.8\linewidth]{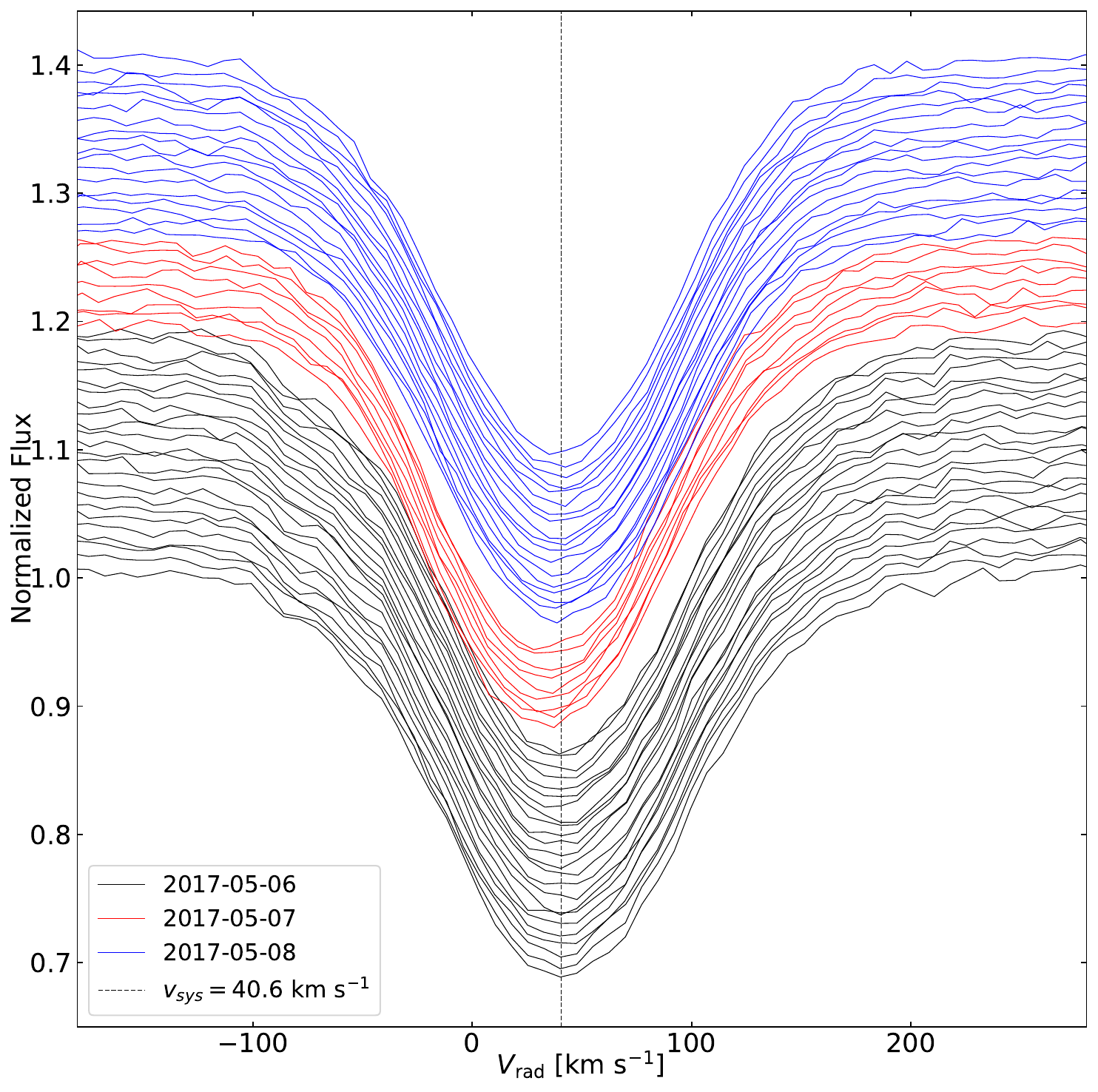}
    \caption{Variability of the \ion{He}{I} $\lambda$6678 line profile during three consecutive TO observing nights. For clarity, successive spectra are vertically offset. The vertical dashed line marks the systemic radial velocity determined in this work from the mean value of the first moment.}
    \label{fig:l_p_3}
\end{figure}

\begin{figure}[H]
%    \isPreprints{\centering}{}
    \includegraphics[width=1\linewidth]{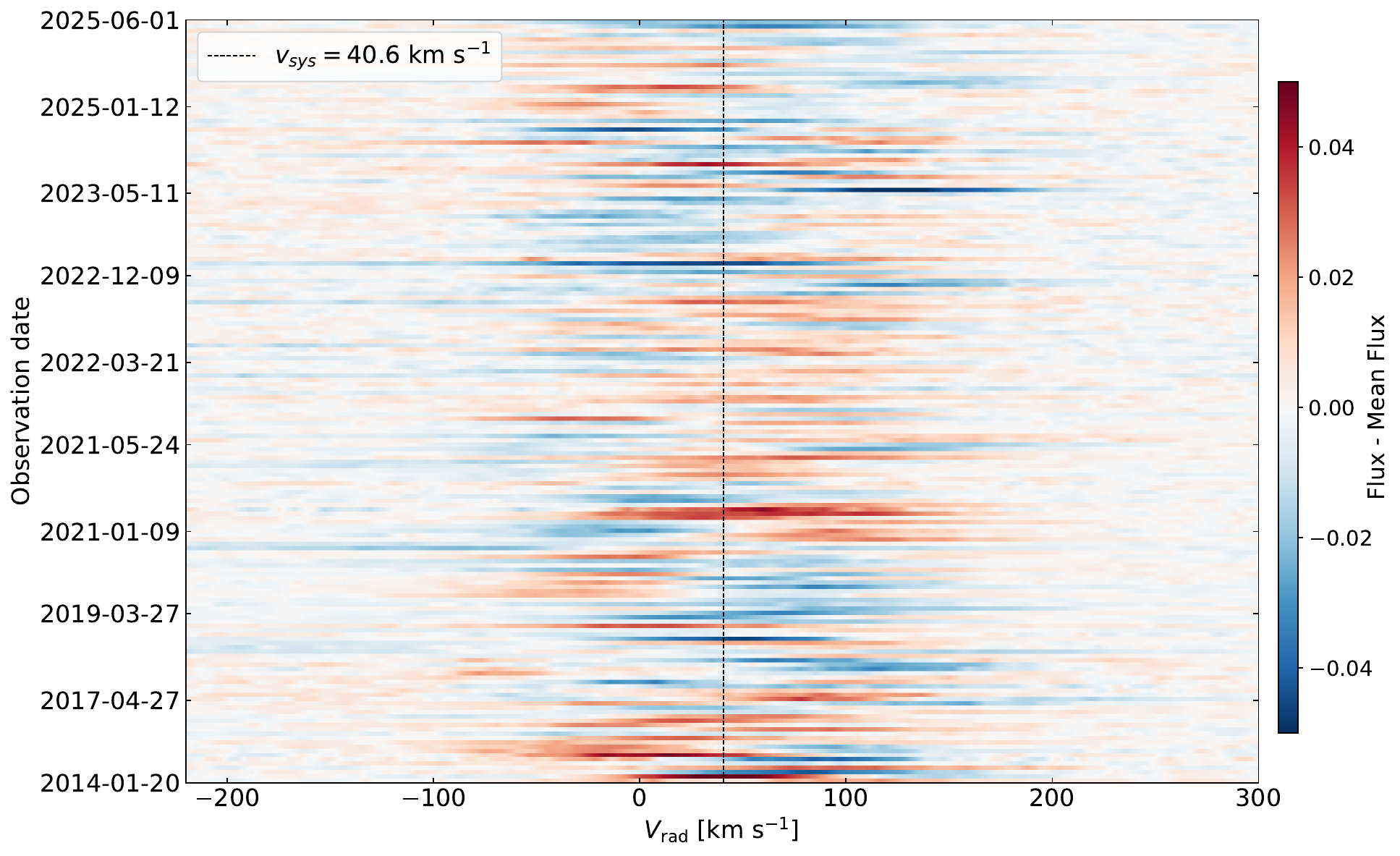}
    \caption{Residual dynamic spectrum of the line-profile variability of the photospheric \ion{He}{I} $\lambda6678$ line during the 11.5-year TO observing period. The horizontal axis shows Doppler velocity relative to the line centre, while time increases upwards. Note that the observing dates are not equally spaced. To illustrate the long-term variability, the first spectrum from each observing night is included. Residual profiles are calculated by subtracting the time-averaged profile from each normalised line profile. Blue and red colours indicate residual flux below and above the average profile, respectively.}
    \label{fig:dyn_l_p_12}
\end{figure}

\textls[-25]{The variability of the \ion{He}{I} $\lambda$6678 line profile is illustrated in Figures\,\ref{fig:l_p_12}--\ref{fig:dyn_l_p_12}. Figure\,\ref{fig:l_p_12} shows the mean line profile together with one spectrum from each observing night, providing an overview of the variability over the entire 11.5-year time span. Figure\,\ref{fig:l_p_3} presents a stacked sequence of spectra obtained during three consecutive nights, illustrating the short-term line-profile variability. The long-term temporal evolution of the residual line profiles is shown as a residual dynamic spectrum in Figure\,\ref{fig:dyn_l_p_12}. The figures illustrate the changes in the line-profile shape that are investigated quantitatively in this work through an analysis of the line-profile moments.}

\vspace{-9pt}

\begin{figure}[H]
\subfloat[First moment $\langle v^1 \rangle$]{
\includegraphics[width=0.8\linewidth]{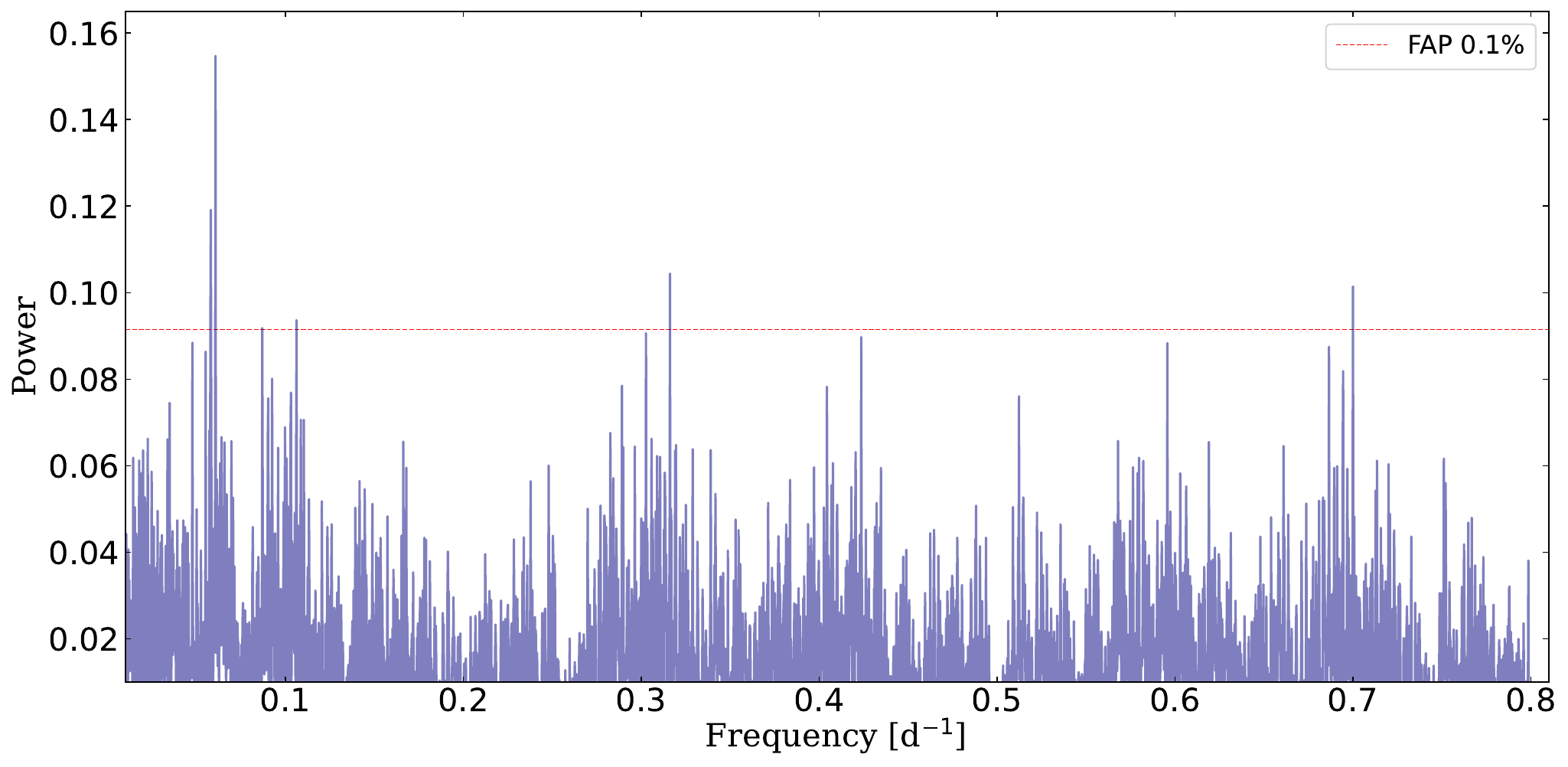}\label{fig:GLS_M1_all}}

\subfloat[Second moment $\langle v^2 \rangle$]{
\includegraphics[width=0.8\linewidth]{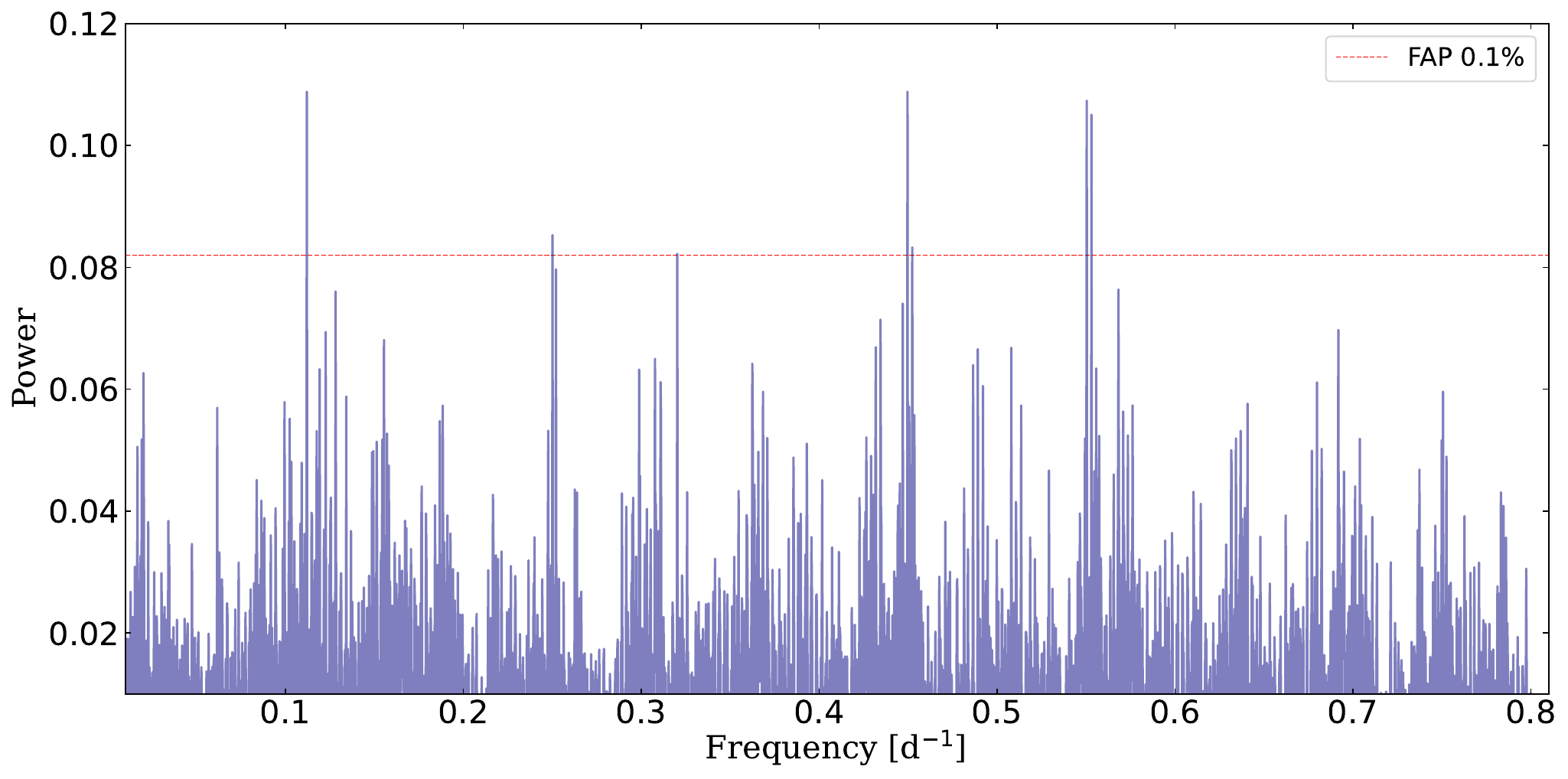}\label{fig:GLS_M2_all}}

\subfloat[Central third  moment $\langle v^3 \rangle$]{
\includegraphics[width=0.8\linewidth]{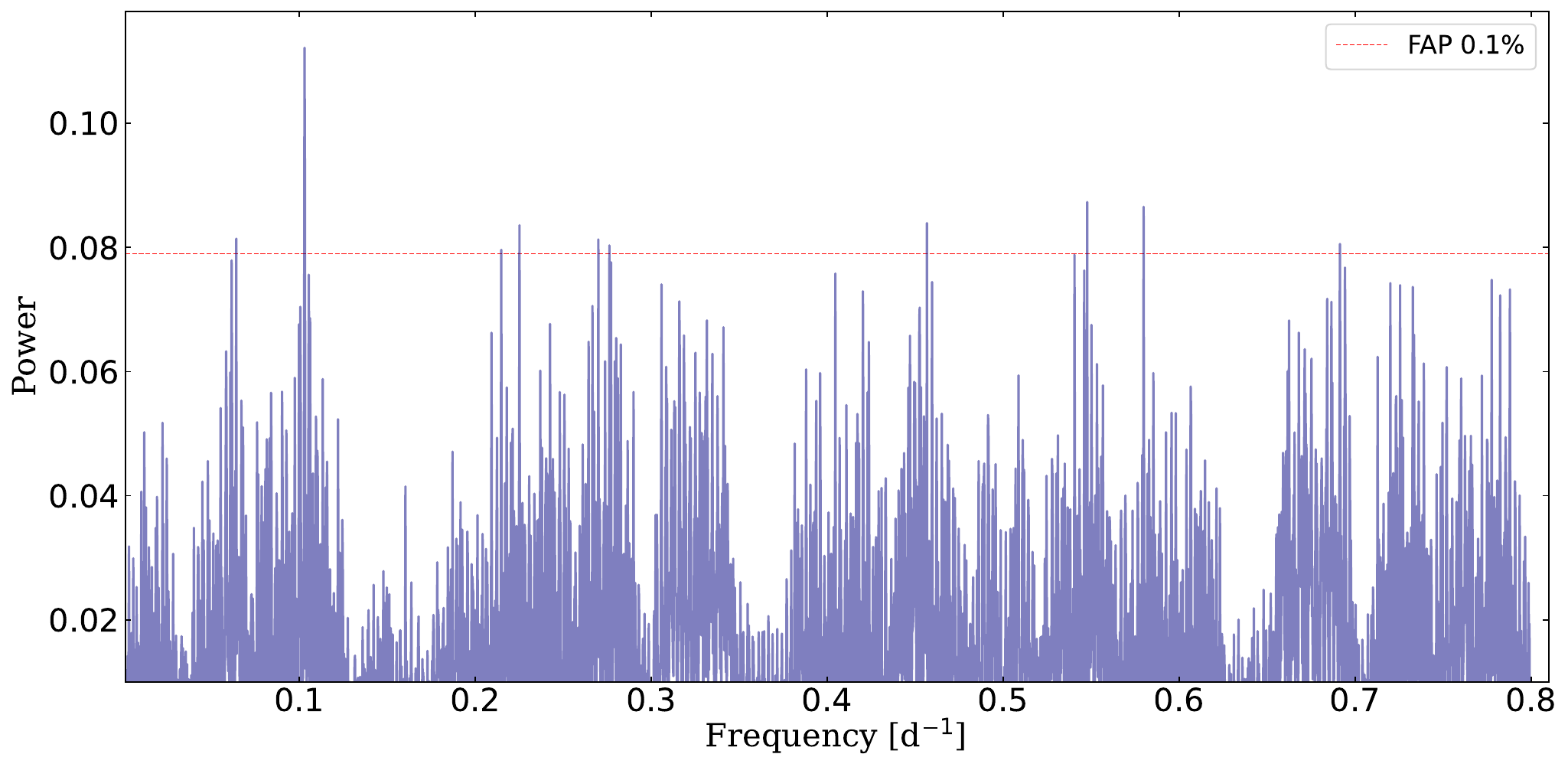}\label{fig:GLS_M3_all}}
\caption{GLS periodograms of the first three moments based on the full 11.5-year TO monitoring~dataset. 
\label{fig:GLS_M123_all}}
\end{figure}

First, the GLS method was applied. The power spectra of the first three normalised moments are presented in Figure~\ref{fig:GLS_M123_all} and the corresponding periods in Tables~\ref{tab:M1_all}--\ref{tab:M3_all}. No significant peaks were detected in the 0.8--1.25\,d$^{-1}$ frequency range. Therefore, the figure presents a detailed view of the dominant frequencies below 0.8~d$^{-1}$ together with the corresponding FAP level.

The first prominent feature of the periodogram is the large number of peaks with varying power. Given the long time span of the observations and the intrinsic variability of this star, such a complex frequency spectrum is expected. Several peaks exceed the stringent FAP = 0.1\% level, demonstrating statistically significant variability on multiple timescales. The first and third moments both show appreciable power at low frequencies, whereas the frequency distribution of the second moment is markedly different, with less low-frequency power and a set of more isolated, narrow high-frequency peaks. A distinctive feature is a signal at $F\approx 0.11$ d$^{-1}$ ($P\approx 8.9$ d), which is not present in the first and third moments. Several prominent peaks around 0.45 d$^{-1}$ and 0.55 d$^{-1}$ occur in more than one moment and may be related to daily aliasing. \ion{He}{I} $\lambda$6678.

Despite the continuous appearance and disappearance of different periodicities within individual observing seasons, one dominant frequency ($F\approx0.0607$ d$^{-1}$, $P\approx16.46$ d) clearly stands out in the first moment and persists throughout the entire 11.5-year time base. A similar period, $P\approx16.26$ d, is detected in the third moment after prewhitening the dominant $\approx9.7$-d periodicity. 
In contrast, no corresponding signal is detected in the second moment. Thus, the $\sim$16-d variability is manifested most prominently as changes in the line centroid, with a weaker signature in the line asymmetry, while it is not apparent in the line width.

\begin{figure}[H]
\subfloat[First normalised moment $\langle v^1 \rangle$ (radial velocity)]{
\includegraphics[width=0.95\linewidth]{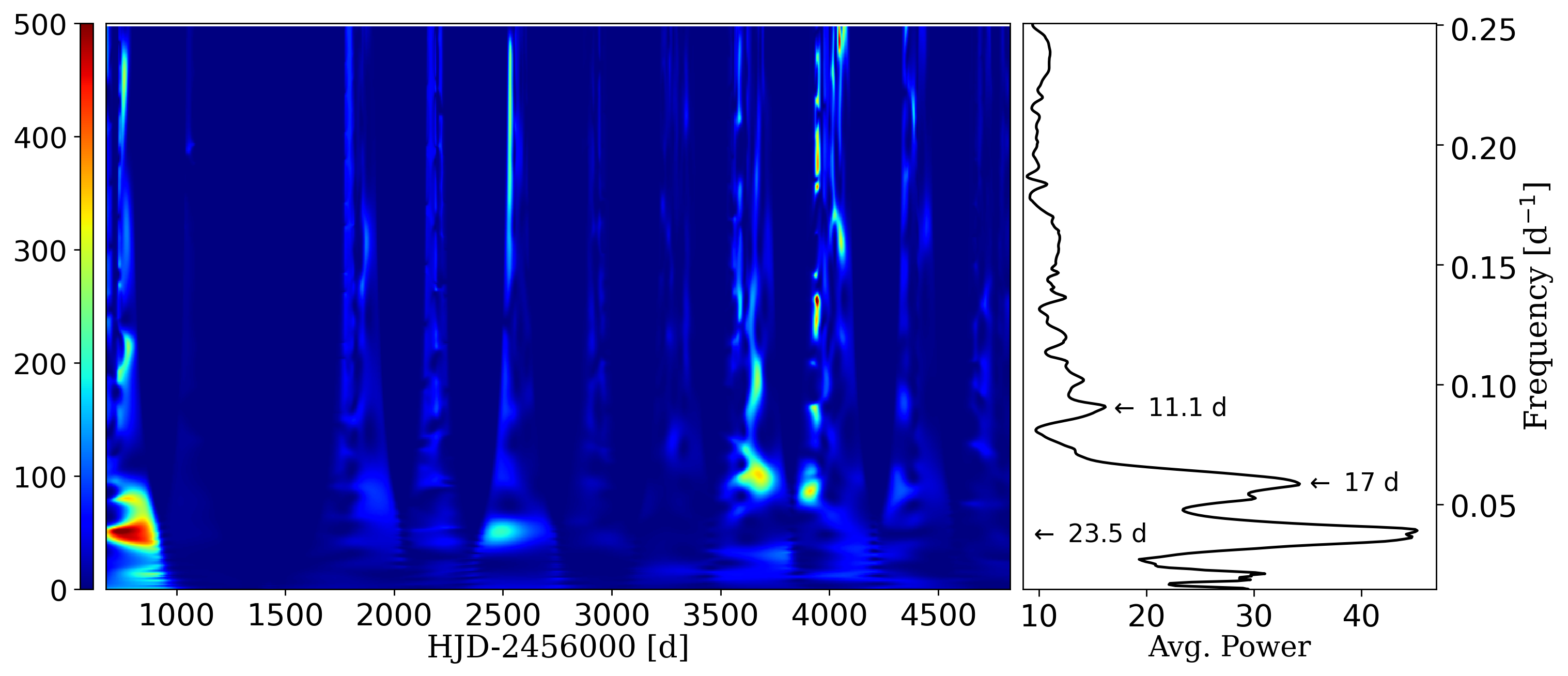}
\label{fig:WWZ_M1_all}}

\vspace{-5pt}
\subfloat[Second normalised moment $\langle v^2 \rangle$ (width)]{
\includegraphics[width=0.95\linewidth]{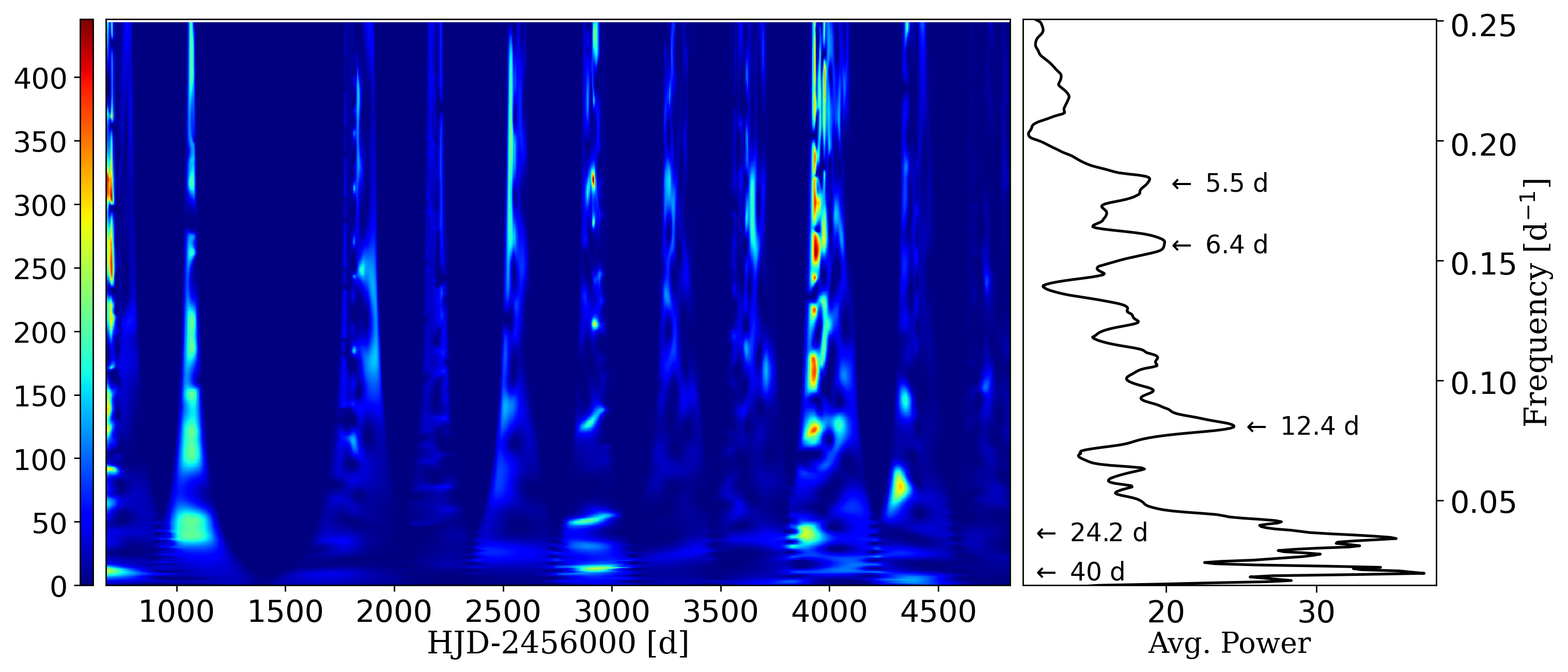}
\label{fig:WWZ_M2_all}}

\vspace{-5pt}
\subfloat[Third normalised central moment $\langle v^3 \rangle$ (skewness)]{
\includegraphics[width=0.95\linewidth]{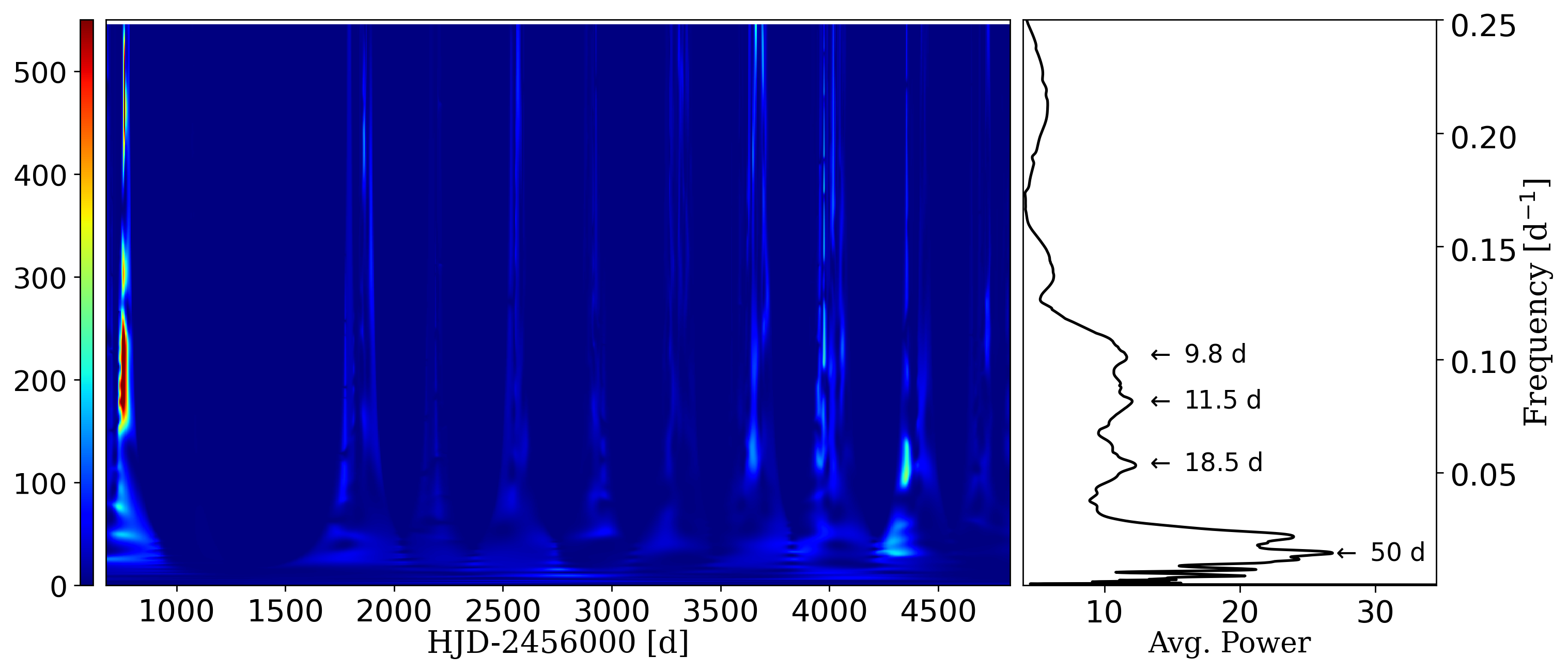}
\label{fig:WWZ_M3_all}}

\caption{WWZ analysis of the full TO dataset for the first three moments of the \ion{He}{I} $\lambda$6678 line. The left panels show the time-frequency map with colour-coded wavelet power. The black solid line in the right panels shows the time-averaged wavelet power, with the most significant periods indicated. The right-hand frequency axis applies to both panels.}

\label{fig:WWZ_all_moments}
\end{figure}

Lastly, we applied the WWZ analysis to investigate the temporal evolution of the detected frequencies and to identify quasi-periodic variability. The period range was set from 4 to 2000 days, as this interval was found to be the most informative for this type of analysis. The WWZ scalograms for the full 11.5-year dataset, computed for the first three normalised moments $\langle v^1 \rangle$, $\langle v^2 \rangle$, and $\langle v^3 \rangle$, are shown in Figure~\ref{fig:WWZ_all_moments}a--c. Because the long gaps between observing seasons reduce the visibility of the intra-seasonal behaviour in the full-dataset scalograms, we additionally present season-by-season WWZ analyses in Appendix~\ref{wwz-all}. These include the first three normalised moments for the densely sampled 2022 observing season and the first moment for all observing seasons, allowing the temporal evolution of the detected frequency components to be examined on intra-seasonal~timescales.

\begin{table}[H]
\setlength{\tabcolsep}{2.75mm}
\caption{Periods detected in the full TO spectroscopic data based on the \ion{He}{I} $\lambda$6678 line; the first moment $\langle v^1 \rangle$ (centroid radial velocity). Results of the analysis are arranged by decreasing period in each column. The number in brackets [N] denotes the rank of the period by detected power. Uncertainties smaller than 0.01 are shown as 0.00.}\label{tab:M1_all}

\isPreprints{\centering}{} 

\begin{tabularx}{\textwidth}{ccccccc}
\toprule

\multicolumn{3}{c}{GLS} &
\multicolumn{4}{c}{GLSp} \\

\cmidrule(lr){1-3}
\cmidrule{4-7}

N & Frequency & Period & N & Frequency & Period & Amplitude \\
& [d$^{-1}$] & [d] &
& [d$^{-1}$] & [d] & [km/s] \\

\midrule

$F_1$ & 0.0026 [6] & $377.22 \pm 34.30$ &
$F_{p1}$ & 0.0226 [4] & $44.15 \pm 0.47$ & $2.35 \pm 0.13$ \\

$F_2$ & 0.0581 [2] & $17.22 \pm 0.30$ &
$F_{p2}$ & 0.0434 [5] & $23.05 \pm 0.13$ & $2.22 \pm 0.22$ \\

$F_3$ & 0.0607 [1] & $16.46 \pm 0.08$ &
$F_{p3}$ & 0.0607 [1] & $16.46 \pm 0.08$ & $3.94 \pm 0.19$ \\

$F_4$ & 0.0870 [7] & $11.49 \pm 0.10$ &
$F_{p4}$ & 0.1663 [2] & $6.01 \pm 0.01$ & $2.77 \pm 0.18$ \\

$F_5$ & 0.1063 [5] & $9.41 \pm 0.00$ &
$F_{p5}$ & 0.2480 [6] & $4.03 \pm 0.00$ & $1.98 \pm 0.24$ \\

$F_6$ & 0.3027 [4] & $3.30 \pm 0.00$ &
$F_{p6}$ & 0.7511 [7] & $1.33 \pm 0.00$ & $1.94 \pm 0.23$ \\

$F_7$ & 0.3162 [3] & $3.16 \pm 0.00$ &
$F_{p7}$ & 0.7000 [3] & $1.43 \pm 0.00$ & $2.60 \pm 0.33$ \\

\bottomrule
\end{tabularx}

\end{table}

\vspace{-9pt}
\begin{table}[H]
\setlength{\tabcolsep}{3.05mm}
\caption{Periods detected in the second moment $\langle v^2 \rangle$ (width) in the full TO dataset based on the \ion{He}{I} $\lambda$6678 line using GLS and GLSp analysis. Results of the analysis are arranged by decreasing period in each column. The number in brackets [N] denotes the rank of the period by detected power. Uncertainties smaller than 0.01 are shown as 0.00.}\label{tab:M2_all}

\isPreprints{\centering}{} 

\begin{tabularx}{\textwidth}{ccccccc}
\toprule

\multicolumn{3}{c}{GLS} &
\multicolumn{4}{c}{GLSp} \\

\cmidrule(lr){1-3}
\cmidrule{4-7}

N & Frequency & Period  & N & Frequency & Period & Amplitude \\

& [d$^{-1}$] & [d] &
& [d$^{-1}$] & [d] &
$10^{3}\,(\mathrm{km/s})^2$ \\
\midrule
$F_1$ & 0.1121 [1] & $8.92 \pm 0.02$ &
$F_{p1}$ & 0.0202 [3] & $49.50 \pm 0.59$ & $119 \pm 10$ \\

$F_2$ & 0.2500 [4] & $4.00 \pm 0.02$ &
$F_{p2}$ & 0.0342 [6] & $29.24 \pm 0.06$ & $~~89 \pm 10$ \\

$F_3$ & 0.3203 [7] & $3.12 \pm 0.00$ &
$F_{p3}$ & 0.1121 [1] & $8.92 \pm 0.02$ & $149 \pm 11$ \\

$F_4$ & 0.4497 [2] & $2.22 \pm 0.00$ &
$F_{p4}$ & 0.1884 [5] & $5.31 \pm 0.01$ & $98 \pm 5$ \\

$F_5$ & 0.4524 [5] & $2.21 \pm 0.00$ &
$F_{p5}$ & 0.4496 [2] & $2.22 \pm 0.00$ & $144 \pm 13$ \\

$F_6$ & 0.5531 [3] & $1.81 \pm 0.00$ &
$F_{p6}$ & 0.6316 [4] & $1.58 \pm 0.00$ & $105 \pm 8~~$ \\

$F_7$ & 0.5683 [6] & $1.76 \pm 0.00$ &
$F_{p7}$ & 0.6946 [7] & $1.44 \pm 0.00$ & $86 \pm 9$ \\

\bottomrule
\end{tabularx}

\end{table}

\vspace{-9pt}
\begin{table}[H]
\setlength{\tabcolsep}{2.95mm}
\caption{Periods detected in the full TO spectroscopic dataset for the central third moment $\langle v^3 \rangle$ (skewness). Results of the analysis are arranged by decreasing period in each column. The number in brackets [N] denotes the rank of the period by detected power. Uncertainties smaller than 0.01 are shown as 0.00.}
\label{tab:M3_all}

\isPreprints{\centering}{}

\begin{tabularx}{\textwidth}{ccccccc}
\toprule

\multicolumn{3}{c}{GLS} &
\multicolumn{4}{c}{GLSp} \\

\cmidrule(lr){1-3}
\cmidrule{4-7}

N & Frequency & Period  & N & Frequency & Period & Amplitude \\

& [d$^{-1}$] & [d] &
& [d$^{-1}$] & [d] &
$10^{3}\,(\mathrm{km/s})^2$ \\
\midrule

$F_1$ & 0.0641 [6] & $15.60 \pm 0.07$ &
$F_{p1}$ & 0.0221 [7] & $45.17 \pm 0.50$ & $5.0 \pm 0.6$ \\

$F_2$ & 0.1029 [1] & $9.72 \pm 0.02$ &
$F_{p2}$ & 0.0614 [2] & $16.29 \pm 0.06$ & $5.4 \pm 0.4$ \\

$F_3$ & 0.2251 [5] & $4.44 \pm 0.00$ &
$F_{p3}$ & 0.0786 [5] & $12.71 \pm 0.10$ & $4.5 \pm 0.5$ \\

$F_4$ & 0.2699 [7] & $3.71 \pm 0.00$ &
$F_{p4}$ & 0.1029 [1] & $9.72 \pm 0.02$ & $6.8 \pm 0.7$ \\

$F_5$ & 0.4567 [4] & $2.19 \pm 0.00$ &
$F_{p5}$ & 0.1680 [4] & $5.95 \pm 0.01$ & $2.3 \pm 0.3$ \\

$F_6$ & 0.5477 [2] & $1.83 \pm 0.02$ &
$F_{p6}$ & 0.2696 [6] & $3.71 \pm 0.00$ & $2.0 \pm 0.3$ \\

$F_7$ & 0.5798 [3] & $1.72 \pm 0.00$ &
$F_{p7}$ & 0.4596 [3] & $2.18 \pm 0.00$ & $6.3 \pm 0.8$ \\

\bottomrule
\end{tabularx}

\end{table}

\paragraph*{First Moment}

The first moment $\langle v^1 \rangle$ traces radial velocity variations.
The period of $\approx$16.5 days is of particular interest, as it is detected by all three methods of frequency analysis and appears to be the most prominent signal (Table~\ref{tab:M1_all}). It is clearly visible in the frequency spectrum (Figure~\ref{fig:GLS_M123_all}a) and produces a well-defined phase-folded diagram (Figure~\ref{fig:M1andM3-phase-third}), indicating a high degree of temporal stability. The low-amplitude period of 377.22 days in Table~\ref{tab:M1_all} should not be interpreted as a physical signal. It is an observational artefact introduced by the one-year gaps between consecutive observing seasons. The period $F_{p1} \approx 0.02\,\mathrm{d^{-1}}$ ($P\approx44.2$ d) is also likely an artefact resulting from imperfect removal of multiple periodic components, as it appears at low amplitude only after sequential prewhitening and is absent from the original periodogram.

\begin{figure}[H]
%    \isPreprints{\centering}{}
    \includegraphics[width=1\linewidth]{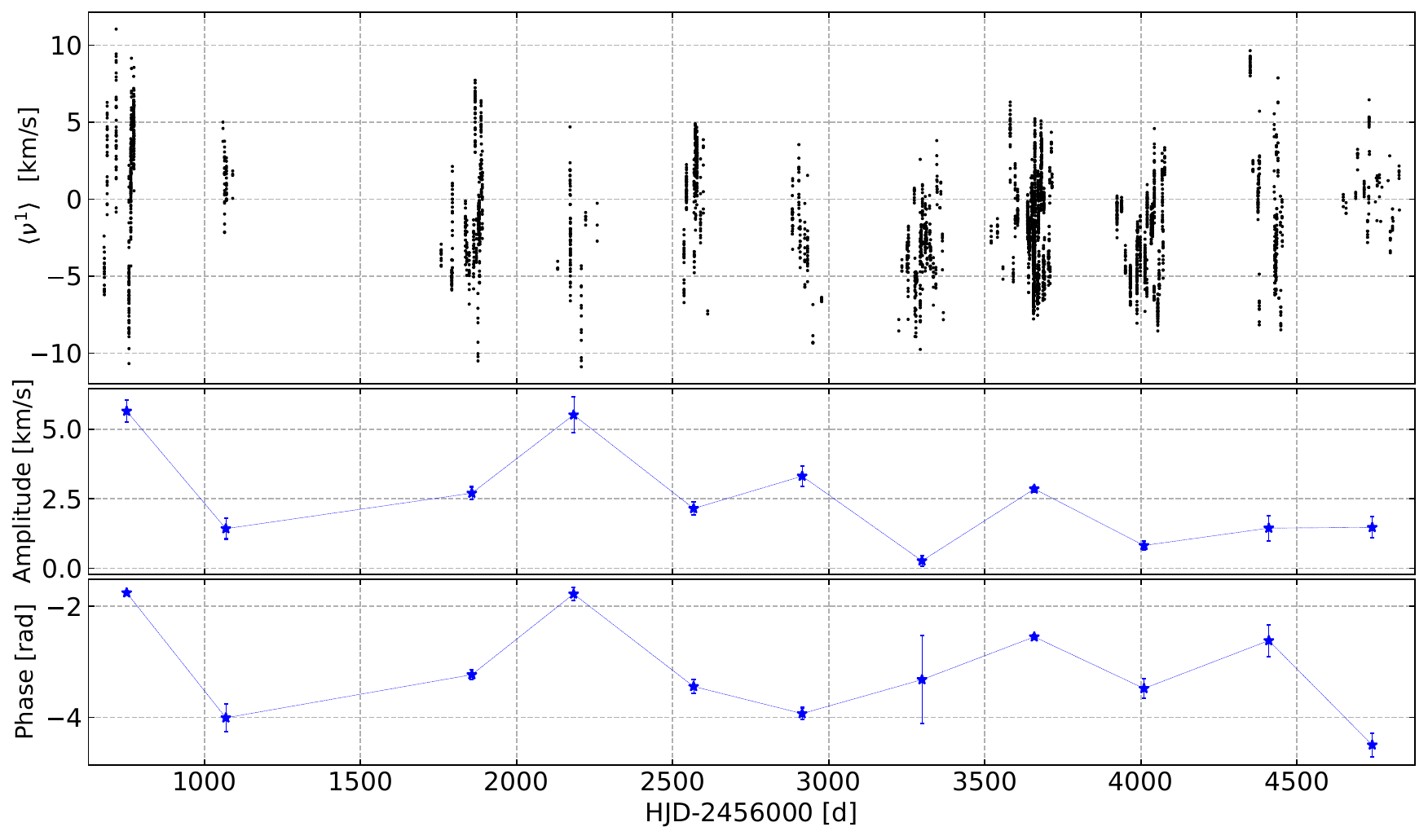}
    \caption{Top panel: First moment $\langle v^1 \rangle$ (radial velocity) variations of the \ion{He}{I}\,$\lambda$6678 line profile over the full 11.5-year TO monitoring period. The zero-point of the plot corresponds to $v_0=42$\,km/s. Middle and lower panels: Amplitude and phase modulation of the dominant 16.46-d period in $\langle v^1 \rangle$.}
    \label{fig:amp-phase_TO}
\end{figure}

To investigate the long-term stability of the dominant periodic signal, we examined the temporal evolution of its amplitude and phase (Figure\,\ref{fig:amp-phase_TO}). The dominant period was fixed at $P = 16.46$\,d, as determined from the frequency analysis of the complete radial-velocity dataset. Because the observations are naturally divided into individual observing seasons separated by long gaps, each season was analysed independently.
For every observing season, the radial-velocity measurements were fitted with a sinusoidal model at the fixed period using least-squares minimisation, from which the amplitude and phase of the signal were derived together with their uncertainties. The resulting seasonal amplitudes and phases are shown in Figure\,\ref{fig:amp-phase_TO}. Unlike transient frequencies that appear only during particular observing seasons, the 16.46-d signal is recovered throughout the dataset, although its amplitude and phase evolve with time. The WWZ scalograms presented in Appendix \ref{wwz-all} also reveal variations in the frequency and relative strength of this signal within individual observing seasons.

A prominent feature in the WWZ scalogram for the full TO dataset (Figure~\ref{fig:WWZ_all_moments}a) is enhanced power at periods in the range 23--25 days (frequencies 0.04--0.043 d$^{-1}$) detected in 2014 with a semi-amplitude of $k$ = 5.60 km/s and in 2019 (HJD$'$ = 2536--2612, where HJD$'$~=~HJD--2\,456\,000 days) with $k$ = 3.15 km/s. This signal is visible with variable amplitude in almost all seasons (Figure\,\ref{fig:WWZ_M1_seasons}). The $P\approx~$12\,d is only weakly detected in the first moment.

\paragraph*{Second Moment}

In the second moment, the GLS analysis (Table~\ref{tab:M2_all}) revealed a significant peak at $F_{1} = 0.112\,\mathrm{d^{-1}}$ ($P \approx 8.92$ d). This frequency is close to twice the dominant frequency identified in the first moment, $F_3 \approx 0.0607\,\mathrm{d^{-1}}$ ($P \approx 16.46$ d), i.e., $F_{1}\langle v^2 \rangle \approx 2F_3\langle v^1 \rangle$. This might suggest that the detected signal represents the first harmonic of the primary variability. However, the observed frequency is about 8\% lower than the expected harmonic, a difference that is significantly larger than the frequency uncertainties, particularly given the 11.5-year time baseline. A more likely interpretation is that the 8.92 d signal represents an independent mode that primarily affects the line width.

The GLS periodogram (Figure \ref{fig:GLS_M123_all}b) shows two prominent peaks at $F_4 \approx 0.45~\mathrm{d^{-1}}$ ($P \approx 2.22$ d) and $F_6 \approx 0.55~\mathrm{d^{-1}}$ ($P \approx 1.82$ d). These frequencies satisfy the daily sampling relation, $f_{\mathrm{alias}} = 1 - f$, indicating that they form a pair of one-day aliases produced by the observing cadence rather than two independent periodicities. Prewhitening either of these frequencies removes the other peak, confirming that they arise from the same underlying~signal.

A low-frequency signal at $F_{p1} \approx 0.02\,\mathrm{d^{-1}}$ ($P\approx49.5$ d) appears only after sequential prewhitening and is not present in the original periodogram. We, therefore, interpret this feature as an artefact caused by imperfect removal of multiple periodic components and spectral window effects, rather than a genuine physical signal. 

The prominence of the harmonics in the second moment suggests that the variability deviates significantly from a purely sinusoidal form. Such behaviour is expected in the presence of non-linear processes, for example rotational modulation by asymmetric structures or non-linear pulsational effects.

The wavelet analysis of the second moment (Figure~\ref{fig:WWZ_all_moments}b) reveals four dominant periods. In particular, the $P\approx~$12-day signal is clearly detected in 2023 with a semi-amplitude of $k=1.3\,\times10^{3}\,(\mathrm{km/s})^2$. By contrast, the dominant GLS frequency at 0.112 d$^{-1}$ produces only a weak feature in the WWZ power spectrum averaged over the full 11.5-year baseline. The corresponding signal is detected only during a few observing seasons and exhibits strongly variable amplitude (Figure~\ref{fig:WWZ_moments_2022}b), suggesting that it represents a transient mode rather than a persistent periodicity.

\paragraph*{Third Moment}

\begin{figure}[H]
%    \isPreprints{\centering}{}
    \includegraphics[width=0.65\linewidth]{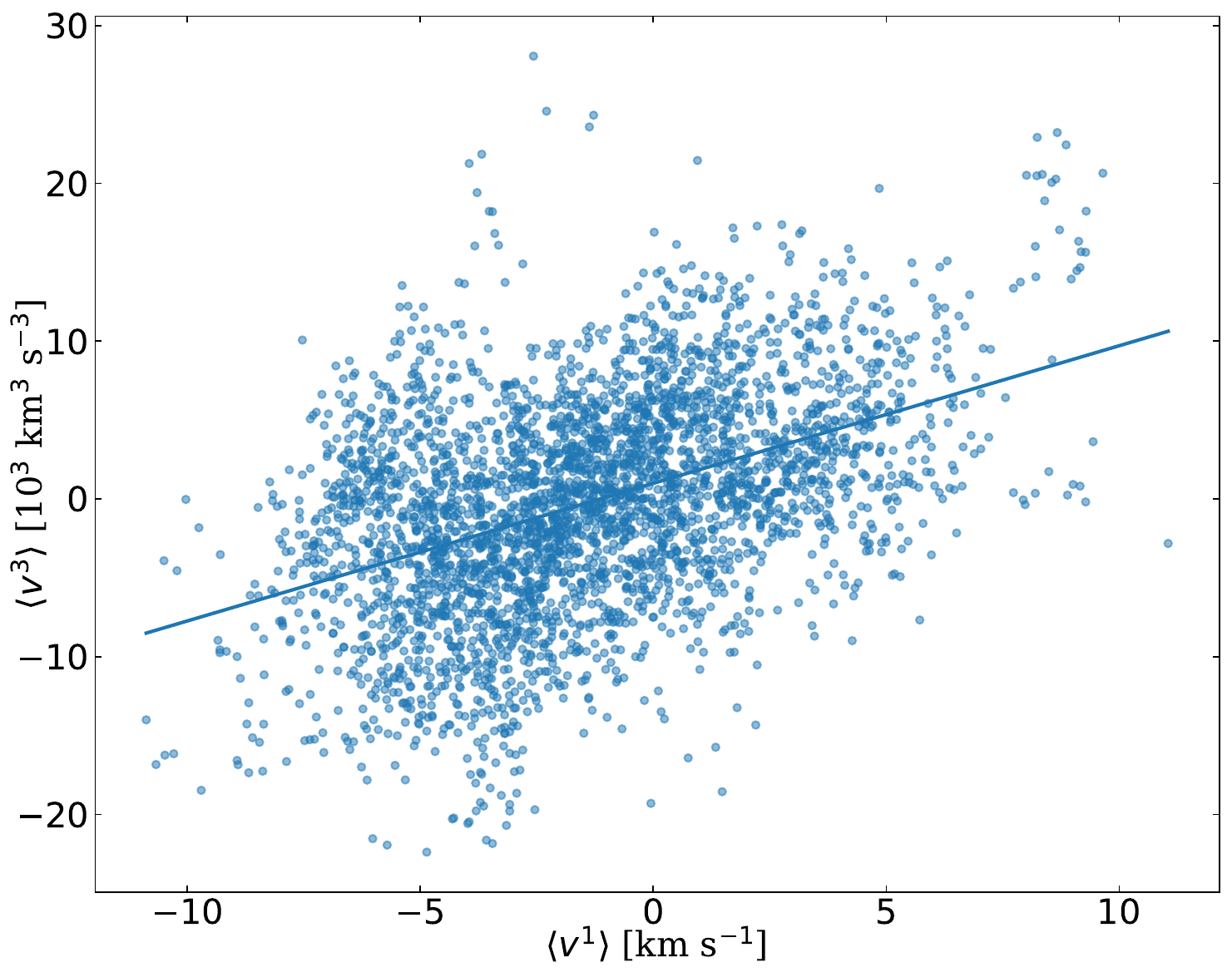}
    \caption{Correlation between the normalised first  moment $\langle v^1 \rangle$ and the normalised central third moment $\langle v^3 \rangle$ of the \ion{He}{I} $\lambda6678$ line for the complete TO dataset. The solid line represents the linear regression fit.}
    \label{fig:M1vsM3_third}
\end{figure}

The third moment, which quantifies the asymmetry (skewness) of the line profile (Figure~\ref{fig:GLS_M123_all}c), shows 
    substantial power at low frequencies, similar to the first moment. The $\sim$16-d variability that is most prominently manifested in the first moment also has a weaker signature in the line asymmetry. The phase-folded variations of the first and third moments also show a clear correspondence when folded with the $\approx16.46$-d period (Figure~\ref{fig:M1andM3-phase-third}). To quantify this relation, we directly compared $\langle v^1 \rangle$ and $\langle v^3 \rangle$ values (Figure~\ref{fig:M1vsM3_third}). The Pearson and Spearman correlation coefficients are $r=0.446$ and $\rho=0.436$, respectively, while the linear regression yields $R^2=0.2$, indicating a moderate correlation between the two~moments.

  As the central third moment is calculated relative to the instantaneous line centroid, this correlation cannot be attributed simply to a common response to line-profile displacement. Instead, it suggests that the centroid variations associated with the $\approx16$-d variability are accompanied by systematic changes in the line-profile asymmetry. The fact that the corresponding signal is absent from the second moment further indicates that this variability is not primarily associated with changes in the overall line width.

The WWZ analysis reveals that the $\approx$16.5-day period is present, to varying degrees, in nearly all seasons. In particular, that period found in 2022 has a semi-amplitude of $k = 4.3\pm0.9\,\times10^{3}\,(\mathrm{km/s})^3$  (Figure~\ref{fig:WWZ_moments_2022}c).

\subsubsection{SONG High-Resolution Spectroscopy}\label{SONG}
The SONG telescope observed {\rholeo} in high-cadence mode, with spectra overlapping Season 2017 of the TO dataset. The first moment of \ion{Si}{III} 4552\,{\AA} and \ion{He}{I} 5875\,{\AA} lines together with contemporaneous TO data are presented in Figure~\ref{fig:TO&SONG-short}. For consistency with the TO data, the same radial-velocity zero point $v_0=42$\,km/s was used.

\begin{figure}[H]
    \isPreprints{\centering}{}
    \includegraphics[width=0.92\linewidth]{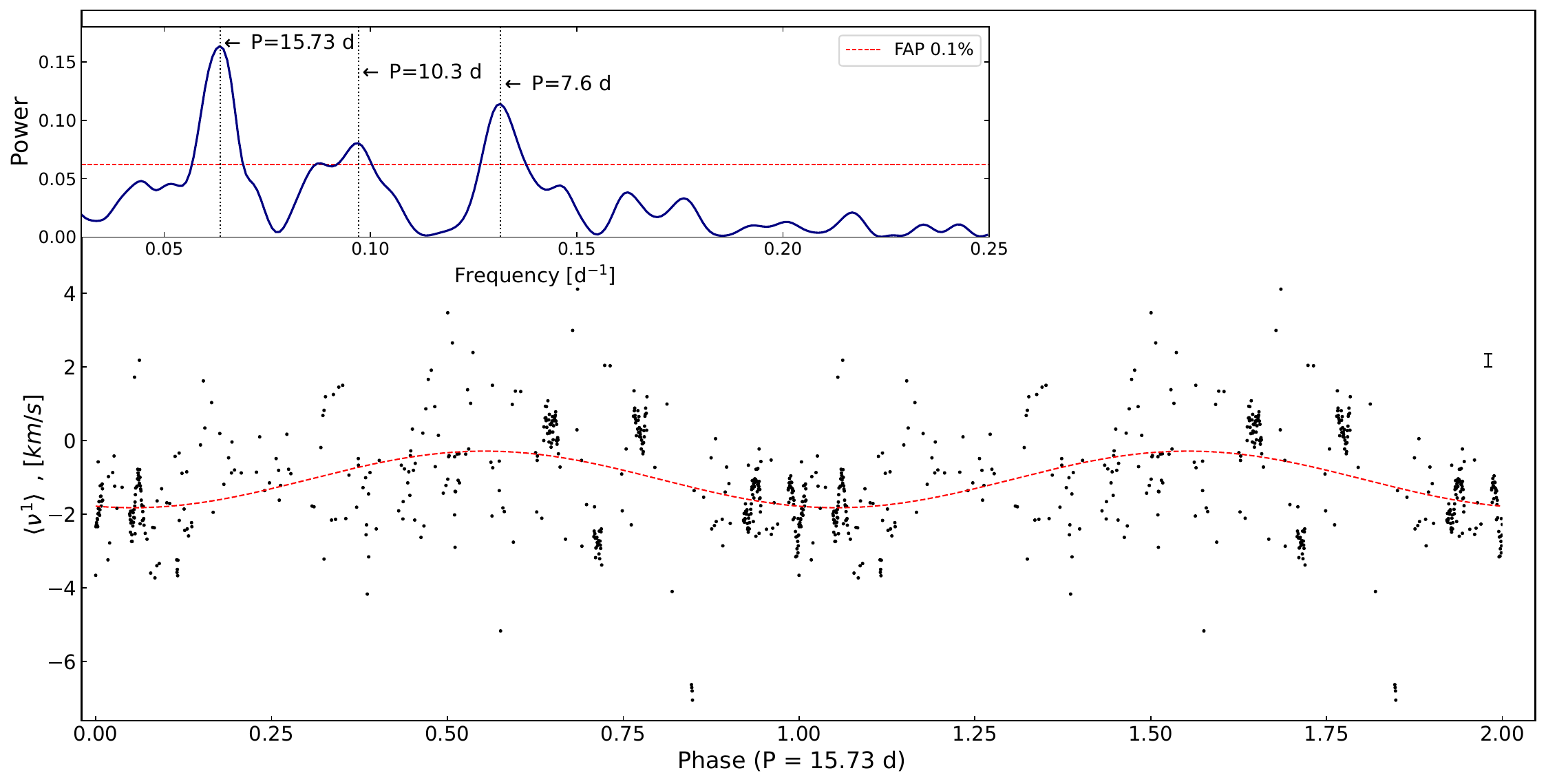}
    \caption{SONG spectroscopy: phase curve of the first moment of the \ion{Si}{III} $\lambda$4552 line with a period of 15.73 d. GLS power spectrum is shown in the upper-left panel. Typical error bar (0.18 km/s) is shown on the right side.} 
    \label{fig:SiIII_GLS+Phase}
\end{figure}

The GLS analysis revealed three significant periods that lie above the FAP level: 15.73~days, 10.3 days, and 7.6 days. The phase diagram of the first moment of the \ion{Si}{III} $\lambda$4552 line, folded with a period of 15.73 d and amplitude A = 0.77 km/s, is shown in Figure~\ref{fig:SiIII_GLS+Phase}, with the GLS power spectrum in the inset.

\begin{figure}[H]
    \isPreprints{\centering}{}
    \includegraphics[width=\linewidth]{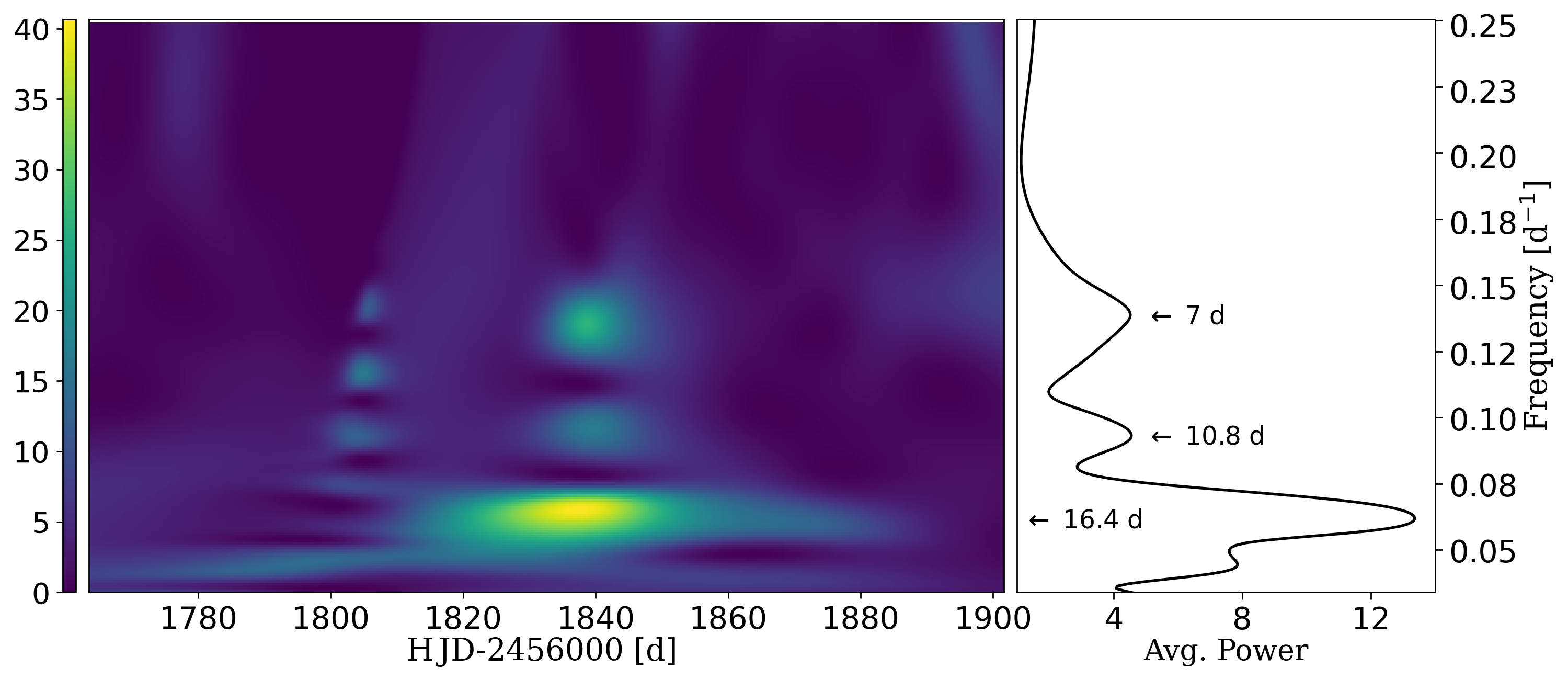}
    \caption{WWZ scalogram for the first moment of \ion{Si}{III} $\lambda$4552 line in SONG spectra. The colour-coding is the wavelet power (\textbf{left panel}). The lower bright spot corresponds to the period of $\approx$16.4 days. The black solid line in the \textbf{right panel} shows the time-averaged wavelet power. The right-hand frequency axis applies to both panels.} 
    \label{fig:WWZ_SiIII} 
\end{figure}

As can be seen in Figure~\ref{fig:TO&SONG-short}, the first-moment time series exhibits both long-period and short-period variability, as the dataset includes several sequences of consecutive nightly observations. Therefore, the WWZ analysis enables us not only to identify variability periods but also to investigate their temporal evolution by tracking the appearance, disappearance, and relative strength of individual frequency components. Applying the WWZ analysis to the \ion{Si}{III} $\lambda$4552 line moments (Figure~\ref{fig:WWZ_SiIII}), we detect the same dominant period of $\approx$16.4\,d as found in the TO data. The densest observations around HJD$'\approx1840$ also reveal two additional shorter periods of approximately 10.8\,d and 7\,d. These shorter-period signals are present only during a relatively brief interval, approximately in the middle of the observing campaign, whereas the 16.4-day period persists for a much longer time and is characterised by substantially higher power.

\subsubsection{Comparison of TO and SONG Data}

\begin{figure}[H]
    \isPreprints{\centering}{}
    \includegraphics[width=0.65\linewidth]{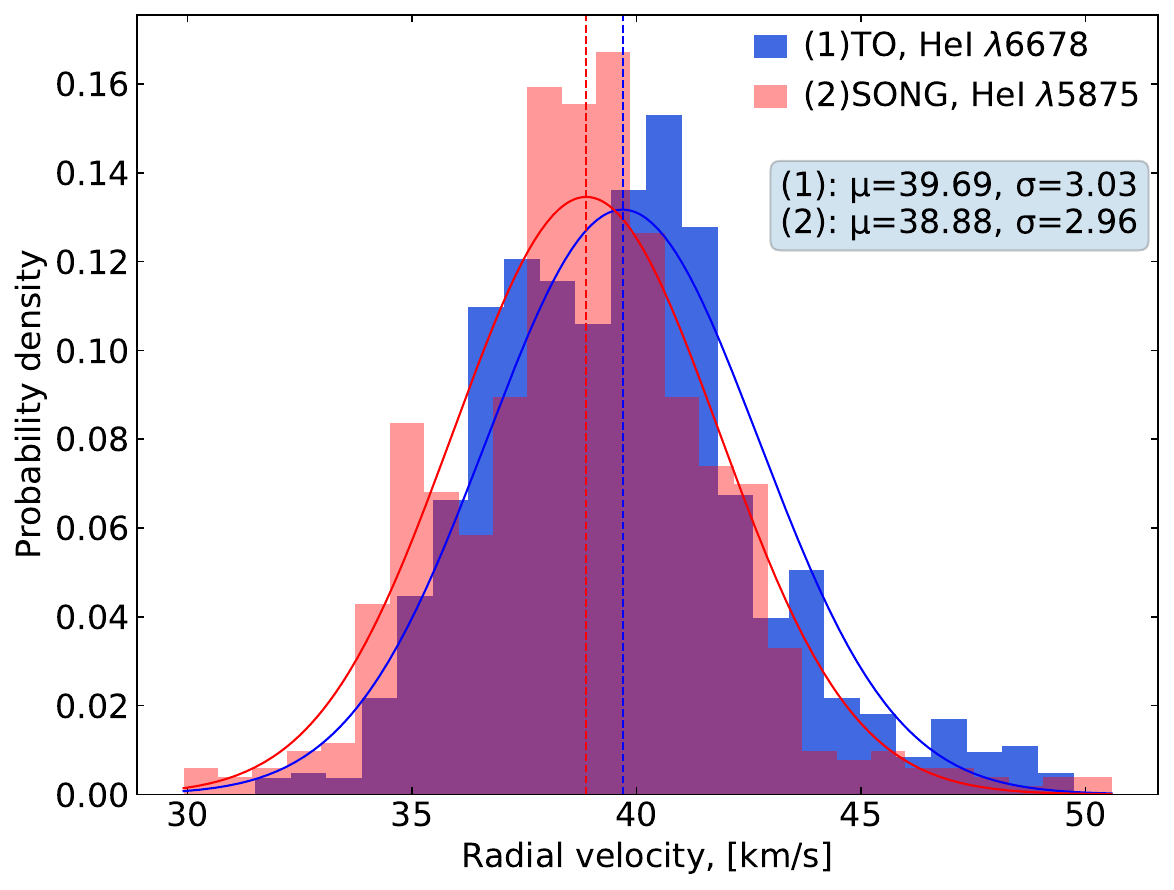}
    \caption{Statistics of the first moment $\langle v^{1} \rangle$ for He lines during the overlapping epoch of the TO and SONG datasets. TO observations shown in blue and SONG data in red.} 
    \label{fig:M1_stat_TO+SONG}
\end{figure}

Comparison of the first-moment statistics, $\langle v^1 \rangle$, derived from the TO \ion{He}{I} $\lambda$6678 line and the SONG \ion{He}{I} $\lambda$5875 line over the same epoch (season 2017) demonstrates the consistency between the medium-resolution TO and high-resolution SONG observations (Figure\,\ref{fig:M1_stat_TO+SONG}). The mean velocity in this season is lower than the previously reported value of $42\pm 0.8$~km/s \citep{https://doi.org/10.1002/asna.200710776}. An offset of the mean relative to the zero level is also clearly visible in Figure~\ref{fig:TO&SONG-short}.

\begin{figure}[H]
    \isPreprints{\centering}{}
    \includegraphics[width=0.65\linewidth]{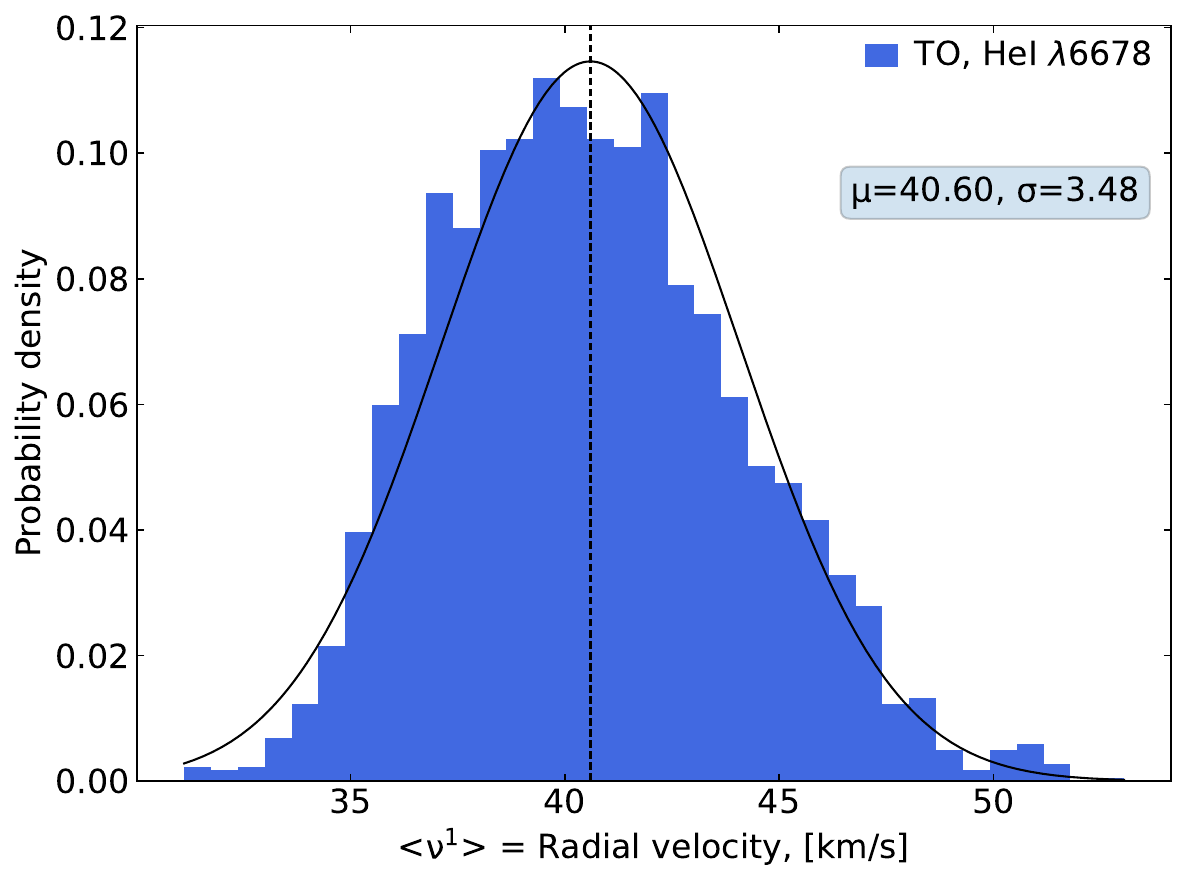}
    \caption{Statistics of the first moment $\langle v^{1} \rangle$ of the \ion{He}{I} $\lambda$6678 line over the full 11.5-year TO dataset.} 
    \label{fig:M1_stat_TO}
\end{figure}

The long time span of TO observations of $\rho$~Leo, combined with the large number of available spectra, allows a reliable estimate of the systemic radial velocity. From the analysis of the \ion{He}{I} $\lambda$6678 line, based on 3494 TO spectra obtained over 11.5 years, we derive $v_{\rm sys} = 40.60 \pm 0.06$~km/s (error of the mean), with a scatter of 3.48~km/s  (Figure~\ref{fig:M1_stat_TO}).

\subsection{Analysis of the TESS Light Curve}\label{S-anaphotTESS}

The extensive spectroscopic analysis is complemented by a photometric dataset. In our previous study \citep{2026A&A...706A.200C}, we analysed only the combined consecutive TESS Sectors 45 and 46, obtaining a dominant period of approximately 18 days. However, analysing each sector separately (Sectors 45, 46, or 72) significantly limits the sensitivity to long-period variability because of the relatively short duration of the individual datasets. Therefore, despite the large temporal gap between Sectors 46 and 72, we also analysed the combined dataset comprising all three sectors. The extended time baseline improves the sensitivity to long-period variability that could not be detected from the individual sectors alone.

\begin{figure}[H]
    \isPreprints{\centering}{}
    \includegraphics[width=\linewidth]{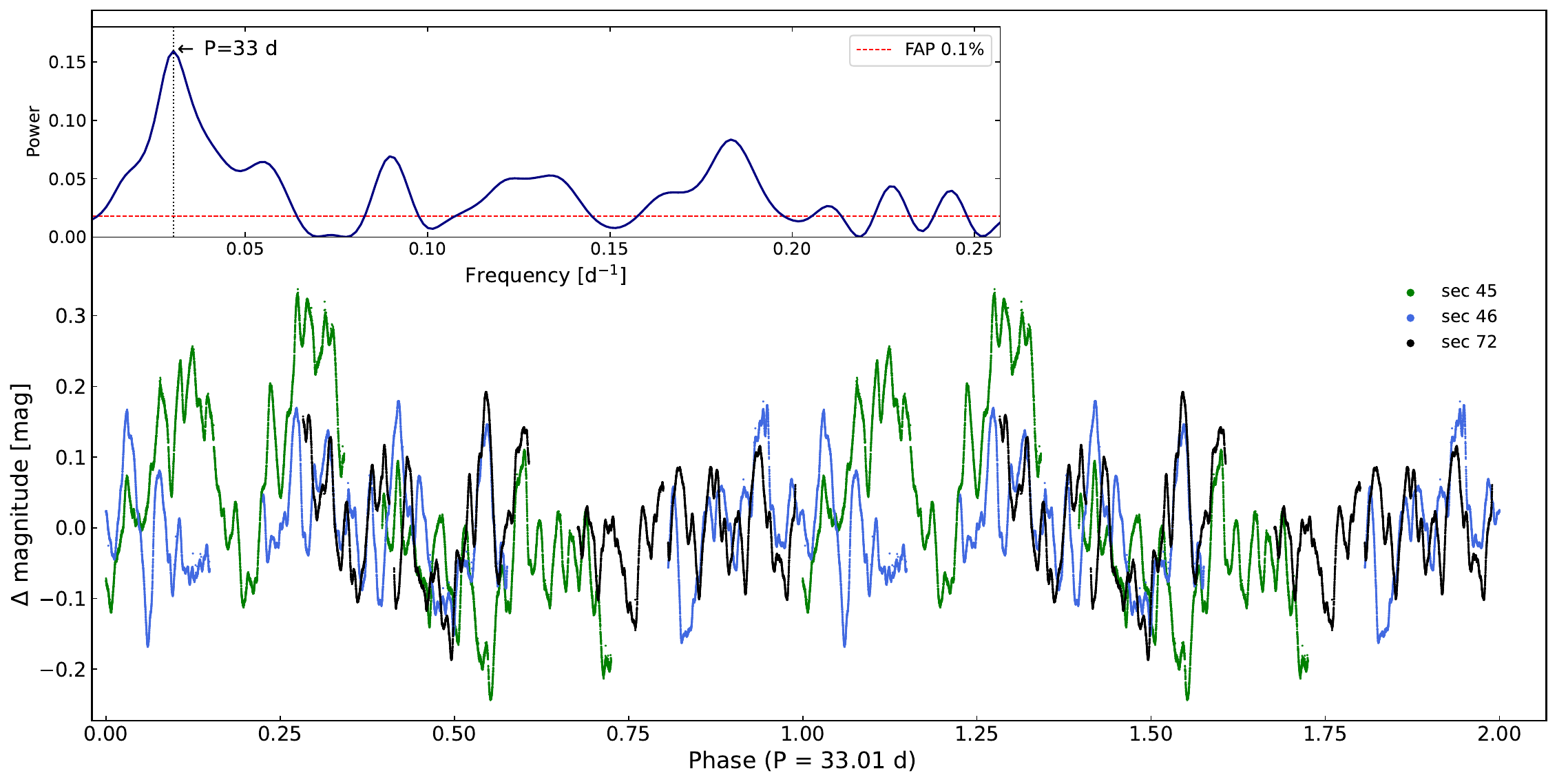}
    \caption{TESS photometry: 45, 46, 72 sectors (baseline 758.88 days), phase curve with period 33.01~days. Result of GLS analysis is shown on the left upper part.}
    \label{TESS-GLS+Phase}
\end{figure}

\textls[-15]{The GLS power spectrum of the combined dataset is shown in the inset panel of Figure~\ref{TESS-GLS+Phase}. No significant peaks were detected in the 0.25--1.25\,d$^{-1}$ frequency range. Therefore, the figure presents a detailed view of the dominant frequencies below 0.25~d$^{-1}$ together with the corresponding FAP level. The strongest signal occurs near 0.03~d$^{-1}$, corresponding to a period of  $P \approx 33$ days.}

\begin{table}[H]
\setlength{\tabcolsep}{2.85mm}
\caption{Periods detected in TESS photometry, combining Sectors 45, 46, and 72, using GLS and GLSp analysis.
Results of the analysis are arranged by decreasing period in each column. The number in brackets [N] denotes the rank of the period by detected power. The amplitudes were determined from the phase-folded light curves after each successive pre-whitening step in the GLSp analysis.}\label{tab:TESS}

\isPreprints{\centering}{}

\begin{tabularx}{\textwidth}{ccccccc}

\toprule

\multicolumn{3}{c}{GLS} &
\multicolumn{4}{c}{GLSp} \\

\cmidrule(lr){1-3}
\cmidrule{4-7}

N & Frequency & Period &
N & Frequency & Period & Amplitude \\

& [d$^{-1}$] & [d] &
& [d$^{-1}$] & [d] & [mag] \\

\midrule

$F_1$ & 0.0303 [1] & $33.01 \pm 1.43$ &
$F_{p1}$ & 0.0303 [1] & $33.01 \pm 1.43$ & $0.118~\pm~0.003$ \\

$F_2$ & 0.0554 [4] & $18.06 \pm 0.50$ &
$F_{p2}$ & 0.0514 [2] & $19.45 \pm 0.50$ & $0.076~\pm~0.004$ \\

$F_3$ & 0.0896 [3] & $11.16 \pm 0.74$ &
$F_{p3}$ & 0.0938 [5] & $10.66 \pm 0.15$ & $0.066~\pm~0.002$ \\

$F_4$ & 0.1344 [5] & $7.44 \pm 2.02$ &
$F_{p4}$ & 0.1239 [4] & $8.07 \pm 0.12$ & $0.070~\pm~0.002$ \\

$F_5$ & 0.1674 [7] & $5.97 \pm 1.94$ &
$F_{p5}$ & 0.1663 [6] & $6.01 \pm 0.05$ & $0.056~\pm~0.002$ \\

$F_6$ & 0.1832 [2] & $5.46 \pm 0.57$ &
$F_{p6}$ & 0.1832 [3] & $5.46 \pm 0.57$ & $0.069~\pm~0.002$ \\

$F_7$ & 0.2280 [6] & $4.39 \pm 0.18$ &
$F_{p7}$ & 0.2295 [7] & $4.36 \pm 0.03$ & $0.053~\pm~0.002$ \\

\bottomrule

\end{tabularx}

\end{table}

\begin{figure}[H]
    \isPreprints{\centering}{}
    \includegraphics[width=\linewidth]{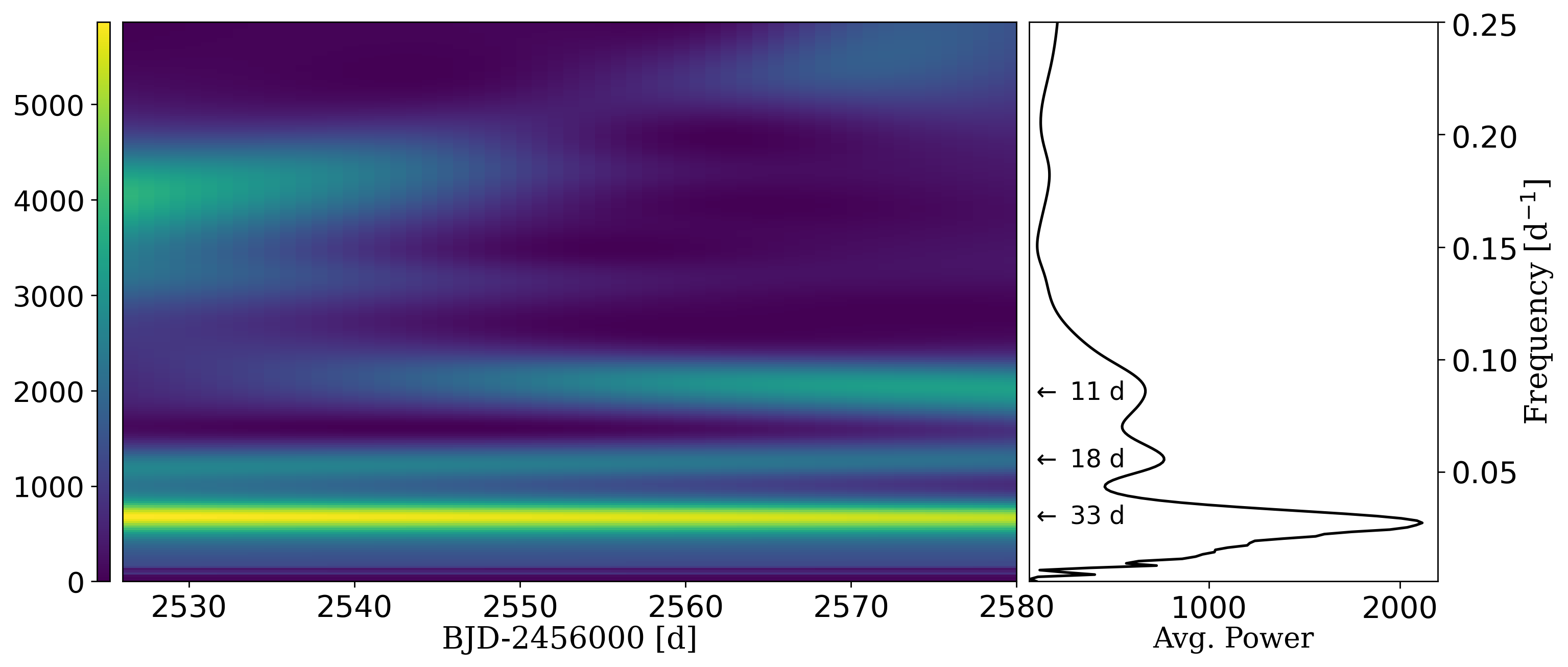}
\caption{WWZ scalogram for TESS photometry. Colour-coded is the wavelet power. The lower bright stripe corresponds to the $P\approx$\,33 days. The black solid line on the right panel shows the time-averaged wavelet power. The right-hand frequency axis applies to both panels.}
\label{fig:TESS_WWZ}
\end{figure}

By constructing a phase diagram, we can illustrate how the period aligns with the data. Moreover, it enables us to pinpoint a more accurate value for the period. Hence, the next step was GLSp. In Table~\ref{tab:TESS} we present a set of frequencies that we found. The dominant frequency is $F_{p1}$~=~0.0303 d$^{-1}$, corresponding to a period of $P\approx 33$ days. This result should be treated with caution for the reasons given in Section \ref{phot_var}.
The frequency $F_{p3} = 0.0938$ d$^{-1}$ ($P = 10.66$ days) is close to the GLS-identified $F_3$. The remaining signals can be interpreted as subharmonics or combinations of independent frequencies: $F_{p4} \approx F_{p1} + F_{p3}$. 
In the GLSp analysis, the amplitudes decrease rapidly after pre-whitening the dominant frequency.  Nevertheless, these low-amplitude signals are clearly visible in the WWZ scalogram (Figure~\ref{fig:TESS_WWZ}).

The WWZ analysis also shows a dominant period of approximately 33\,days (Figure~\ref{fig:TESS_WWZ}), along with two less pronounced periods at about 18\,days and 11\,days. Owing to the large temporal gap between the first two sectors (45 and 46) and the third sector (72), the figure displays only the time interval covering the first two sectors for clarity. We note that the 33-day period is approximately twice the spectroscopic period of 16.46\,days.

\section{Discussion}

The TO spectroscopic time series reveal persistent variability in \ion{He}{I} $\lambda$6678 spectral line with period $P = 16.46$~d, manifested most prominently in the radial velocity, with a weaker correlated signature in the line asymmetry. When the well-sampled observing seasons from 2017, 2019, 2021, 2022 and 2023 are folded using a common ephemeris, the resulting phase diagrams overlap remarkably well (Figure~\ref{fig:Chosen_phase}). Despite a moderate scatter, the combined dataset forms a coherent phase pattern extending over approximately six years. This long-term phase stability provides a strong constraint on the physical origin of the variability.

A further recurrent signal is found at a period of $\sim$11--12\,d, although with variable strength and not consistently with all analysis methods. It is detected in the TO \ion{He}{I} $\lambda$6678 line in the first moment (GLS and WWZ), second moment (WWZ), and third moment (GLSp and WWZ), as well as in the first moment of the SONG \ion{Si}{III} $\lambda$4552 line (WWZ) and in the TESS photometry (GLS and WWZ). Its recurrence in independent spectroscopic datasets and in photometry suggests that this timescale is intrinsic to the star rather than an artefact of a particular dataset or analysis method. The $\sim$11--12-d variability may be associated with stellar rotation, as previously suggested in our study of $\rho$ Leo \cite{2026A&A...706A.200C}.

\subsection{Pulsational Interpretation}

The star is classified as an $\alpha$~Cygni--type variable, that is, a luminous blue supergiant known to exhibit non-radial pulsations. Such stars commonly show multi-periodic variability, semi-coherent timescales, cycle-to-cycle frequency and amplitude modulation, and slow variations of the mean radial velocity. The observed behaviour is fully consistent with this phenomenology.

First, although the 16.46-d signal persists over several years, the phase-folded radial-velocity curve is not strictly sinusoidal and exhibits intrinsic scatter. This behaviour is typical of pulsational variability, where the superposition of multiple modes and amplitude modulation lead to deviations from a repeatable waveform \citep{2006MNRAS.368..571B}.

Second, seasonal mean radial velocities vary by several km/s without displaying a monotonic trend (Figure\,\ref{fig:amp-phase_TO}). Such irregular long-term shifts are naturally explained by atmospheric variability and mode beating in multi-periodic pulsators \citep{2006MNRAS.368..571B}.

Third, the simultaneous presence of the 16.46-d period in first and third moments (Figure~\ref{fig:M1andM3-phase-third}) provides supporting evidence. Non-radial pulsations produce travelling surface velocity fields that distort line profiles, leading to correlated variations in centroid velocity and skewness, as predicted by the moment method \citep{1998ASPC..135..149T}.
The stability of the period over several years does not contradict a pulsational origin. In evolved massive stars, dominant pulsation modes may remain coherent for long timescales despite amplitude and waveform variability \citep{2013MNRAS.433.1246S}.

If the dominant 16.46\,d periodicity is assumed to originate from stellar pulsations, its pulsation constant can be estimated as
\begin{equation}
Q = P \sqrt{\frac{\bar{\rho}}{\bar{\rho}_{\odot}}}
  = P \sqrt{\frac{M/M_{\odot}}{(R/R_{\odot})^{3}}},
\end{equation}
where $P$ is the pulsation period, $M$ is the stellar mass, and $R$ is the stellar radius.
Using the adopted stellar parameters of $\rho$ Leo ($M = 22\,M_{\odot}$ and $R = 32\,R_{\odot}$) together with the dominant period of $P = 16.46$\,d, we obtain
\begin{equation}
Q \approx 0.43~\mathrm{d}.
\end{equation}
This value is substantially larger than the pulsation constants expected for radial and pressure modes ($Q \lesssim$ 0.05--0.1\,d), indicating that the observed periodicity is inconsistent with radial or pressure-mode pulsations \citep{ChristensenDalsgaard2014,Handler_2013}. If the 16.46\,d signal is of pulsational origin, the derived pulsation constant is more consistent with a high-order gravity mode in an evolved massive star \citep{2012AN....333..946O,2015MNRAS.447.2378O}.

However, the pulsation constant alone cannot distinguish between a pulsational and an orbital origin of the observed radial-velocity variability. Thus, we have considered both possibilities. While the derived value of $Q$ disfavours an interpretation in terms of radial pulsation, it remains consistent with either a non-radial gravity mode or radial-velocity modulation associated with a binary companion.

The detection of a significant 8.92 d period at approximately half the dominant radial-velocity timescale in the second moment (Figure~\ref{fig:M1+M2-All}) 
provides additional evidence for intrinsic line-profile variability. This signal may represent either an independent transient mode that primarily affects the line-profile width or, less likely, the first harmonic of the dominant 16.46 d variability. In either case, its presence is more naturally explained by pulsational line-profile variability than by a purely Keplerian interpretation.
A Keplerian orbit would produce a rigid
Doppler shift, affecting the first moment primarily, while leaving the second and third moments largely unchanged. 

However, a hypothetical low-mass companion in a close orbit could indirectly influence the observed variability through tidal forcing (see discussion in Section\,\ref{binarity}). A time-dependent tidal potential may perturb pulsation modes, modify their geometry, and enhance harmonic content, leading to detectable power at twice the fundamental frequency. In this framework, the
dominant period at $P \approx 16.5$~d would still be of pulsational origin, while the presence of a companion could contribute to the observed complexity of the frequency spectrum and the phase-locked variability in higher-order moments \citep{2015A&A...581A..75K}.

\subsection{Photometric Variability}\label{phot_var}

In addition to spectroscopic analysis, we investigated space-based photometric data obtained from the TESS mission.
However, the determination of a reliable photometric period is complicated by the limited time coverage and instrumental systematics.
The available TESS observations consist of three sectors obtained with three different CCDs, each spanning approximately 27~days. As a result, longer timescales comparable to or exceeding the sector length are only weakly constrained and are sensitive to the normalisation and de-trending procedure. In particular, possible offsets between sectors may introduce artificial trends that affect the recovery of long periods.

A period analysis of the TESS data yielded different dominant timescales depending on the adopted processing. The light curve of the three sectors considered in this paper shows variability with a period $\approx$33~d, but this signal is not robust to changes in the reduction procedure and is not recovered when only the two consecutive sectors 45 and 46 ($\approx$52\,d baseline) are analysed. The dominant period of $\approx$18.5~d, reported by \citet{2026A&A...706A.200C} based on those two sectors, is also present in our GLS and WWZ analyses of the three-sector dataset, although with significantly lower power.
In addition, K2 photometry with a time span of $\approx$9~d revealed a dominant peak at 
$P\approx$ 25~d, accompanied by a secondary peak at 
$P\approx$ 16.5~d \cite{2026A&A...706A.200C}. Given the limited frequency resolution ($\Delta f = 1/T$), such differences are expected, and power can be redistributed between nearby frequencies and their harmonics.

We, therefore, conclude that the photometric data clearly indicate the presence of long-period variability on timescales of $\approx$15--30~d, but do not allow a unique determination of the dominant period. In contrast, the spectroscopic signal at $P_{\rm spec} = 16.46$~d is robustly detected in the line-profile 
moments and remains stable over several years. The $P_{\rm phot} = 2 P_{\rm spec}$ ratio suggests a harmonic relation, with the photometric variability
likely tracing the first harmonic of the spectroscopic signal.

Consequently, we adopt the spectroscopic period as the primary and most reliable timescale of variability, while the photometric results are treated as supporting evidence for long-term variability rather than as an independent constraint on the exact period.

\subsection{Binarity as a Possible Origin of the Observed Variability}\label{binarity}

The high multiplicity fraction among massive stars provides a strong motivation for considering binarity. Therefore, the 16.46\,d periodicity detected in $\rho$\,Leo may alternatively be interpreted as arising from orbital motion or from pulsations perturbed by binarity. 
Below we summarise the main arguments supporting and challenging this hypothesis.

The dominant spectroscopic period of 16.46\,d is stable over the full 11.5-year time base, which is consistent with a Keplerian origin. Such long-term phase coherence is a typical signature of orbital motion. An additional argument in favour of this hypothesis is that all three spectral 
lines (\ion{He}{I} $\lambda$6678, \ion{Si}{III} $\lambda$4552 and \ion{He}{I} $\lambda$5875) shift synchronously with time. 

In close binary systems, tidal forces can significantly affect stellar pulsations. The equilibrium tide may perturb pulsation frequencies, leading to frequency splitting and multiplets separated by the orbital frequency \citep{1981ApJ...244..299S,2003A&A...404.1051R,2003A&A...409..677R,2005ASPC..333...39S,2018MNRAS.476.4840B}. As a result, a single intrinsic mode can appear as multiple peaks in the observed spectrum, producing complex and evolving frequency patterns, consistent with the transient frequencies seen in our data. However, the available diagnostics do not provide evidence that the $\approx$ 33\,d photometric period is directly related to binarity. Instead, this signal is more consistently interpreted as arising from pulsational variability and its nonlinear manifestations, although the binary hypothesis cannot be excluded as a possible contributor to the complexity of the observed frequency spectrum through tidal perturbations.

Tidal distortion can also modify the geometry of pulsation modes, causing non-spherical oscillations and mode coupling between different angular degrees \citep{2025arXiv250908426S}. In extreme cases, pulsations may be confined near Lagrange points, leading to amplitude and phase modulation linked to the orbital 
period. Such effects could contribute to the coexistence of a stable dominant period and multiple short-lived signals.

Finally, variability in higher-order moments, particularly the third moment, is also compatible with tidally perturbed pulsations, which produce complex line-profile distortions rather than simple Doppler shifts \citep{2026A&A...710A.325R}.

Several observational properties of $\rho$\,Leo appear to argue against a binary interpretation. The radial-velocity curve associated with the 16.46 d signal is not strictly repeatable and exhibits intrinsic scatter exceeding the expected measurement uncertainties. In addition, the seasonal mean velocities show irregular variations from year to year, making it difficult to define a strictly constant systemic velocity. These variations may, at least partly, be related to the stellar wind and intrinsic atmospheric variability, although a detailed assessment of their contribution to the measured radial velocities is beyond the scope of the present work. Nevertheless, within the estimated uncertainty of approximately 3.5\,km/s, the systemic velocity remains consistent with being stable over the full observational baseline. Moreover, theoretical and observational studies of tidally perturbed pulsators indicate that such complex radial-velocity behaviour does not necessarily exclude a binary interpretation \citep{2010aste.book.....A}.
Consequently, the observed lack of long-term phase coherence of the 16.46 d signal should not be regarded as evidence against binarity, but rather as behaviour that is compatible with a tidally perturbed pulsator.

The presence of a persistent periodic signal in the third moment challenges a binary interpretation. Orbital motion is expected to produce a rigid shift of the line profile~\citep{2005oasp.book.....G}, affecting the first moment primarily, without introducing skewness variations. The observed correlation between the first and the third moments, therefore, points to intrinsic line-profile variability rather than orbital motion.

Furthermore, observational surveys of massive-star binaries show that confirmed supergiant systems typically exhibit radial-velocity amplitudes of several tens of km/s, whereas variations of only a few km/s are more commonly associated with pulsations and atmospheric dynamics \citep{2026arXiv260402111M}. The variability pattern observed in $\rho$\,Leo, including the presence of numerous transient frequencies and the lack of strict periodicity, is, therefore, more naturally explained by non-radial pulsations and stochastic variability.

Let us consider binary scenarios consistent with the observed variability.
\begin{enumerate}
\item Assuming that the 16.46~d period arises from orbital motion and adopting amplitude of \(A = 3.94~\mathrm{km/s}\) (Table~\ref{tab:M1_all}), we estimate the system parameters. Mass of the primary is taken to be $M_1=22\,M_\odot$ and radius $R_1=32\,R_\odot$ \citep{10.1111/j.1365-2966.2004.07799.x}. As an example, we assume that the orbital inclination is equal to the inclination of the stellar rotation axis, $21.7^\circ$~\citep{2026A&A...706A.200C}, which seems plausible. 
We assume that the distance to the star is 700~pc \citep{2023A&A...677A.175W}. The resulting mass function yields a companion mass of \(M_2 \approx 0.52\,M_\odot\). The corresponding orbital separation is \(a \approx 0.358\)~AU (\(\approx76.5\,R_\odot \approx2.4R_1\)). The maximum angular separation is \(\theta_{\max} \approx 0.51\)~mas. The apparent visual magnitude of the secondary component is $m_V \approx 16.6$, making it several orders of magnitude fainter than the primary. The expected transit depth is extremely small, \(\Delta m \approx 0.042\)~mmag, rendering eclipses effectively undetectable. While such a configuration is geometrically viable, the photometric signature of the companion would be negligible.
\item The binary scenario proposed by \citet{2023A&A...677A.175W} is also consistent with the observables if the system is viewed nearly pole-on. For a binary system with primary mass of $17\,M_\odot$ and a companion mass of $11\,M_\odot$, an orbital period of 16.46 days and a radial velocity amplitude of $3.94\,\mathrm{km/s}$, the orbit inclination angle should be approximately $2.3^\circ$. The derived orbital separation is $a \approx 0.38$ AU and the corresponding angular separation of the system is approximately $\theta \approx 0.54$ mas. 
This configuration implies that $\rho$ Leo could host a relatively massive companion that remains undetected in spectroscopy due to the extremely low inclination. Given the small orbital separation, of the order of a few stellar radii, such a companion could significantly perturb the inner stellar wind and potentially contribute to the formation of transient features we observed in the H$\alpha$ line profile.
\end{enumerate}

Both scenarios yield an angular separation that is approximately one order of magnitude smaller than the values derived from lunar occultation and interferometric observations (2.9\,mas \citep{1976A&A....48..245D}, 10.3\,mas \citep{1982AJ.....87.1874R}, and 46.1\,mas \citep{2024AJ....168...28T}, respectively). Such large angular separations would correspond to orbital periods substantially longer than 16.46\,days, most likely on the order of several decades. Confirming such long-period orbits would, therefore, require observational time series spanning considerably longer time intervals.

In summary, while the binary hypothesis remains formally consistent with the presence of a stable 16.46\,d period and is supported by the high multiplicity fraction of massive stars and the theoretical framework of tidally perturbed pulsators, the observational evidence does not favour a purely orbital interpretation. The low radial-velocity amplitude, lack of phase stability, and significant line-profile asymmetry variations suggest that intrinsic stellar variability, most likely in the form of non-radial pulsations, provides a more consistent explanation. Nevertheless, the possibility of a low-mass companion influencing the observed variability cannot be entirely excluded and warrants further investigation.

\section{Conclusions}

This work presents the results of intensive long-term spectroscopic monitoring of the star {\rholeo} (from 2014 to 2025), which allowed us to investigate both short- and long-period variability, as well as the temporal evolution of these periods.
The main outcome of this study is the identification of a coherent set of periodicities derived from extensive spectroscopic and photometric time-series analysis. The periods obtained using different frequency-analysis techniques are in very good mutual agreement.

\begin{itemize}
\item	The long-living spectroscopic periodicity is detected at $P_{\rm spectr} = 16.46$~d. It appears most prominently in the radial velocity (first moment of the line profile) with a weaker correlated signature in the line asymmetry (central third moment).
The observed phase-locked variability of both moments indicates periodic line-profile distortions caused by surface velocity fields. 
\item A recurrent $\sim$11--12-d periodicity is detected with variable strength in several independent spectroscopic diagnostics and in TESS photometry. This timescale may be associated with stellar rotation, as suggested in our previous study of $\rho$\,Leo \cite{2026A&A...706A.200C}.
\item	Independent TESS photometry reveals a dominant period of $P_{\rm phot} \approx 33$~d, almost exactly twice the persistent spectroscopic period.
\item	Based on 11.5-year average we derive systemic velocity of $v_{\rm sys} = 40.6 \pm 3.5$~km/s .
\item	An alternative scenario involving binarity cannot be entirely excluded. 
However, given that the primary is a blue supergiant exhibiting numerous persistent and transient pulsation modes, any orbital signal would be strongly masked by intrinsic atmospheric variability.
\end{itemize}

In conclusion, the persistent $P=16.46$\,d radial-velocity variability, accompanied by a weaker correlated signature in the line asymmetry, together with the 33-day photometric modulation, provides strong evidence that the observed variability is dominated by long-term non-radial pulsations. Binarity remains a possible alternative hypothesis, but it is not required to explain the observed spectroscopic and photometric behaviour.

%%%%%%%%%%%%%%%%%%%%%%%%%%%%%%%%%%%%%%%%%%
\vspace{6pt} 

%%%%%%%%%%%%%%%%%%%%%%%%%%%%%%%%%%%%%%%%%%

\supplementary{The following supporting information can be downloaded at:  \linksupplementary{s1}.}

\authorcontributions{Conceptualisation, V.C. and A.A.; methodology, V.C.; software, V.C.; validation, A.A., T.L. and I.K.; formal analysis, V.C.; investigation, V.C., A.A., V.M., A.K., T.E., S.P.D.B. and H.R.; resources, A.A. and T.L.; data curation, V.C.; writing---original draft preparation, V.C.; writing---review and editing, V.C., A.A., T.L., I.K., V.M., T.E. and S.P.D.B.; visualisation, V.C.; supervision, A.A. and T.L.; project administration, A.A. and T.L.; funding acquisition, A.A. and T.L. All authors have read and agreed to the published version of the manuscript.}

\funding{This work has used the ESA Estonia research infrastructure of Tartu Observatory, funded by the Estonian Research Council grants TT8 and TARISTU24-TK3, and by the project KosEST funded by the EU Regional Development Fund. The authors acknowledge support from the Estonian Research Council grant IUT-40. This project has received funding from the European Union's Horizon Europe research and innovation program under grant agreement No. 101079231 (EXOHOST), and from the United Kingdom Research and Innovation (UKRI) Horizon Europe Guarantee Scheme (grant number 10051045). This project has received funding from the European Union’s Framework Programme for Research and Innovation Horizon 2020 under the Marie Skłodowska-Curie Grant Agreement No.~823734 (POEMS). AA acknowledges support from the Estonian Research Council grant PRG~2159.}

\dataavailability{The full TO dataset containing one-dimensional continuum-normalised spectra of \rholeo\ centred on the photospheric \ion{He}{I} $\lambda$6678 line is available as Supplementary Material. Additional observational data from Tartu Observatory are available from the corresponding author upon reasonable request.
The SONG spectroscopic data used in this study can be obtained from the SODA archive: \url{https://soda.phys.au.dk/} (accessed on 26.08.2026).
The TESS light curves can be retrieved from the MAST archive:
\url{https://archive.stsci.edu/missions-and-data/tess} (accessed on 26.08.2026).}

\acknowledgments{This research made use of the NASA Astrophysics Data System (ADS) and of the SIMBAD database, which is operated at CDS, Strasbourg, France.
This paper includes data collected with the TESS mission, obtained from the MAST data archive at the Space Telescope Science Institute (STScI). Funding for US Institutions for the TESS mission is provided by the NASA Explorer Program. STScI is operated by the Association of Universities for Research in Astronomy, Inc., under NASA contract NAS 5–26555.
A.A. used ChatGPT (GPT-5.5, OpenAI) for assistance with LaTeX compilation issues and grammar checks of several sentences. The authors thank Üllar Kivila and Tarvi Verro for assistance with the spectroscopic observations and acquisition of part of the observational data used in this study.}

\conflictsofinterest{The authors declare no conflicts of interest. The funders had no role in the design of the study; in the collection, analyses, or interpretation of data; in the writing of the manuscript; or in the decision to publish the results.} 

%%%%%%%%%%%%%%%%%%%%%%%%%%%%%%%%%%%%%%%%%%

\appendixtitles{yes} 
\appendixstart
\appendix
\section[\appendixname~\thesection]{Spectroscopic and Photometric Time Series}\label{timeseries}

This appendix presents a set of five figures illustrating the results of the data analysis
and the resulting time-series behaviour. 

 \begin{figure}[H]
    \isPreprints{\centering}{}
    \includegraphics[width=\linewidth]{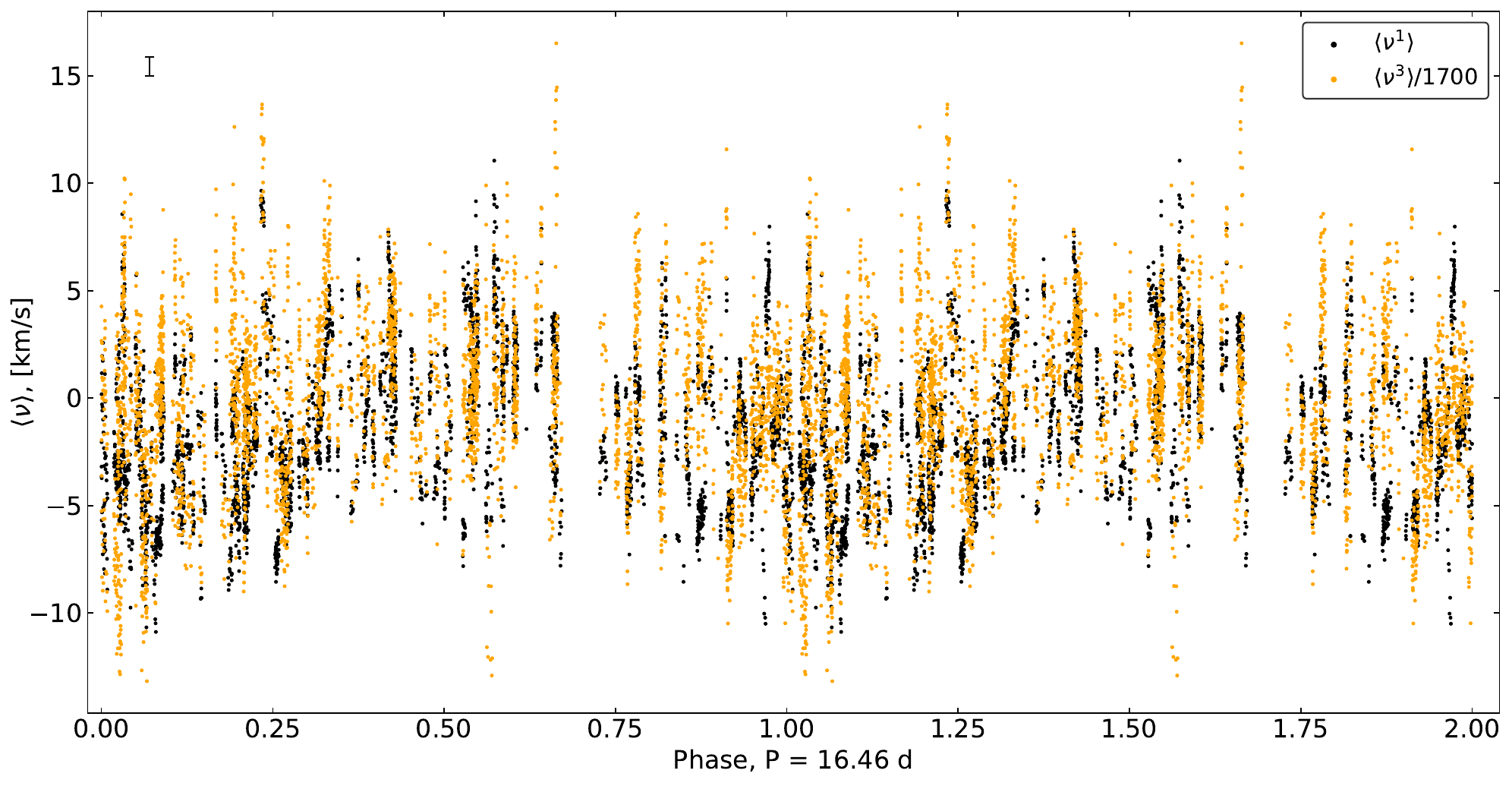}
    \caption{Phase diagram for the first $\langle v^1 \rangle$, [km/s] and central third $\langle v^3 \rangle$, [km/s]$^3$ moments of the \ion{He}{I}\,$\lambda$6678 line profile during an 11.5-year period of observations at TO (21~January~2014 to 1~June~2025, 3492 spectra in total) with period 16.46 days. Typical error bar is shown in the top left~corner.}
    \label{fig:M1andM3-phase-third}
 \end{figure}
\vspace{-6pt}
\begin{figure}[H]
    \isPreprints{\centering}{}
    \includegraphics[width=\linewidth]{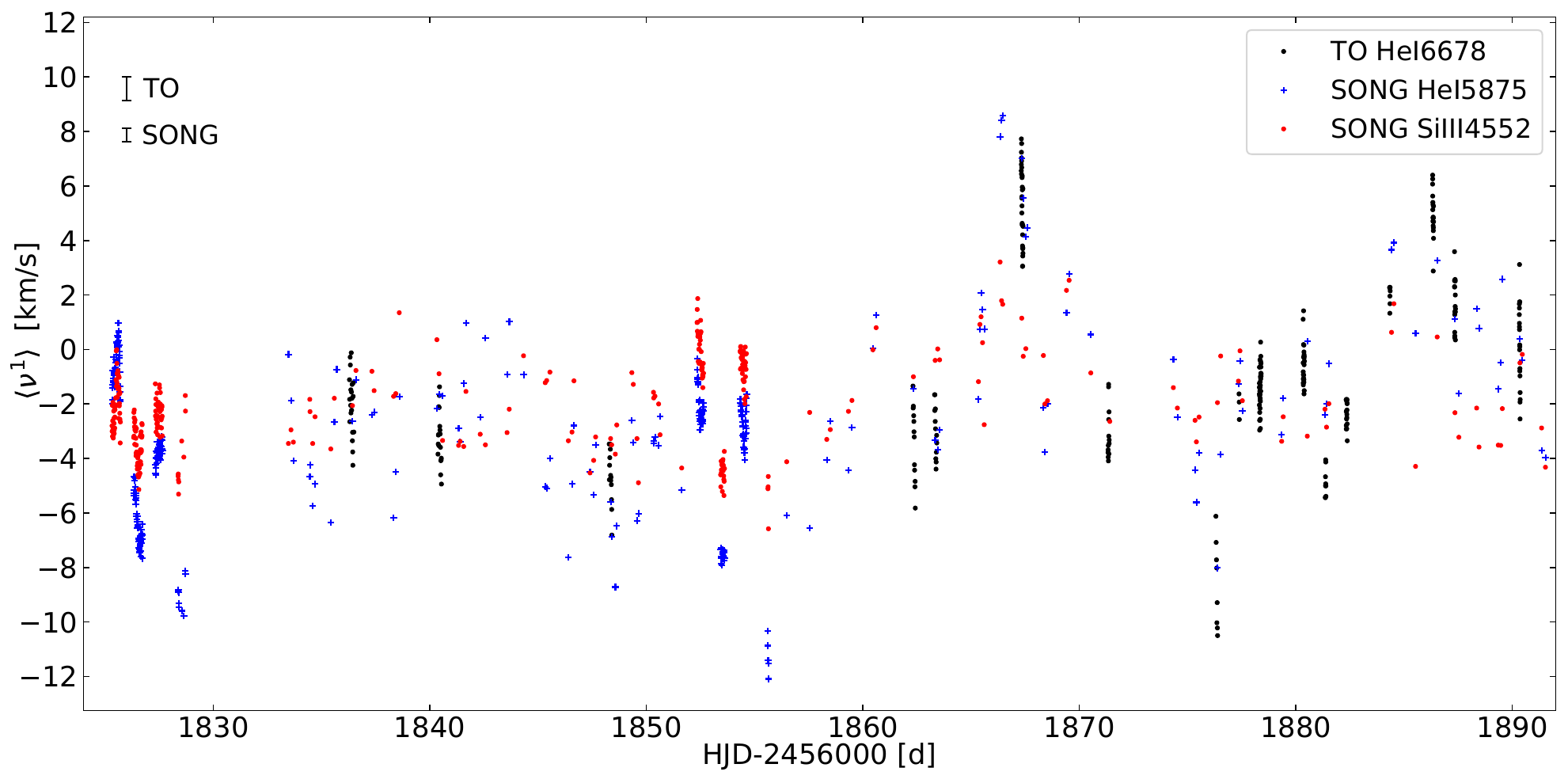}
\caption{The first moment $\langle v^1 \rangle$ (radial velocity) calculated for TO spectra, \ion{He}{I}\,$\lambda$6678 spectral line, and for SONG spectra, \ion{Si}{III} $\lambda$4552 and \ion{He}{I} $\lambda$5875 lines, respectively. The most dense part of the season 2017 is shown. The zero level corresponds to  $v_0=42$ km/s. Typical error bars are shown in the top left corner.}
\label{fig:TO&SONG-short}
\end{figure}

\begin{figure}[H]
    \isPreprints{\centering}{}
    \includegraphics[width=0.85\linewidth]{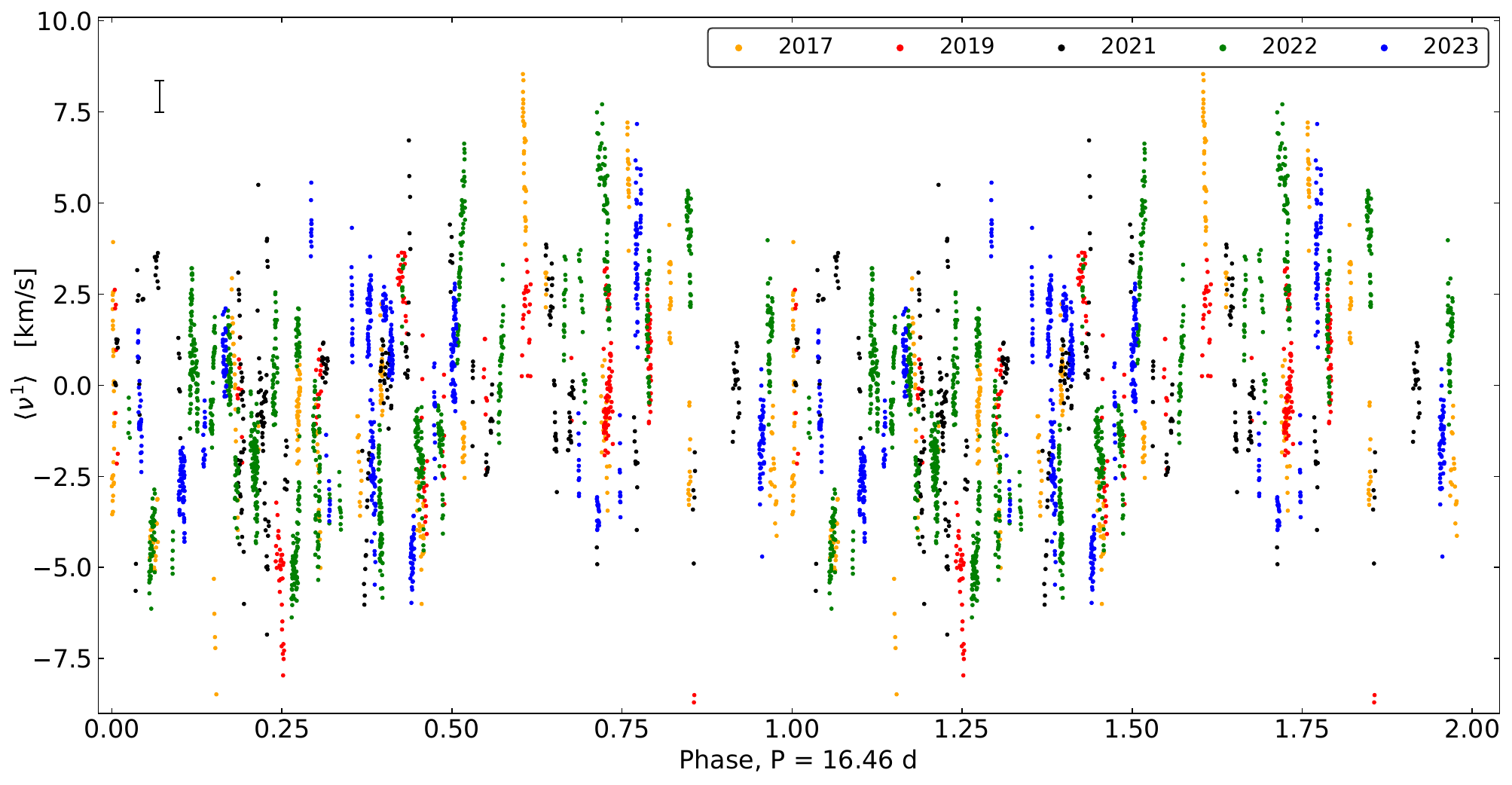}
    \caption{Phase diagram for the first $\langle v^1 \rangle$ moment of the \ion{He}{I}\,$\lambda$6678 line profile with period 16.464~days. Five TO observing seasons with the best observational coverage are shown. Typical error bar is shown in left upper corner. Zero level corresponds to $v_0=42$ km/s.}
    \label{fig:Chosen_phase}
\end{figure}
\vspace{-6pt}
\begin{figure}[H]
    \isPreprints{\centering}{}
    \includegraphics[width=0.85\linewidth]{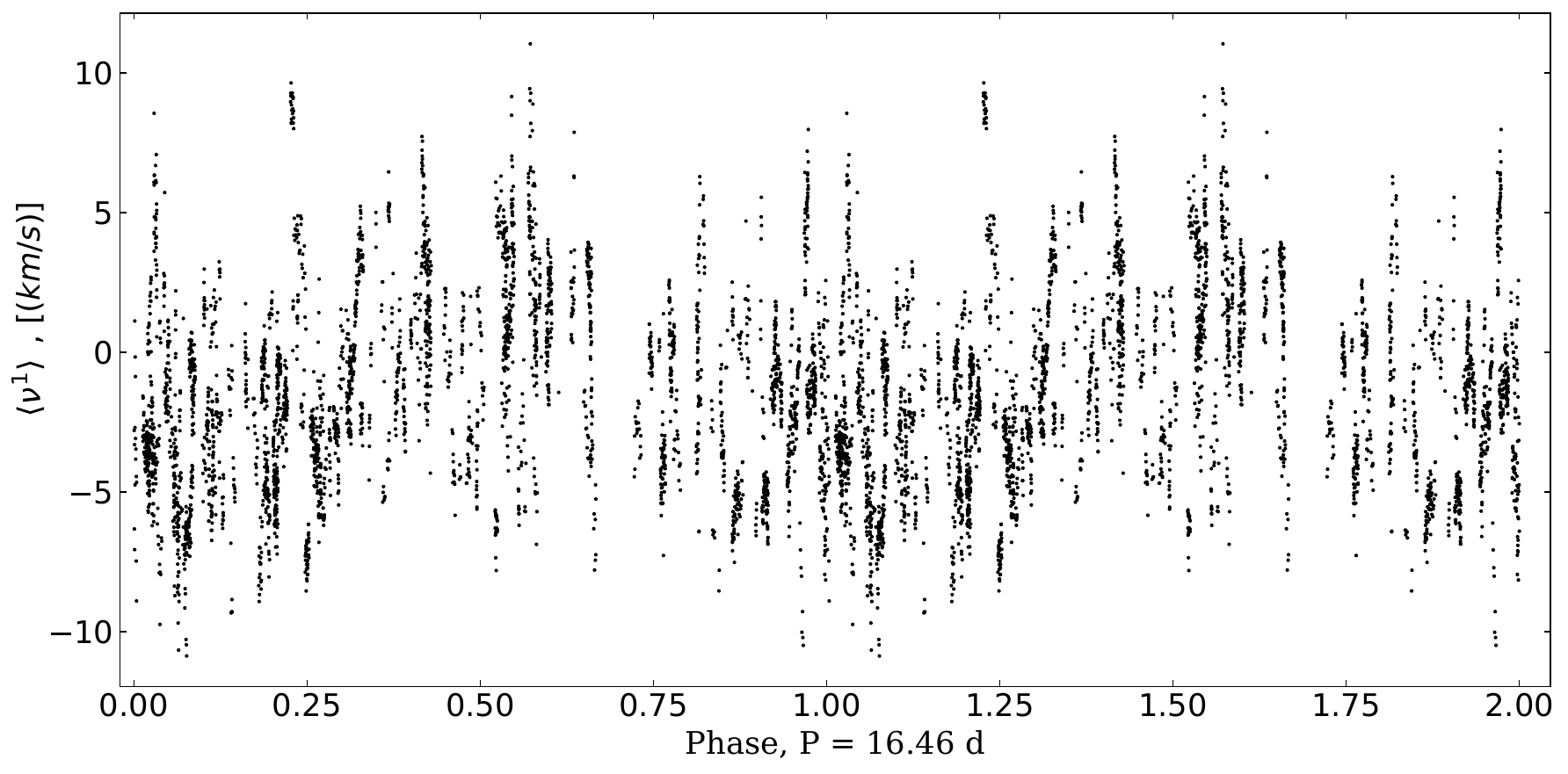}

    \includegraphics[width=0.85\linewidth]{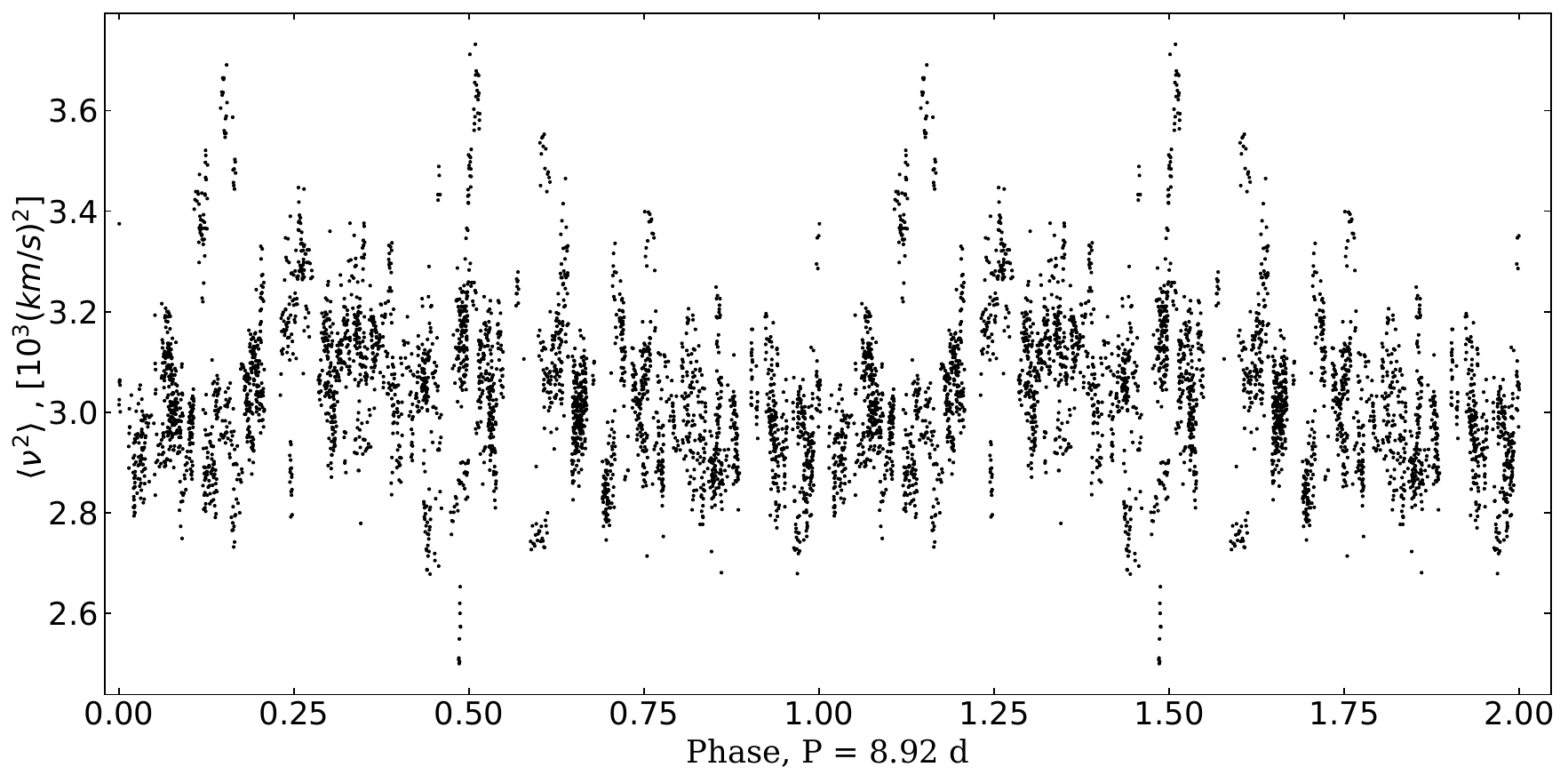}

    \caption{Results of the first pre-whitening step of the GLSp analysis for the $\langle v^1 \rangle$ (\textbf{upper panel}) and $\langle v^2 \rangle$ (\textbf{lower panel}) of the \ion{He}{I}\,$\lambda$6678 line profile. The dominant periods are 16.46 days \hbox{($k = 1.96 \,\mathrm{km/s}$)} and 8.92 days ($k = 0.74 \cdot 10^{3}\,\mathrm{km/s}$) for the first and second moments, respectively. The analysis is based on the full TO spectroscopic dataset spanning 4149.8 days. Zero level in the upper panel corresponds to $v_0=42$\,km/s.}
    \label{fig:M1+M2-All}
\end{figure}

\section[\appendixname~\thesection]{WWZ Analysis}\label{wwz-all}

\textls[-15]{This appendix presents the results of the WWZ analysis of the TO spectroscopic data on a season-by-season basis. To illustrate the temporal behaviour of the detected frequency components on intra-seasonal timescales, we first present WWZ scalograms for the first three normalised moments obtained for the 2022 observing season, which represents the most densely sampled part of our dataset. This is followed by a series of 11 figures showing the WWZ analysis of the first normalised moment, $\langle v^1 \rangle$, for each individual observing~season.}

\begin{figure}[H]

\subfloat[First normalised moment $\langle v^1 \rangle$ (radial velocity).%
\label{fig:WWZ_M1_2022}]{
\includegraphics[width=0.85\linewidth]{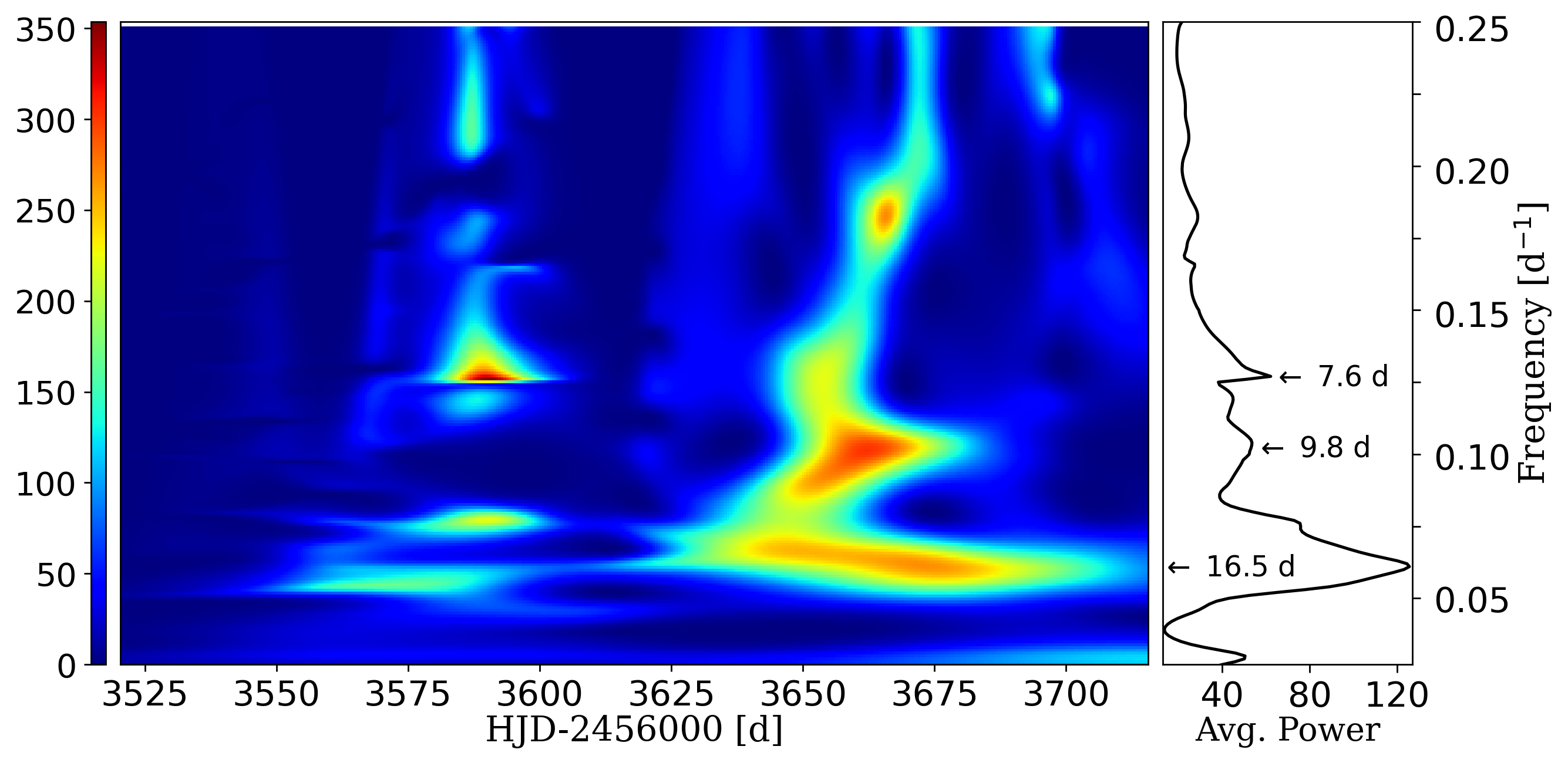}}%

\vspace{-13pt}

\subfloat[Second normalised moment $\langle v^2 \rangle$ (line width).%
\label{fig:WWZ_M2_2022}]{
\includegraphics[width=0.85\linewidth]{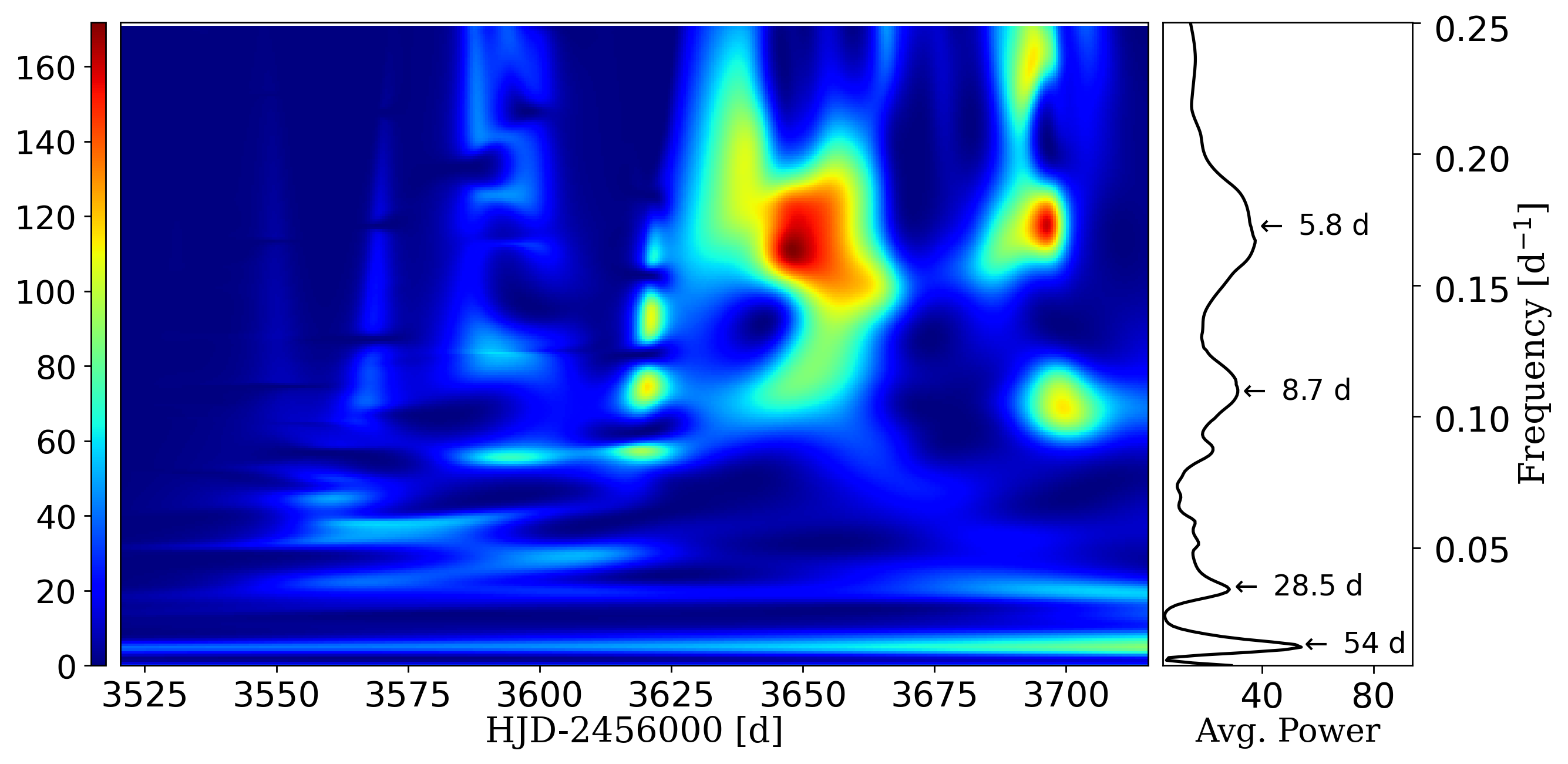}}%

\vspace{-13pt}

\subfloat[Third normalised central moment $\langle v^3 \rangle$ (skewness).%
\label{fig:WWZ_M3_2022}]{
\includegraphics[width=0.85\linewidth]{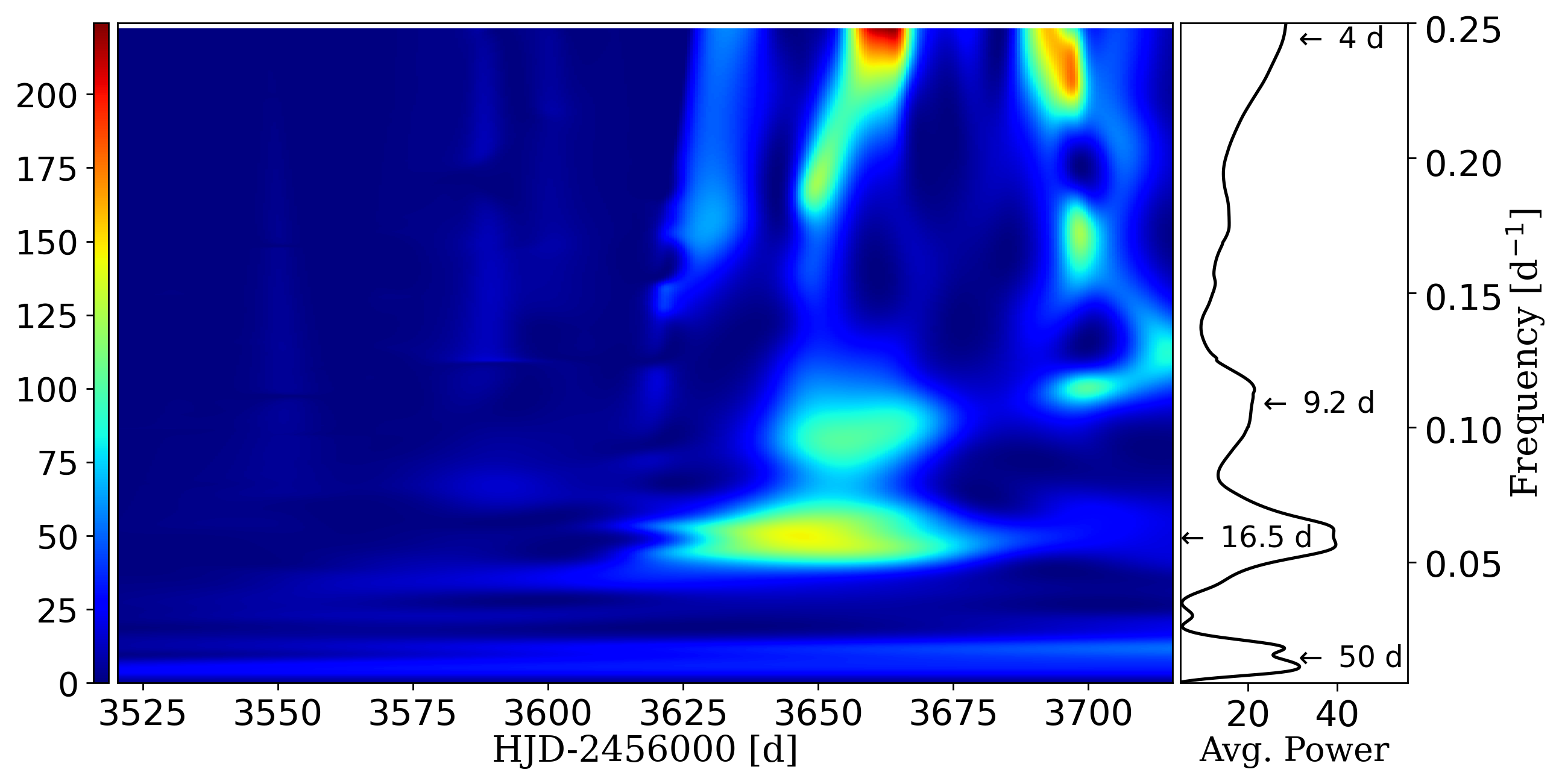}}

\caption{\textls[-20]{WWZ scalograms of the \ion{He}{I} $\lambda6678$ line moments for the 2022 observing season. From top to bottom: the first normalised moment $\langle v^1 \rangle$ (radial velocity), the second normalised moment $\langle v^2 \rangle$ (line width), and the third normalised central moment $\langle v^3 \rangle$ (skewness). The left panels show the time-frequency maps with colour-coded wavelet power, while the right panels show the time-averaged wavelet power, with the most significant periods indicated. The right-hand frequency axis applies to both panels.}}
\label{fig:WWZ_moments_2022}
\end{figure}

\begin{figure}[H]
\centering

\subfloat[2014 observing season.\label{fig:WWZ_2014}]{
\includegraphics[width=0.95\linewidth]{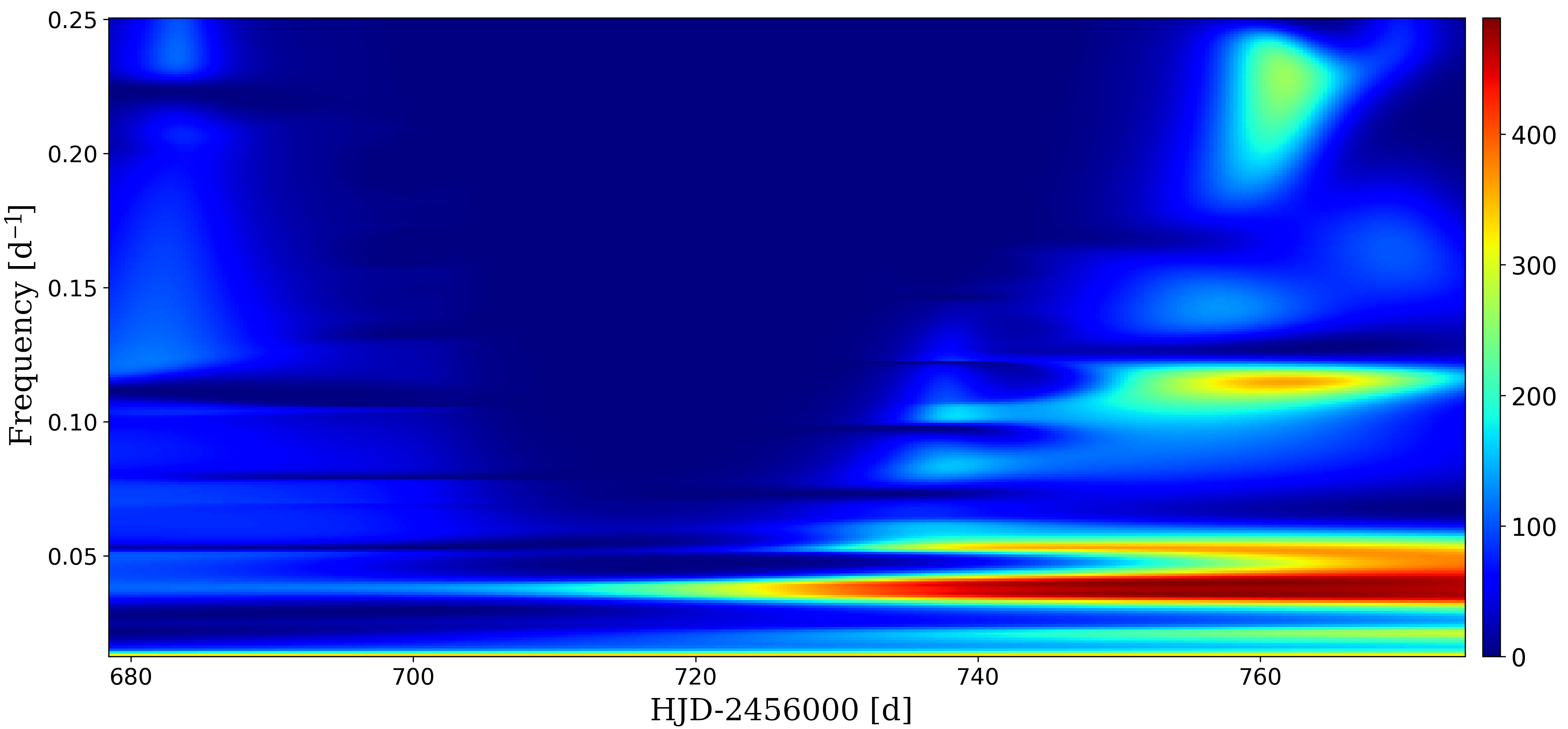}}

\vspace{-5pt}

\subfloat[2015 observing season.\label{fig:WWZ_2015}]{
\includegraphics[width=0.95\linewidth]{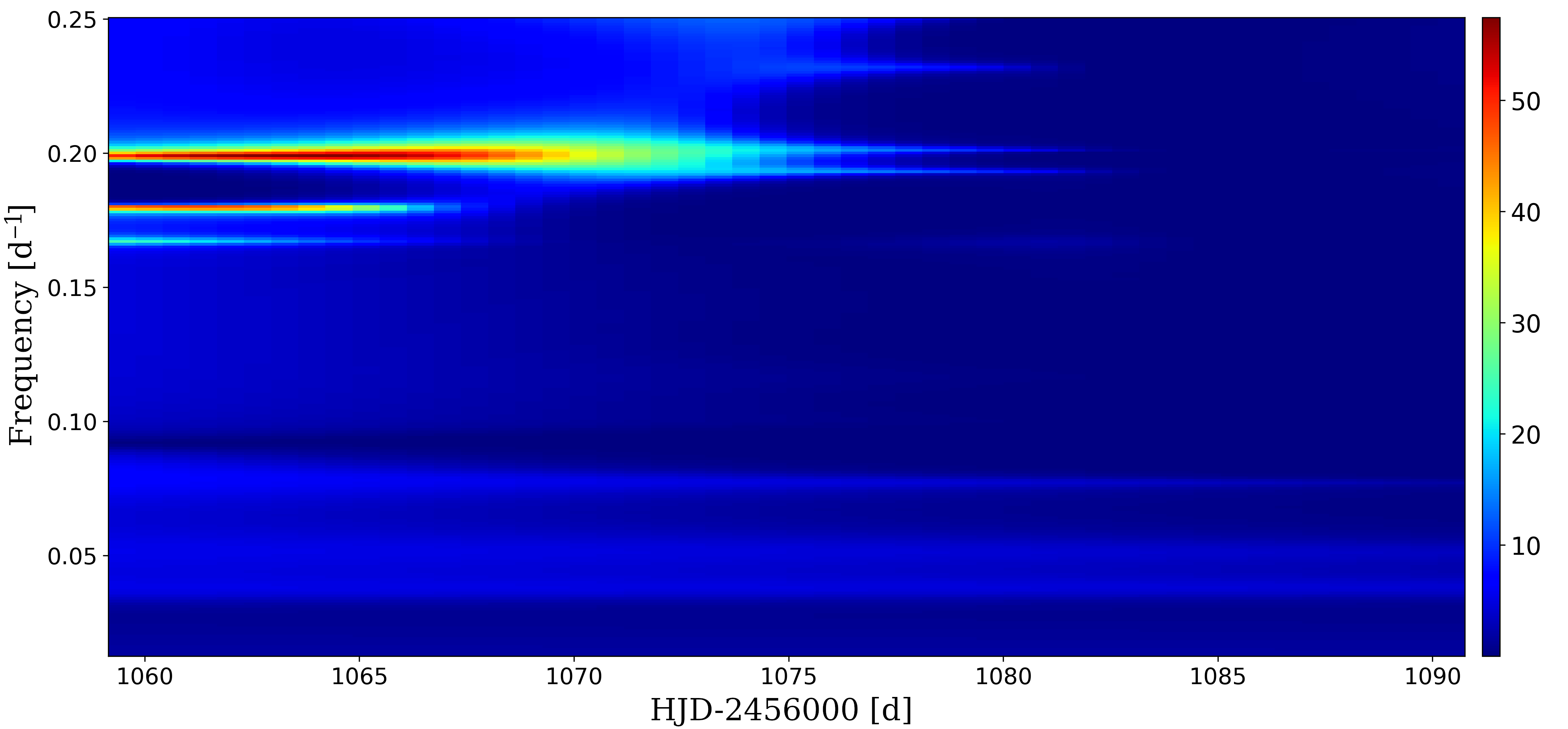}}

\vspace{-5pt}

\subfloat[2017 observing season.\label{fig:WWZ_2017}]{
\includegraphics[width=0.95\linewidth]{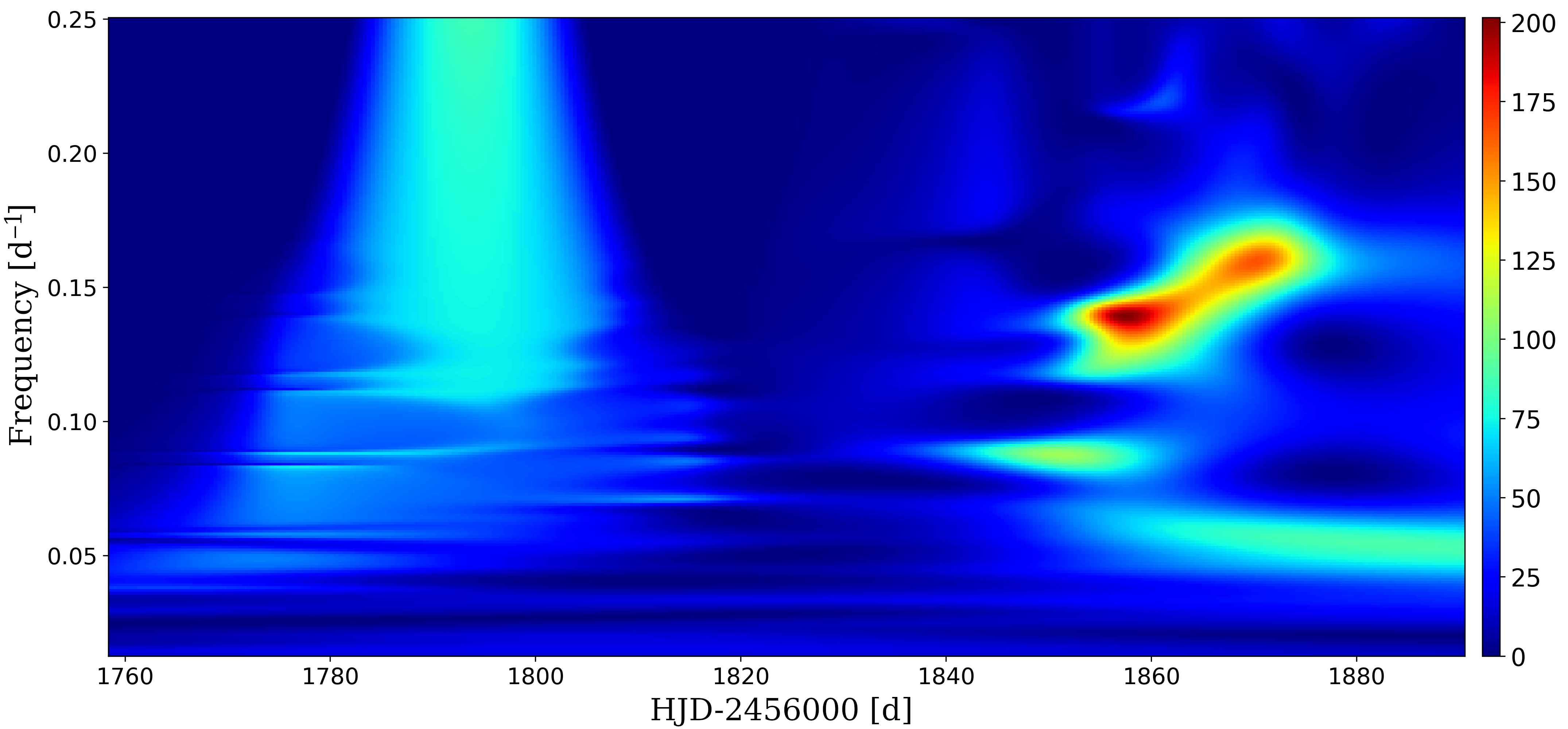}}

\caption{\textit{Cont}.}

\label{fig:WWZ_M1_seasons}

\end{figure}

\begin{figure}[H]
\ContinuedFloat
\centering

\subfloat[2018 observing season.\label{fig:WWZ_2018}]{
\includegraphics[width=0.95\linewidth]{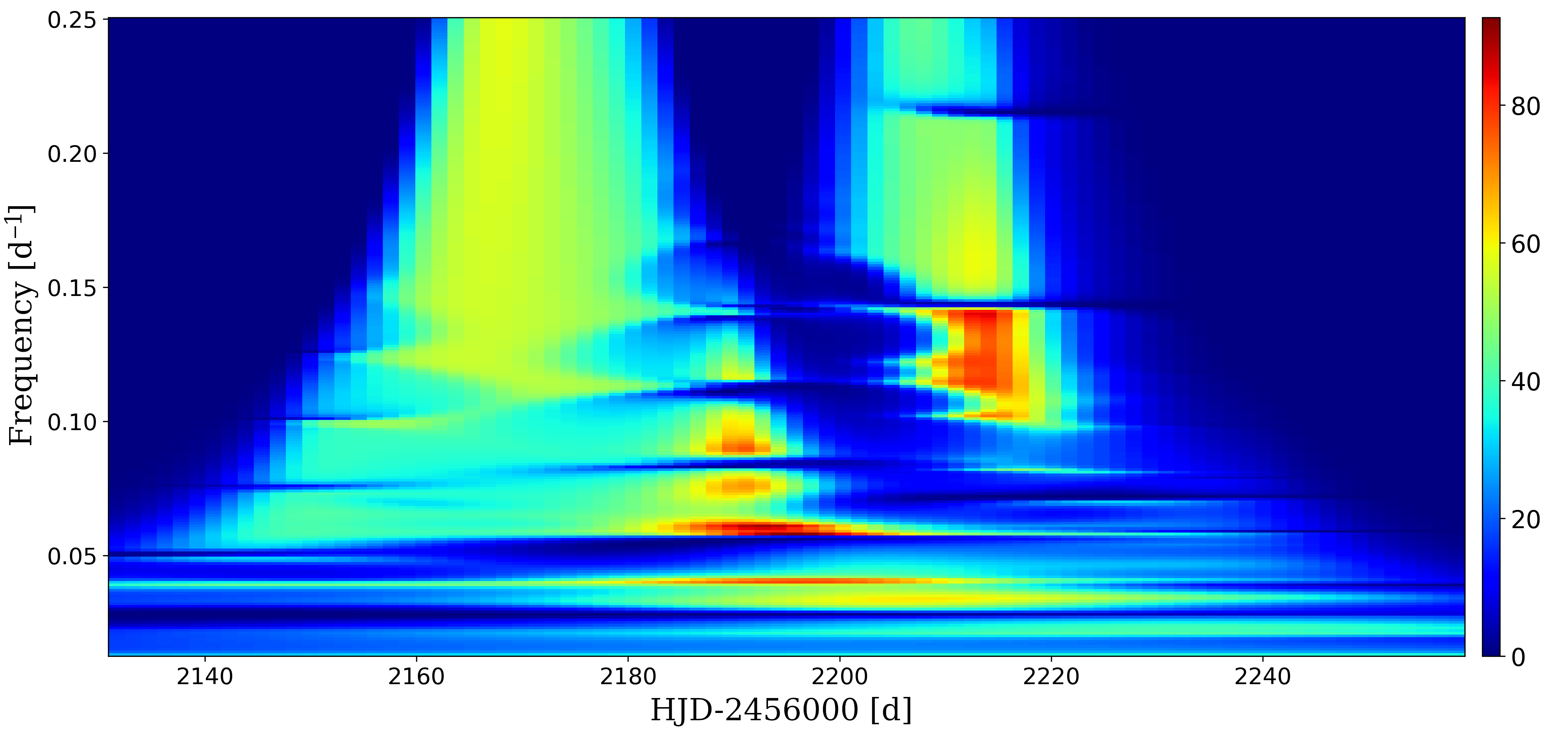}}

\vspace{-5pt}

\subfloat[2019 observing season.\label{fig:WWZ_2019}]{
\includegraphics[width=0.95\linewidth]{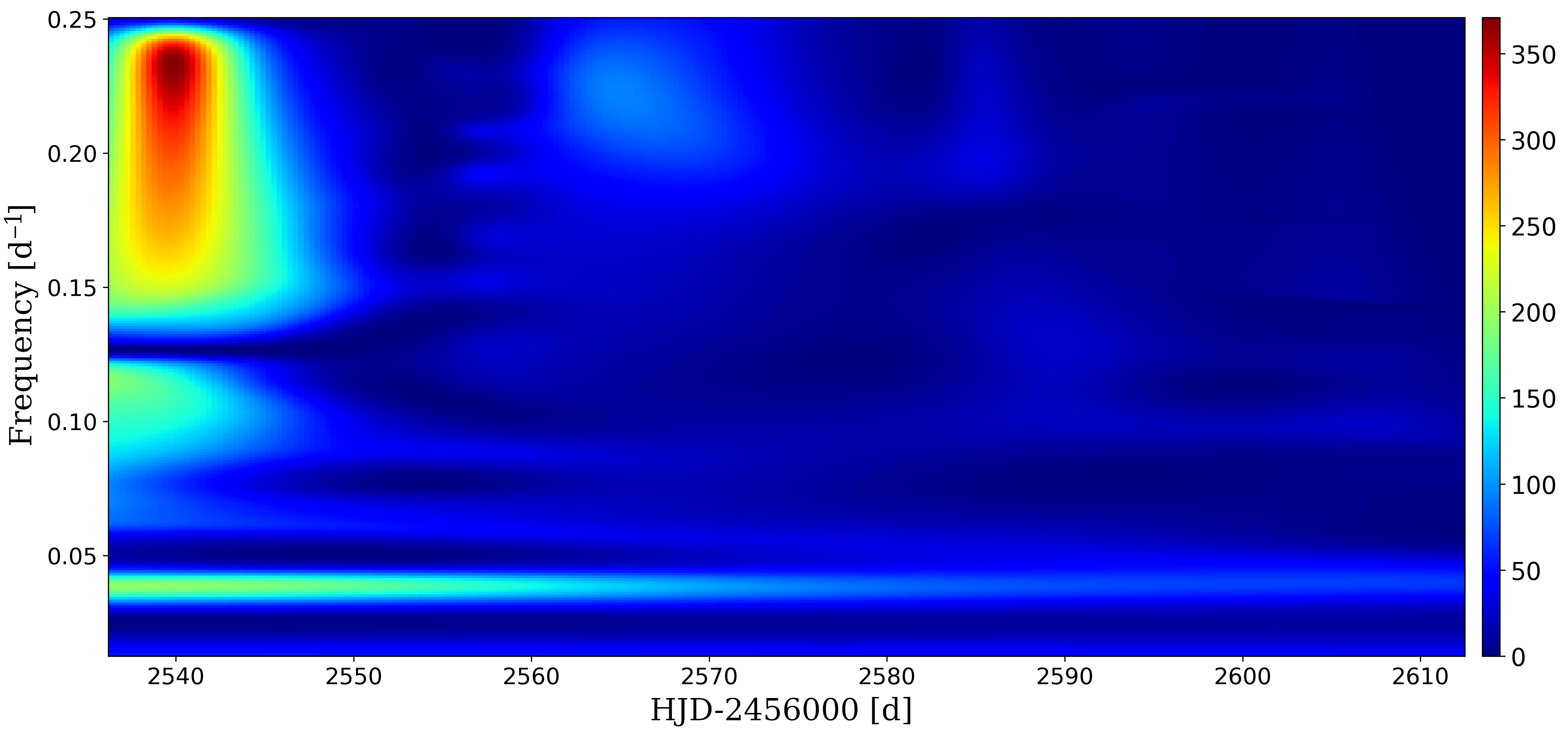}}

\vspace{-5pt}

\subfloat[2020 observing season.\label{fig:WWZ_2020}]{
\includegraphics[width=0.95\linewidth]{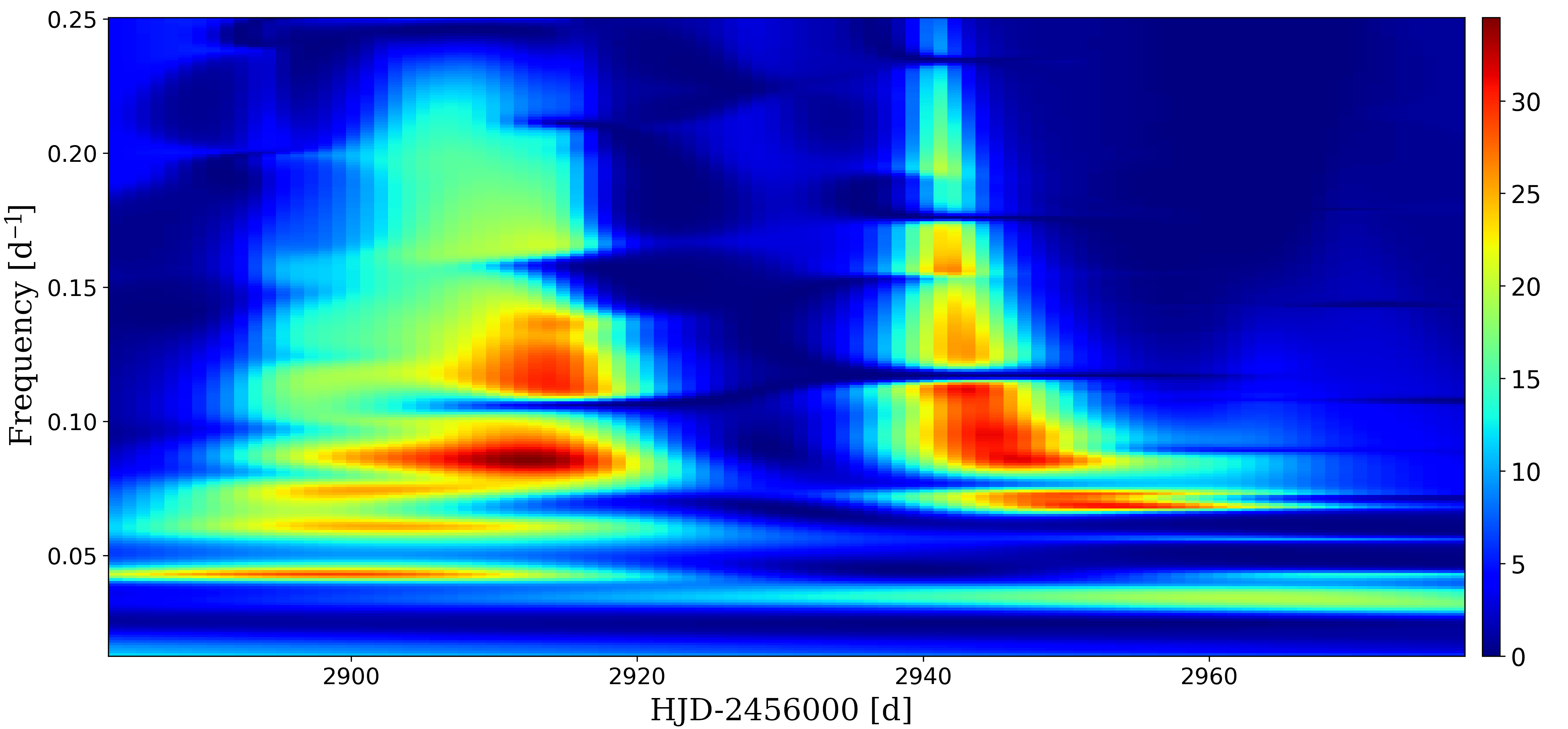}}

\caption{\textit{Cont}.}

\end{figure}

\begin{figure}[H]
\ContinuedFloat
\centering

\subfloat[2021 observing season.\label{fig:WWZ_2021}]{
\includegraphics[width=0.95\linewidth]{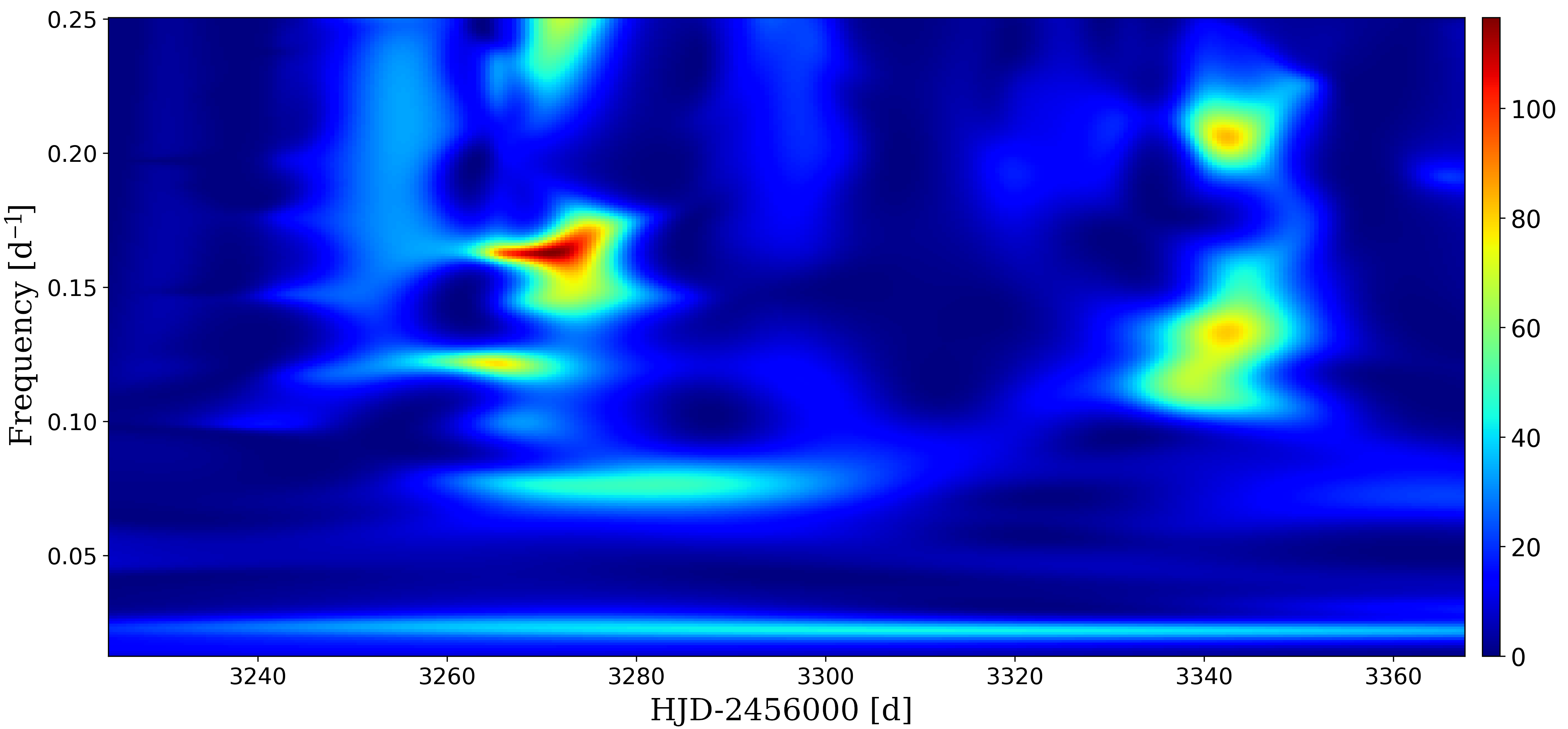}}

\vspace{-5pt}

\subfloat[2022 observing season.\label{fig:WWZ_2022}]{
\includegraphics[width=0.95\linewidth]{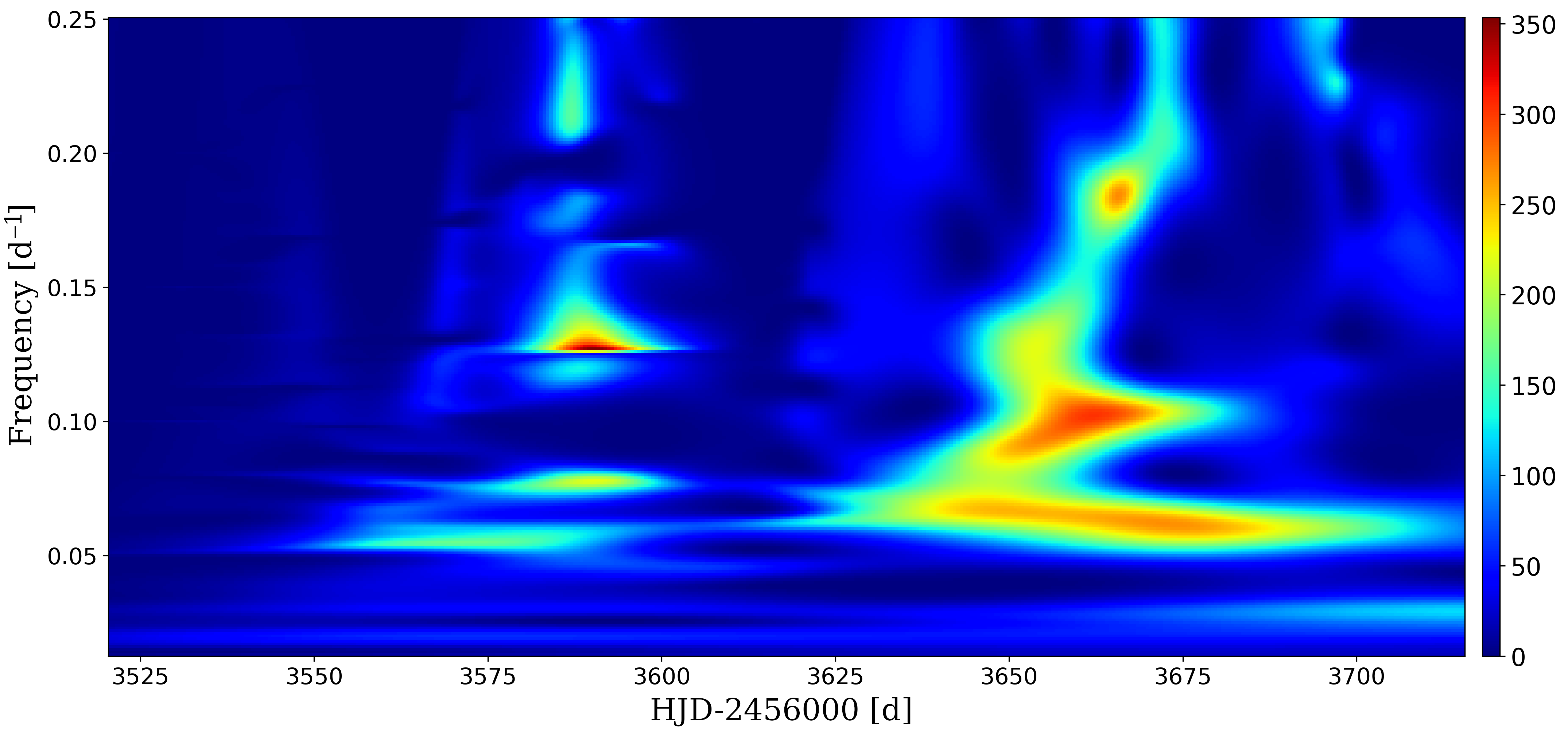}}

\vspace{-5pt}

\subfloat[2023 observing season.\label{fig:WWZ_2023}]{
\includegraphics[width=0.95\linewidth]{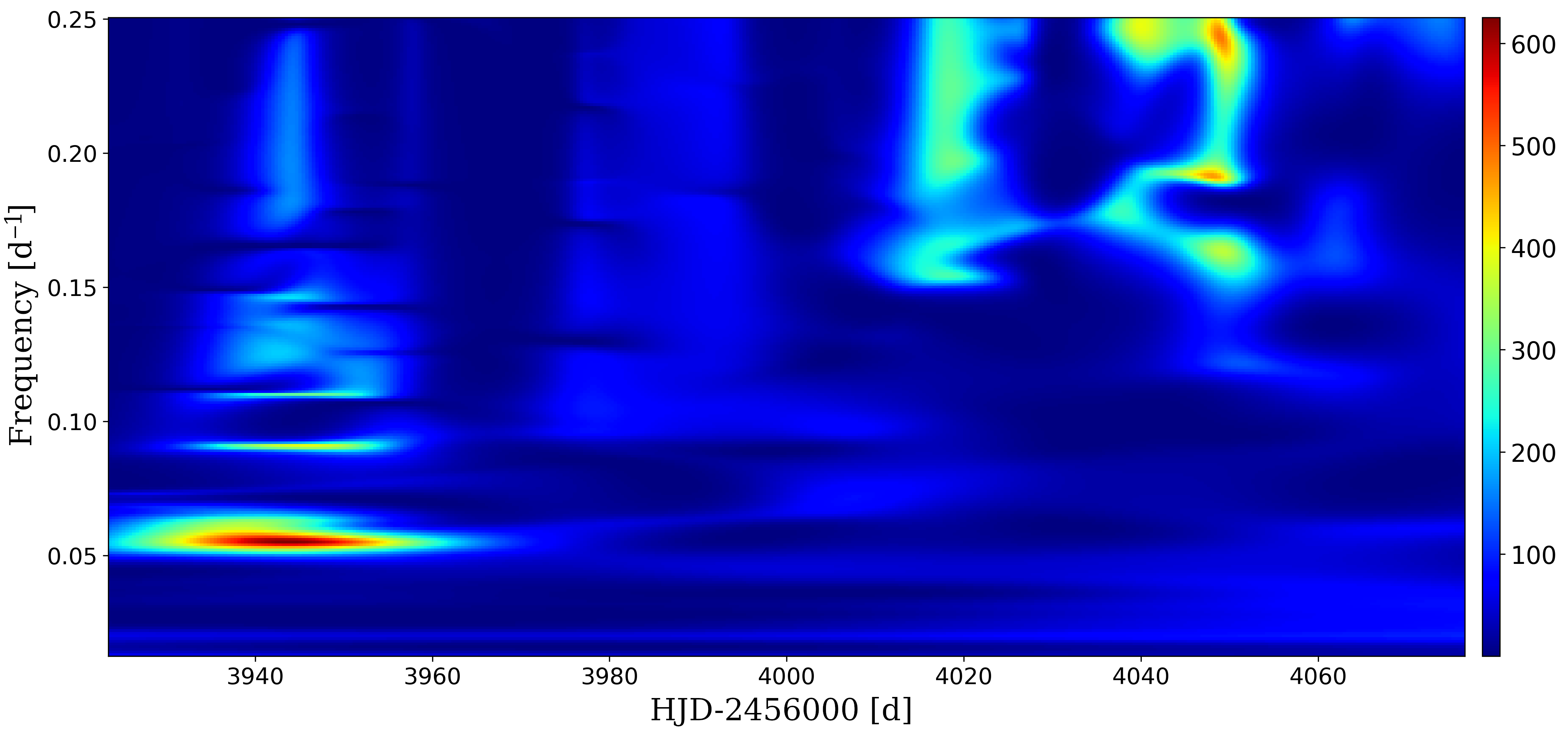}}

\caption{\textit{Cont}.}

\end{figure}

\begin{figure}[H]
\ContinuedFloat
\centering

\subfloat[2024 observing season.\label{fig:WWZ_2024}]{
\includegraphics[width=0.95\linewidth]{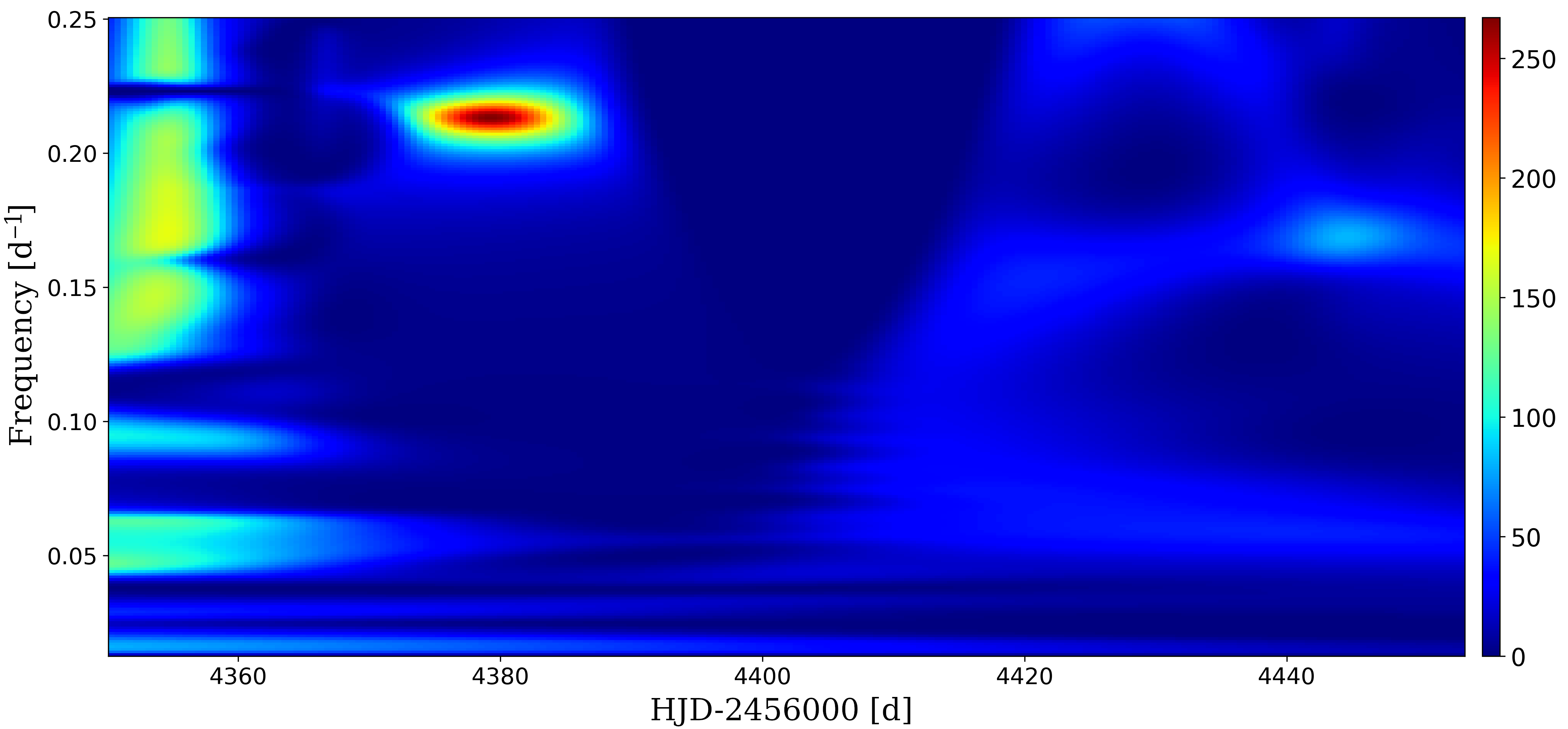}}

\vspace{-5pt}

\subfloat[2025 observing season.\label{fig:WWZ_2025}]{
\includegraphics[width=0.95\linewidth]{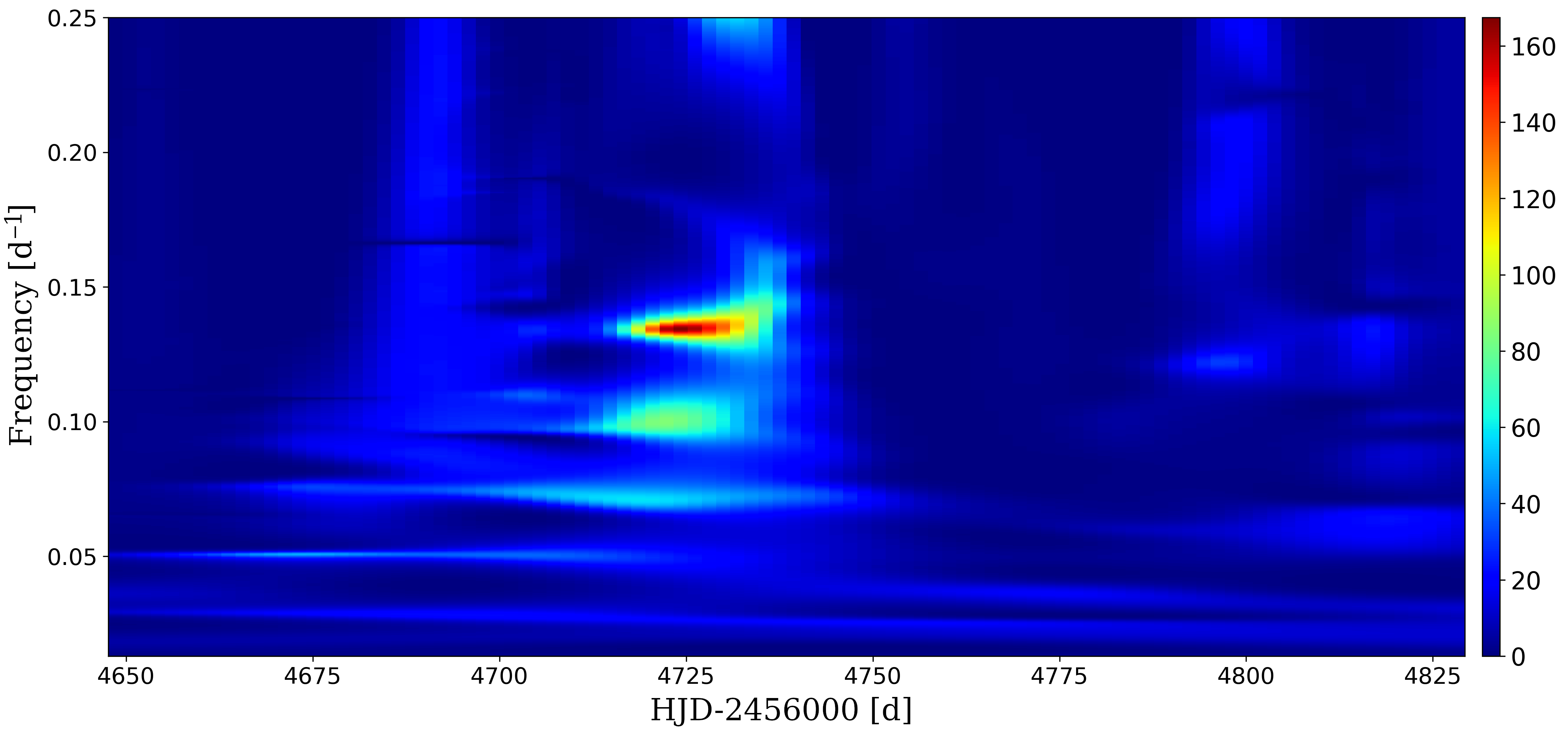}}

\caption{Season-by-season WWZ scalograms of the TO dataset for the first normalised moment, $\langle v^1 \rangle$ (radial velocity), of the \ion{He}{I} $\lambda$6678 line. Colour-coding presents the wavelet power.}

\end{figure}

\section[\appendixname~\thesection]{Log of Spectroscopic TO Observations}\label{S-appendixobs}
\vspace{-6pt}
\setcounter{table}{0}

\begin{table}[H]
\
\caption{TO spectroscopic observations from the 2014--2025 seasons. Number of spectra per night is shown in column N. S/N is measured near the \ion{He}{I} line at 6678\,\AA{}. S/N and exposure values are averaged per night. 3492 spectra in total.\label{tab:Spectra_sorted_full}}
	\begin{adjustwidth}{-\extralength}{0cm}
\setlength{\tabcolsep}{3pt}
\begin{tabularx}{\fulllength}{lccclccclccclccc}
\toprule
\textbf{Night} & \textbf{N} & \textbf{S/N} & \textbf{Exp. [s]} &
\textbf{Night} & \textbf{N} & \textbf{S/N} & \textbf{Exp. [s]} &
\textbf{Night} & \textbf{N} & \textbf{S/N} & \textbf{Exp. [s]} &
\textbf{Night} & \textbf{N} & \textbf{S/N} & \textbf{Exp. [s]} \\
\midrule
2014-01-20 & 2  & 430 & 360 & 2018-05-19 & 3  & 320 & 130 & 2022-04-09 & 16 & 310 & 170 & 2024-03-09 & 8  & 340 & 180  \\
2014-01-21 & 15 & 300 & 260 & 2019-02-21 & 23 & 520 & 720 & 2022-04-10 & 34 & 360 & 210 & 2024-03-10 & 10 & 370 & 180  \\
2014-01-30 & 24 & 440 & 400 & 2019-03-01 & 47 & 370 & 600 & 2022-04-11 & 25 & 330 & 160 & 2024-04-26 & 7  & 360 & 180  \\
2014-02-28 & 36 & 350 & 150 & 2019-03-25 & 9  & 600 & 690 & 2022-04-12 & 38 & 360 & 190 & 2024-04-28 & 30 & 310 & 160  \\
2014-04-10 & 58 & 210 & 90  & 2019-03-26 & 10 & 620 & 720 & 2022-04-16 & 6  & 330 & 250 & 2024-05-01 & 17 & 330 & 140  \\
2014-04-16 & 78 & 360 & 90  & 2019-03-27 & 14 & 620 & 710 & 2022-04-18 & 32 & 340 & 160 & 2024-05-02 & 13 & 310 & 130  \\
2014-04-18 & 43 & 290 & 80  & 2019-03-29 & 22 & 590 & 700 & 2022-04-19 & 30 & 330 & 200 & 2024-05-03 & 8  & 350 & 180  \\
2014-04-25 & 46 & 320 & 90  & 2019-03-30 & 8  & 310 & 170 & 2022-04-21 & 20 & 340 & 160 & 2024-05-04 & 22 & 300 & 180  \\
2014-04-26 & 32 & 330 & 80  & 2019-03-31 & 8  & 600 & 670 & 2022-05-04 & 25 & 260 & 200 & 2024-05-08 & 10 & 370 & 190  \\
2015-02-05 & 3  & 340 & 170 & 2019-04-01 & 22 & 580 & 680 & 2022-05-06 & 9  & 330 & 185 & 2024-05-11 & 2  & 280 & 240  \\
2015-02-10 & 23 & 370 & 280 & 2019-04-02 & 2  & 430 & 400 & 2022-05-08 & 10 & 360 & 250 & 2024-05-16 & 5  & 330 & 260    \\
2015-02-16 & 20 & 400 & 280 & 2019-04-03 & 21 & 510 & 710 & 2022-05-09 & 6  & 340 & 165 & 2024-05-17 & 10 & 280 & 415    \\
2015-03-08 & 4  & 340 & 280 & 2019-04-04 & 40 & 340 & 160 & 2022-05-13 & 9  & 340 & 210 & 2024-05-21 & 5  & 330 & 220    \\

\bottomrule
\end{tabularx}
	\end{adjustwidth}
\end{table}

\begin{table}[H]\ContinuedFloat
\
\caption{{\em Cont.}}
	\begin{adjustwidth}{-\extralength}{0cm}
\setlength{\tabcolsep}{3pt}
\begin{tabularx}{\fulllength}{lccclccclccclccc}
\toprule
\textbf{Night} & \textbf{N} & \textbf{S/N} & \textbf{Exp. [s]} &
\textbf{Night} & \textbf{N} & \textbf{S/N} & \textbf{Exp. [s]} &
\textbf{Night} & \textbf{N} & \textbf{S/N} & \textbf{Exp. [s]} &
\textbf{Night} & \textbf{N} & \textbf{S/N} & \textbf{Exp. [s]} \\
\midrule
2017-01-04 & 11 & 330 & 350 & 2019-04-15 & 15 & 550 & 680 & 2022-05-15 & 10 & 270 & 220 & 2024-05-22 & 5  & 330 & 190    \\
2017-02-07 & 25 & 410 & 600 & 2019-04-24 & 7  & 580 & 710 & 2022-12-09 & 40 & 330 & 130 & 2024-12-03 & 5  & 360 & 290    \\
2017-02-09 & 21 & 390 & 460 & 2019-05-08 & 2  & 610 & 790 & 2022-12-24 & 22 & 400 & 300 & 2024-12-12 & 4  & 340 & 280    \\
2017-03-23 & 21 & 450 & 640 & 2020-02-03 & 11 & 650 & 730 & 2023-01-05 & 16 & 370 & 240 & 2025-01-12 & 4  & 420 & 380    \\
2017-03-27 & 18 & 450 & 660 & 2020-02-04 & 12 & 560 & 660 & 2023-01-21 & 60 & 340 & 190 & 2025-01-18 & 7  & 340 & 260    \\
2017-04-04 & 13 & 520 & 760 & 2020-02-24 & 17 & 560 & 710 & 2023-02-11 & 55 & 350 & 140 & 2025-02-08 & 10 & 340 & 190    \\
2017-04-18 & 12 & 530 & 640 & 2020-02-28 & 14 & 400 & 740 & 2023-02-13 & 30 & 360 & 190 & 2025-02-19 & 7  & 380 & 240  \\
2017-04-19 & 12 & 440 & 690 & 2020-03-14 & 14 & 510 & 740 & 2023-02-22 & 30 & 390 & 210 & 2025-02-23 & 7  & 380 & 460  \\
2017-04-23 & 30 & 320 & 200 & 2020-03-23 & 18 & 430 & 730 & 2023-03-09 & 52 & 290 & 140 & 2025-02-24 & 11 & 340 & 230  \\
2017-04-27 & 13 & 260 & 250 & 2020-04-09 & 5  & 510 & 630 & 2023-03-15 & 10 & 330 & 260 & 2025-02-25 & 2  & 370 & 250  \\
2017-05-02 & 35 & 220 & 180 & 2020-05-07 & 6  & 490 & 780 & 2023-03-16 & 44 & 230 & 140 & 2025-03-16 & 1  & 480 & 800  \\
2017-05-03 & 10 & 230 & 180 & 2021-11-01 & 5  & 440 & 640 & 2023-03-29 & 36 & 370 & 150 & 2025-03-19 & 1  & 390 & 740  \\
2017-05-04 & 36 & 270 & 160 & 2021-11-21 & 8  & 530 & 790 & 2023-04-02 & 39 & 350 & 140 & 2025-03-21 & 1  & 420 & 770  \\
2017-05-06 & 24 & 260 & 170 & 2021-12-09 & 3  & 310 & 500 & 2023-04-07 & 20 & 340 & 180 & 2025-03-22 & 5  & 340 & 210  \\
2017-05-07 & 9  & 330 & 230 & 2022-01-01 & 26 & 400 & 620 & 2023-04-08 & 44 & 330 & 90  & 2025-03-26 & 1  & 560 & 690  \\
2017-05-08 & 18 & 370 & 210 & 2022-01-11 & 9  & 460 & 390 & 2023-04-19 & 40 & 370 & 150 & 2025-03-29 & 4  & 310 & 190  \\
2017-05-10 & 7  & 300 & 300 & 2022-01-17 & 18 & 500 & 580 & 2023-04-23 & 10 & 350 & 150 & 2025-04-02 & 1  & 510 & 1400 \\
2017-05-12 & 18 & 350 & 200 & 2022-01-25 & 30 & 340 & 240 & 2023-04-24 & 11 & 320 & 120 & 2025-04-06 & 1  & 660 & 730  \\
2017-05-13 & 18 & 260 & 190 & 2022-02-26 & 50 & 360 & 220 & 2023-05-03 & 11 & 340 & 150 & 2025-04-26 & 1  & 700 & 990  \\
2017-05-16 & 19 & 250 & 160 & 2022-02-27 & 20 & 390 & 185 & 2023-05-04 & 19 & 330 & 110 & 2025-05-01 & 1  & 480 & 600  \\
2018-01-12 & 4  & 530 & 750 & 2022-03-22 & 36 & 270 & 190 & 2023-05-06 & 15 & 370 & 130 & 2025-05-03 & 7  & 310 & 160  \\
2018-02-21 & 13 & 480 & 850 & 2022-03-30 & 49 & 330 & 170 & 2023-05-11 & 10 & 290 & 140 & 2025-05-10 & 8  & 370 & 230  \\
2018-02-22 & 13 & 480 & 560 & 2022-03-31 & 37 & 350 & 150 & 2024-02-09 & 20 & 390 & 290 & 2025-05-12 & 1  & 580 & 170  \\
2018-02-22 & 44 & 330 & 250 & 2022-04-01 & 2  & 250 & 275 & 2024-02-18 & 8  & 210 & 230 & 2025-05-31 & 5  & 340 & 260  \\
2018-03-29 & 14 & 690 & 850 & 2022-04-02 & 36 & 350 & 180 & 2024-03-05 & 30 & 380 & 140 & 2025-06-01 & 1  & 520 & 240    \\
2018-04-12 & 6  & 520 & 850 & 2022-04-03 & 24 & 350 & 160 & 2024-03-07 & 10 & 250 & 210 &            &    &     &      \\
\bottomrule
\end{tabularx}
	\end{adjustwidth}
\end{table}

%%%%%%%%%%%%%%%%%%%%%%%%%%%%%%%%%%%%%%%%%%
%\isPreprints{}{% This command is only used for ``preprints''.
\begin{adjustwidth}{-\extralength}{0cm}
%} % If the paper is ``preprints'', please uncomment this parenthesis.
\printendnotes[custom] % Un-comment to print a list of endnotes

\reftitle{References}

\PublishersNote{}
%\isPreprints{}{% This command is only used for ``preprints''.
\end{adjustwidth}
%} % If the paper is ``preprints'', please uncomment this parenthesis.

\end{document}